\documentclass[ba]{imsart}

\usepackage{amsthm,amsmath,amsfonts,amssymb,bm}
\usepackage{multirow}
\usepackage{xcolor}
\usepackage{mathtools}
\usepackage{comment}
\usepackage{graphicx}
\usepackage{subcaption}
\usepackage[utf8]{inputenc}
\usepackage{xstring}
\usepackage{nameref}
\usepackage{natbib}
\usepackage{tabularx}
\usepackage{listings}
\usepackage{booktabs}
\usepackage{multicol}
\usepackage{enumitem}
\usepackage{xr}
\usepackage{hyperref}

\usepackage{tikz}
\usetikzlibrary{bayesnet}

\makeatletter
\newcommand*{\addFileDependency}[1]{
	\typeout{(#1)}
	\@addtofilelist{#1}
	\IfFileExists{#1}{}{\typeout{No file #1.}}
}
\makeatother

\DeclareMathOperator*{\argmax}{arg\,max}

\newtheorem{result}{Result}
\newtheorem{theorem}{Theorem}
\newtheorem{definition}{Definition}
\newtheorem{lemma}{Lemma}

\DeclarePairedDelimiterX{\infdivx}[2]{[}{]}{%
	#1\;\delimsize\|\;#2%
}

\usepackage{framed}

\newcommand{\kl}[2]{\mathrm{KL}\infdivx{#1}{#2}}

\usepackage{algpseudocode}
\usepackage{algorithm}

\algtext*{EndFor}

\begin{document}
	
	\begin{frontmatter}
		\title{Fast and Accurate Estimation of Non-Nested Binomial Hierarchical Models Using Variational Inference}
		\runtitle{Fast Estimation of Binomial Hierarchical Models}
		\thankstext{T1}{
			Open-source software to implement the models in this paper in \texttt{R} can be downloaded from \href{https://github.com/mgoplerud/vglmer}{github.com/mgoplerud/vglmer}. I thank the anonymous reviewers, Michael Auslen, Naoki Egami, Shusei Eshima, Justin Grimmer, June Hwang, Kosuke Imai, Pierre Jacob, Gary King, Shiro Kuriwaki, Marc Ratkovic, Sun Young Park, Casey Petroff, Marc Ratkovic, Tyler Simko, Diana Stanescu, Dustin Tingley, Soichiro Yamauchi, and participants at PolMeth 2020 and the University of Pittsburgh's Statistics Seminar for helpful comments on earlier versions of this paper. All remaining errors are my own.}
		
		\begin{aug}
			\author{\fnms{Max} \snm{Goplerud}}
			\runauthor{M. GOPLERUD}
			
			\address[addr1]{Assistant Professor, Department of Political Science, University of Pittsburgh; \href{mailto:mgoplerud@pitt.edu}{mgoplerud@pitt.edu}.}

			
		\end{aug}
		
		\begin{abstract}
			
			Non-linear hierarchical models are commonly used in many disciplines. However, inference in the presence of  non-nested effects and on large datasets is challenging and computationally burdensome. This paper provides two contributions to scalable and accurate inference. First, I derive a new mean-field variational algorithm for estimating binomial logistic hierarchical models with an arbitrary number of non-nested random effects. Second, I propose ``marginally augmented variational Bayes'' (MAVB) that further improves the initial approximation through a step of Bayesian post-processing. I prove that MAVB provides a guaranteed improvement in the approximation quality at low computational cost and induces dependencies that were assumed away by the initial factorization assumptions. 
			
			I apply these techniques to a study of voter behavior using a high-dimensional application of the popular approach of multilevel regression and post-stratification (MRP). Existing estimation took hours whereas the algorithms proposed run in minutes. The posterior means are well-recovered even under strong factorization assumptions. Applying MAVB further improves the approximation by partially correcting the under-estimated variance. The proposed methodology is implemented in an open source software package. 
		\end{abstract}
		
	\end{frontmatter}
	
	\section{Introduction and Motivating Example}
	
Hierarchical models, often known as multilevel, mixed, or random effects models, are ubiquitous in the social sciences (\citealt{gelman2006multi,skrondal2008multilevel}). In political science alone, these models are used for addressing unobserved heterogeneity, explicitly modeling dependence between observations, allowing effects to vary across space or time, and many other applications (e.g. \citealt{clark2015fe,bell2015explaining,steenbergen2002multilevel,stegmueller2013multilevel}). They are also popular in other fields such as educational research and psychology.

The benefits and challenges of these models can be illustrated by an increasingly popular application for survey research in social science: Multilevel Regression and Post-Stratification (MRP; \citealt{gelman1997mrp,park2004mrp,gao2019improving}). Described in more detail in Section~\ref{section:gg_app}, the core purpose of this method is to extrapolate outcomes from nationally representative surveys to small geographic areas with limited data (e.g. city, state, or legislative district) using (i) a rich hierarchical model fit on the national survey and the (usually) binary or binomial outcome and (ii) post-stratification of predicted values based on the underlying population. This method has been widely applied to a variety of questions such as measuring public opinion on a wide variety of policies, examining ideology at the city level, and exploring determinants of vote choice and turnout decisions (e.g. \citealt{ghitza2013mrp,lax2009gay,lax2012deficit,buttice2013mrp,tausanovitch2014representation}). 

Early applications of these models usually additively included reasonable number of non-nested effects (e.g. four), but subsequent work noted the inability of such models to capture the rich complexity of the data (\citealt{ghitza2013mrp}). That paper increased the complexity of the model substantially by using \emph{eighteen} mostly non-nested random effects and thus specifying a model with thousands of parameters. More broadly, the idea of using a more complex model has led to a variety of papers implementing more complex hierarchical models (\citealt{gelman2016using,gao2019improving}) or relying on machine learning methods (\citealt{bisbee2019barp,ornstein2019stacked,goplerud2018sparse}). Regardless of whether one relies on a ``traditional'' MRP or a recent extension, it is clear that comparing multiple specifications in a principled way is fundamental to performing reliable inference. Given the long history and popularity of using traditional hierarchical models when performing MRP, it is essential that there is a method to fit those models reliably and quickly given computational constraints for many practitioners.

Unfortunately, inference for non-linear hierarchical models---especially at the complexity needed to be competitive with machine learning alternatives---can be challenging as the likelihood function contains an intractable, high-dimensional, integral. There are two popular methods for applied researchers (\citealt{stegmueller2013multilevel}): First, one can approximate the integral numerically (e.g. \citealt{bates2015lmer,rabehesketh2004gllamm}). Second, one can use a fully Bayesian approach and sample from the joint distribution of all of the parameters of the model (e.g. \citealt{carpenter2017stan}). The key downside of these methods is that they can be slow even on modestly sized problems, and thus it is challenging to get estimates of reasonable quality in a modest period of time. This is a problem of ``scalability'' to the large and complex models required for many empirical applications. A key downside of non-scalable models is that common techniques such as $K$-fold cross-validation or bootstrapping are prohibitively expensive.

This paper makes two contributions to tackling this problem. First, I outline a series of new variational algorithms based on Polya-Gamma augmentation that allow coordinate ascent variational inference to be implemented for binomial logistic regression for an arbitrary number of (non-nested) random effects while imposing only a mean-field factorization assumption. This extends existing work on variational methods for this class of model, as there does not appear to be a tailored algorithm to estimate models with more than two non-nested random effects.\footnote{Generic methods for variational inference, e.g. stochastic variational inference or automatic differentiation variational inference (ADVI; \citealt{kuckelbir2017advi}), can be applied to most models, including hierarchical ones. I compare ADVI against my ``tailored'' algorithms and show it performs worse.} Further, the algorithm can be implemented without assuming independence between the ``fixed'' (i.e. fully pooled) and random effects.

Second, I outline a generic procedure for improving an initial variational approximation when a parameter expansion of the underlying Bayesian model exists. I do this by drawing a connection to ``marginal augmentation'' from the Markov Chain Monte Carlo literature (e.g. \citealt{liu1999parameter,van2001art}) and showing that this parameter expansion often permits a nearly costless improvement of the initial approximation. The method (``marginally augmented variational Bayes''---MAVB) transforms the initial approximation by sampling the expansion parameter and re-transforming the original samples while maintaining the stationarity of the target posterior. This induces dependencies between the parameters that were assumed away in estimating the initial procedure and provides a provable \emph{guaranteed} improvement upon the original approximation.

Methodologically, this pushes forward the literature on variational inference for hierarchical models by extending work in the case of a single random effect (\citealt{hall2011poisson,ormerod2012gva,tan2013variational,hall2019fast}) or two non-nested random effects (\citealt{jeon2017variational,menictas2019streamlined}) to the general case. The proposed method requires no integration, unlike many existing methods for binary outcomes (\citealt{ormerod2012gva,tan2013variational,jeon2017variational}). It further provides a link to existing work that seeks to combine Markov Chain Monte Carlo and variational inference by stochastic optimization (e.g. \citealt{salimans2015markov,ruiz2019contrastive,yin2018semi}). Instead of optimizing the transformed density, MAVB transforms the samples from the initial approximation with a partial step of MCMC using marginal augmentation that, in practice, appears as performing a stochastic location/scale transformation of the sampled parameters. This leverages a sampler that is known to mix well in the case of fully Bayesian MCMC and lacks internal tuning parameters as its primary goal is to find a computationally inexpensive way to improve an initial approximation. While it bears some similarities to work on re-parameterization in hierarchical models for variational algorithms (e.g. \citealt{tan2013variational,tan2021rvb}), it does not fix the re-parameterization in advance of estimation. It differs from other approaches that seek to improve an initial approximation (e.g. linear response variational Bayes; \citealt{giordano2015lrvb}) in that it has a guarantee on improving the approximation quality. Future work could examine how such methods work alongside Polya-Gamma data augmentation.

The remainder of the paper proceeds as follows. Section~\ref{section:CAVI} states multiple factorization assumptions under which Polya-Gamma augmentation can be used to estimate a variational approximation for a binomial logistic hierarchical model. Section~\ref{section:MAVB} links parameter expansion to variational Bayes and explains MAVB formally.

Sections~\ref{section:simulations} and~\ref{section:gg_app} conduct simulations and examine performance on the empirical example (\citealt{ghitza2013mrp}). The latter shows dramatic gains in speed: Even after applying MAVB and drawing 4,000 samples, the fastest variational algorithm is nearly 60 times faster than Laplace approximation and nearly 350 times faster than Hamiltonian Monte Carlo for the most complex models. This reduces the run time from hours to minutes. All variational methods well-recover the posterior means. While the strongest factorization assumptions have poor performance in terms of estimating the posterior variance, applying MAVB corrects a large amount of the problem.

Section~\ref{section:gg_app} then uses this algorithm to engage in model comparison that was computationally infeasible in \cite{ghitza2013mrp}. I perform 10-fold cross-validation across nine models ranging from having four to 18 random effects and thousands of parameters. The process takes around 30 minutes compared to the hours needed to fit even a single model once using existing approaches. The results provide some evidence of over-fitting in the original specification suggesting that the most complex model does not outperform models of intermediate complexity. I use this to draw out some guidance for practitioners of MRP in other substantive domains.

\section{Mean-Field Variational Inference for Binomial Hierarchical Models}
\label{section:CAVI}

I focus on the following generative model that is broader than MRP but also captures the majority of applications. For each observation $i \in \{1, \cdots, N\}$, the researcher observes $y_i$ ``successes'' out of $n_i$ trials (e.g. how many individuals in a population of size $n_i$ turn out to vote). I model this using a binomial distribution with probability of success $p_i$ defined via a linear predictor ($\psi_i$) put through a logistic link. Equation~\ref{eq:model_gendesign} expresses this model using a ``general design'' notation (\citealt{zhao2006mlm}). Appendix~\ref{section_app:pg} shows the model using \citet{gelman2006multi}'s notation and a plate diagram.

\begin{subequations}
	\label{eq:model_gendesign}
	\begin{alignat}{2}
	&y_i | \bm{\beta}, \bm{\alpha} \sim \mathrm{Binom}(n_i, p_i), \quad p_i = \frac{\exp(\psi_i)}{1+\exp(\psi_i)}, \quad \psi_i = \bm{x}_i^T\bm{\beta} + \bm{z}_i^T\bm{\alpha} \\
	&\bm{\alpha}_{j} | \bm{\Sigma}_j \sim N\left(\bm{0}, \bm{I}_{g_j} \otimes \bm{\Sigma}_j\right), \quad \bm{\Sigma}_j \sim \mathrm{IW}(\nu_j, \bm{\Phi}_j), \quad p(\bm{\beta}) \propto 1  \\
	& \bm{z}_{i,j} = \bm{m}_{i,j} \otimes \bm{z}^b_{i,j}, \quad \bm{\alpha}^T = [\bm{\alpha}^T_1, \cdots, \bm{\alpha}^T_J], \quad \bm{z}_i^T = [\bm{z}_{i,1}^T, \cdots, \bm{z}_{i,J}^T]
	\end{alignat}
\end{subequations}

As is standard in hierarchical models, the linear predictor consists of $p$ ``fixed'' effects: $\bm{x}_i \in \mathbb{R}^p$. The hierarchical component contains $J$ random effects indexed from $j \in \{1, \cdots, J\}$. For each random effect $j$, there is a $d_j$ dimensional covariate vector indexed by $\bm{z}^b_{i,j}$ where $\bm{z}^{b}_{i,j} = 1$ represents the ubiquitous ``random intercept.'' Each random effect has $g_j$ groups and each observation $i$ is assigned to exactly one group for each random effect; define its membership for random effect $j$ as a one-hot vector $\bm{m}_{i,j} \in \{0, 1\}^{g_j}$. 

The notation in Equation~\ref{eq:model_gendesign} stacks together the hierarchical components as follows; first, for each random effect $j$, $\bm{z}_{i,j}$ represents a $d_j \times g_j$ length vector (mostly sparse) by the Kronecker product ($\otimes$) of the group membership vector $\bm{m}_{i,j}$ and the base covariate. This repeats $\bm{z}^{b}_{i,j}$ once in the position corresponding to the group of which $i$ is a member for random effect $j$. This allows us to model the distribution of the entire parameter vector for random effect $j$ ($\bm{\alpha}_j \in \mathbb{R}^{g_j \cdot d_j}$) as a multivariate normal with a block diagonal matrix where each block is given an identical Inverse Wishart prior as noted in Equation~\ref{eq:model_gendesign}b ($\bm{\Sigma}_j \in \mathbb{R}^{d_j \times d_j}; \bm{\Phi}_j \in \mathbb{R}^{d_j \times d_j}$). Using such priors is standard in the literature on variational inference for hierarchical models (e.g. \citealt{tan2013variational}), although extensions to more weakly informative priors are possible (e.g. \citealt{huang2013simple}). The compact notation in Equation~\ref{eq:model_gendesign}a stacks together all random effects $j$ into a single vector $\bm{z}_i \in \mathbb{R}^{\sum_{j=1}^J g_j \cdot d_j}$ that is highly sparse. It thus accommodates designs with arbitrary patterns of crossing (non-nesting) amongst the $J$ random effects.

A key distinguishing feature of this model as applied to MRP is that $J$ can be large (e.g. greater than ten) and $g_j$ ranges widely from a handful up to over a thousand (e.g. $g_j = 4$ for ethnicity and $g_j = 1,020$ for state-ethnicity-age combinations in \citealt{ghitza2013mrp}). In most applications for MRP, $d_j = 1$ and $\bm{z}^{b}_{i,j} = 1$ (random intercept) but sometimes $d_j = 2$ in the case of a random slope and intercept (\citealt{gelman2006multi}). Regarding the other parameters, for most applications of MRP, $N$ is often relatively modest given post-stratification requirements (see Section~\ref{section:gg_app}) and that surveys can be collapsed into units with identical state-demographic covariates by allowing varying $n_i$. Thus, in many studies, $N$ can be made smaller than 10,000 (e.g. below 5,000 in \citealt{park2004mrp,ghitza2013mrp}). The size of $\bm{\beta}$ ($p$) is also usually modest and below ten.

By using Polya-Gamma augmentation, the model in Equation~\ref{eq:model_gendesign} can be rendered conditionally conjugate, enabling the straightforward application of numerous standard algorithms for Bayesian inference (\citealt{polson2013polyagamma}). Specifically, Equation 2 from \citet{polson2013polyagamma} states that for any $a, b > 0$ the following identity holds, where $f_{PG}(\omega | b, c)$ denotes the Polya-Gamma density with parameters $b$ and $c$. The definition of a Polya-Gamma variable as a weighted infinite convolution of Gamma random variables is also shown.

\begin{subequations}
	\begin{alignat}{2}
	\frac{\exp(\psi)^a}{\left[1+\exp(\psi)\right]^b} = 2^{-b} \int \exp(s \psi - \psi^2/2 \omega) f_{PG}(\omega | b, 0) d\omega,  \quad s = a - b/2 \\
	\omega \sim PG(b,c) \coloneqq \omega = \frac{1}{2\pi^2} \sum_{k=1}^\infty \frac{Z_k}{(k-1/2)^2 + c^2/(4\pi^2)}, \quad Z_k \overset{i.i.d.}{\sim} \mathrm{Gamma}(b, 1)
	\end{alignat}
\end{subequations}

Thus, the complete data likelihood can be expressed as follows where $\bm{\Omega}$ denotes the $N \times N$ diagonal matrix of the corresponding $\omega_i$, $\bm{X}$, $\bm{Z}$ stack the data for each observation into a $N \times p$ and $N \times \sum_{j=1}^{J} d_j g_j$ design matrices, and $\bm{s}$ is a $N \times 1$ vector with $[\bm{s}]_i = y_i - n_i/2$.

\begin{equation}
\begin{split}
p(\bm{y}, \bm{\Omega} | \bm{\alpha}, \bm{\beta}) &\propto \exp\left(\bm{s}^T [\bm{X}\bm{\beta} + \bm{Z} \bm{\alpha}] - \frac{1}{2} \left[\bm{X}\bm{\beta} + \bm{Z}\bm{\alpha}\right]^T\bm{\Omega} \left[\bm{X}\bm{\beta} + \bm{Z}\bm{\alpha}\right]\right) \prod_{i=1}^N f_{PG}(\omega_i | n_i, 0)
\end{split}
\end{equation}

Noting the result from \citet{polson2013polyagamma} that the full conditional of $\omega_i | \bm{y}, \bm{\alpha}, \bm{\beta}, \{\bm{\Sigma}_j\}_{j=1}^J$ has a Polya-Gamma distribution $PG(n_i, \bm{x}_i^T\bm{\beta} + \bm{z}_i^T\bm{\alpha})$, it immediately follows that a Gibbs Sampler exists to sample all of the parameters in the model where the full conditionals on $\bm{\beta}$ and $\bm{\alpha}$ are normal and $\bm{\Sigma}_j$ is Inverse Wishart.

\subsection{Variational Inference}

The first contribution of this paper is to use the Polya-Gamma representation above to find a tractable variational algorithm to approximate the joint posterior of $p(\bm{\beta}, \bm{\alpha}, \{\bm{\Sigma}_j\}, \bm{\Omega} | \bm{y})$ and thus the joint posterior on the parameters excluding $\bm{\Omega}$. \citet{blei2017vi} provides a recent review of these methods. Equation~\ref{eq:vi_def} formulates the problem where $\mathcal{X}$ denotes some (restricted) set of distributions to optimize over. It can be equivalently expressed as finding the closest distribution in $\mathcal{X}$ to the true posterior in terms of KL-divergence. For notational simplicity, denote $\bm{\theta} = \{\bm{\beta}, \bm{\alpha}, \{\bm{\Sigma}_j\}_{j=1}^J, \bm{\Omega}\}$.

\begin{equation}
\label{eq:vi_def}
q^*(\bm{\theta}) = \argmax_{q(\bm{\theta}) \in \mathcal{X}} \mathrm{ELBO}_{q(\bm{\theta})}  \quad \mathrm{where} \quad \mathrm{ELBO}_{q(\bm{\theta})} = E_{q(\bm{\theta})}\left[\ln p(\bm{y}, \bm{\theta})\right] - E_{q(\bm{\theta})}\left[\ln q(\bm{\theta})\right] 
\end{equation}

A common method for solving this problem is known as ``coordinate ascent variational inference'' (CAVI; \citealt{blei2017vi}). It maximizes or increases the target $\mathrm{ELBO}$ with respect to some sub-block of $\bm{\theta}$. By cycling through $\bm{\theta}$ repeatedly, a local optimum can be obtained. The choice of restriction $\mathcal{X}$ is crucial to the accuracy of the approximation method; an extremely popular choice is a ``mean-field'' factorization assumption where blocks of parameters are assumed to be independent. 

Leveraging the existence of a Gibbs Sampler, Result~\ref{result:CAVI} states that the augmented posterior on $q(\bm{\theta})$ can be approximated using a number of mean-field assumptions with no further restrictions on distributional form, all updates having closed analytical forms, and for arbitrary $J, d_j, g_j$. Appendix~\ref{section_app:pg} provides the full derivations as well as noting how to back out the corresponding Gibbs Sampler.

\begin{result}[Existence of CAVI]
	\label{result:CAVI}
	Consider the three factorization assumptions:
	
	\begin{tabular}{llll}
		Scheme I:& ``Strong Factorization'' & --- & $\mathcal{X}_1 = q(\bm{\beta})\prod_{j=1}^{J} q(\bm{\alpha}_j) q(\bm{\Sigma}) q(\bm{\Omega})$ \\
		Scheme II:& ``Partial Factorization'' & --- & $\mathcal{X}_2 = q(\bm{\beta}) q(\bm{\alpha}) q(\bm{\Sigma}) q(\bm{\Omega})$ \\
		Scheme III:& ``Limited Factorization'' & --- & $\mathcal{X}_3 = q(\bm{\beta}, \bm{\alpha}) q(\bm{\Sigma}) q(\bm{\Omega})$
	\end{tabular}
	
	For the model in Equation~\ref{eq:model_gendesign} and for each choice of $\mathcal{X}_k$ above, each step of the CAVI algorithm can be implemented exactly in closed form, with no additional assumptions. For each $\mathcal{X}_k$, the optimal approximation for $q(\bm{\beta}, \bm{\alpha})$ is multivariate normal, $q(\bm{\Sigma})$ is the product of $J$ independent Inverse Wishart densities, and $q(\bm{\Omega})$ is the product of $N$ independent Polya-Gammas.
\end{result}

Algorithm~\ref{alg:CAVI} explicitly outlines the updates for Scheme I. Experiments showed that convergence could be improved at little computational cost by jointly updating the mean parameters of $q(\bm{\beta})$ and $q(\bm{\alpha})$; see Appendix~\ref{section_app:acceleration} for discussion. All models estimated in the paper use this acceleration technique.

\begin{algorithm}[!ht]
	\caption{CAVI for Scheme I}
	\label{alg:CAVI}
	\begin{algorithmic}
		\State{\textbf{Set Priors of Inverse Wishart}: $\{\nu_j, \bm{\Phi}_j\}_{j=1}^J$; \textbf{Set Number of Iterations}: $T$}
		
		\State{\textbf{Initialize Variational Parameters}: $\{\tilde{b}_i, \tilde{c}_i\}_{i=1}^N$ (for Polya-Gamma); $\tilde{\bm{\mu}}_\beta, \tilde{\bm{\Lambda}}_\beta, \tilde{\bm{\mu}}_\alpha, \tilde{\bm{\Lambda}}_\alpha$ (for $\bm{\beta}, \bm{\alpha}$); $\{\tilde{\nu}_j, \tilde{\bm{\Phi}}_j\}_{j=1}^J$ (for $\bm{\Sigma}_j$)}
		
		\For{$t$ in $1, \cdots, T$}
		\State{1. Update Polya-Gammas - $q\left(\{\omega_i\}_{i=1}^N\right)$: $\tilde{b}_i = n_i, \quad \tilde{c}_i = \sqrt{E_{q(\bm{\alpha},\bm{\beta})}\left[(\bm{x}_i^T\bm{\beta} + \bm{z}_i^T\bm{\alpha})^2\right]}$} 
		
		\State{2. Update $q(\bm{\beta}) \sim N(\tilde{\bm{\mu}}_\beta, \tilde{\bm{\Lambda}}_\beta)$}:
		
		$$\tilde{\bm{\Lambda}}_\beta = \left(\sum_{i=1}^N E_{q(\omega_i)}[\omega_i] \bm{x}_i\bm{x}_i^T\right)^{-1}, \quad \tilde{\bm{\mu}}_\beta = \tilde{\bm{\Lambda}}_\beta \bm{X}^T \left(\sum_{i=1}^N \left(y_i - \frac{n_i}{2}\right) - E_{q(\omega_i)}[\omega_i] \cdot \bm{z}_i^T E_{q(\bm{\alpha})}[\bm{\alpha}]\right)$$
		
		\State{3. Update $q\left(\bm{\alpha}_j\right) \sim N(\tilde{\bm{\mu}}_{\alpha,j}, \tilde{\bm{\Lambda}}_{j,\alpha})$, where $\bm{T}_j$ stacks the block diagonal expectation of the precision on the random effects ($\bm{\Sigma}_j^{-1}$):}
		
		$$\tilde{\bm{\Lambda}}_{\alpha,j} = \left(\bm{T}_j + \sum_{i=1}^N E_{q(\omega_i)}[\omega_i] \bm{z}_{i,j}\bm{z}_{i,j}^T\right)^{-1}, \quad \bm{T}_j = E_{q(\bm{\Sigma}_j)}\left[\bm{I}_{g_j} \otimes \bm{\Sigma}_j^{-1}\right] $$
		$$\tilde{\bm{\mu}}_{\alpha,j} = \tilde{\bm{\Lambda}}_{\alpha,j} \bm{Z}_j^T \left[\sum_{i=1}^N \left(y_i - \frac{n_i}{2}\right) - E_{q(\omega_i)}[\omega_i] \cdot \left(\bm{x}_i^T E_{q(\bm{\beta})}[\bm{\beta}] + \sum_{\ell: \{1, \cdots, J\} \setminus j} \bm{z}_{i,\ell}^T E_{q(\bm{\alpha}_\ell)}[\bm{\alpha}_\ell]\right)\right]$$
		\State{4. Update $q\left(\{\bm{\Sigma}_j\}_{j=1}^J\right)$: $\tilde{\nu}_j = \nu_j + g_j, \quad \tilde{\bm{\Phi}}_j = \bm{\Phi}_j + \sum_{g=1}^{g_j} E_{q(\bm{\alpha}_{j,g})}\left[\bm{\alpha}_{j,g}\bm{\alpha}_{j,g}^T\right]$}
		\vspace{0.25em}
		\State{5. Check for convergence, evaluate ELBO (see Appendix~\ref{section_app:pg} for derivation).}
		\EndFor
	\end{algorithmic}
\end{algorithm}

This improves upon existing mean-field schemes for logistic hierarchical models in a number of ways. First, for any of the factorization assumptions, no further distributional assumptions are required (cf. \citealt{ormerod2012gva,tan2013variational} assuming normality). Second, most existing algorithms for binomial outcomes require the repeated evaluation of (low) dimensional integrals at each iteration whose number scales with $g_j$ (cf. \citealt{ormerod2012gva,tan2013variational,jeon2017variational}). Extending these algorithms to $J > 2$ would likely incur significant computational costs as the number of those integrals increases. None of the schemes in Result~\ref{result:CAVI} require integration at any step as the Polya-Gamma augmentation turns inference into iteratively performing weighted ridge regression. In the models considered in this paper, the major bottleneck as one moves from Scheme I to Scheme III is in calculating the variance term of $q(\bm{\beta},\bm{\alpha})$; even relying on a (sparse) Cholesky decomposition, this involves inverting an increasingly dense lower triangular matrix as weaker independence assumptions are imposed. Appendix~\ref{section_app:gg_extra} disaggregates the run-time of Algorithm~\ref{alg:CAVI} by stage and scheme.

Most importantly, the ability to choose between Schemes I, II, and III allows the researcher to smoothly trade-off computational cost and accuracy as in \citet{menictas2019streamlined}'s work on $J=2$ for linear mixed effects models. Scheme I with its strong implied factorization assumptions is immediately scalable to huge datasets with large $J$ or $g_j$. However, the downside is that the strong factorization assumptions will likely degrade performance. Scheme III provides the ability to avoid these strong assumptions at a somewhat increased computational cost. The ability to avoid such factorization assumptions for arbitrary $J > 1$ and binomial outcomes appears to be a new result. The expectation is that it will have the best performance. Scheme II is a compromise between the two extremes, and other hybrid approaches are possible such as applying re-parameterizations to the augmented posterior (e.g. \citealt{tan2013variational,tan2021rvb}).

\section{Marginally Augmented Variational Bayes}
\label{section:MAVB}

The second major contribution of the paper is demonstrating that there is a computationally cheap way of improving the initial approximation resulting from Schemes I, II, or III. The key intuition, formalized below, is that once an initial approximation $q(\bm{\theta})$ is found, one can draw samples from this approximation, perform a single step of Markov Chain Monte Carlo through (some of) the parameters, and thereby ``improve'' the sample. Some existing work in computer science (e.g \citealt{salimans2015markov,ruiz2019contrastive}) has leveraged this point to attempt to \emph{optimize} over the intractable improved density which can be computationally expensive. 

By contrast, this paper explores the idea that if one can find a transition kernel with good mixing, then simply doing a single partial step can provide considerable gains at limited computational cost. While many samplers can be employed for this purpose, initial experiments suggested that the key problem was the independence assumptions in $q(\bm{\beta},\bm{\alpha})$ and thus I chose to focus on marginal augmentation and parameter expansion as it is inexpensive to use in fully Bayesian MCMC to improve a Gibbs Sampler, has demonstrated strong performance in hierarchical models, lacks internal tuning parameters, and was explicitly designed to link the fixed and random effects together (\citealt{liu1999parameter,van2001art,gelman2008px}). I focus on logistic hierarchical models although the procedure is itself much more general; Section~\ref{section:conclusion} discusses some broader implications and Appendix~\ref{section_app:param_x} formulates the results in a more general fashion.

The key idea behind parameter expansion is to create an ``over-parameterized'' model where certain additional parameters ($\bm{\xi}$) are introduced such that they (i) maintain the observed data model but (ii) are not identifiable from the observed data itself. A careful choice of parameter expansion allows the construction of algorithms that have either faster mixing for MCMC (\citealt{liu1999parameter,van2001art}) or faster convergence for deterministic algorithms such as EM (\citealt{liu1998pxem}). The intuition behind its effectiveness is that it allows ``moves'' (either via sampling steps in MCMC or parameter updates in EM) in the un-identified space that can break or escape the strong associations between parameter blocks (e.g. $\bm{\beta}$ and $\bm{\alpha}$) that slow down mixing (\citealt{liu1999parameter}) or lead to the algorithm getting ``stuck'' for many iterations near boundary conditions (e.g. a small sampled $\bm{\Sigma}_j$ shrinking $\bm{\alpha}_{j,g}$ leading to a small $\bm{\Sigma}_j$, etc.; \citealt{gelman2008px}). \cite{liu1998pxem} provide a useful explanation of parameter expansion in the context of EM as a ``covariance adjustment'' to the estimated parameters.

In the case of hierarchical models, the most popular parameter expansion appears as a location and/or scale transformation of the random effects (e.g. \citealt{van2001art}). The location transformation, for example, allows the random effects to have a non-zero mean: $\bm{\alpha}_{j,g} \sim N(\bm{\mu}_j, \bm{\Sigma}_j)$. Note that it is not possible to estimate $\bm{\mu}_j$ from the observed data but that it could be estimated if $\bm{\alpha}_{j,g}$ were known. Implementation is simple and appears as a location or scale transformation of the sampled parameters that leads to very large gains in performance (e.g. \citealt{van2001art,gelman2008px}).

Definition~\ref{def:expanded_hier} generalizes this parameter expansion to the arbitrary $J$ case, where the $\bm{M}_j$ notation is bookkeeping to note which element of $\bm{x}_i$ (and $\bm{\beta}$) corresponds to each element of $\bm{z}^b_{i,j}$ (and $\bm{\alpha}_{j,g}$).

\begin{definition}[Expansions for Hierarchical Models]
	\label{def:expanded_hier}
	
	Define a set of expansion parameters $\bm{\xi}$ that consists, for each $j$, of a mean shift $\bm{\mu}_j \in \mathbb{R}^{d_j}$ and a scale shift $\bm{R}_j \in \mathbb{R}^{d_j \times d_j}$ such that $\bm{R}_j$ is invertible. I use superscript $^X$ to denote the ``expanded'' parameters. 
	
	The mapping between $\bm{\theta}^X$ and $\bm{\theta}$ for a fixed $\bm{\xi}$ is denoted as $t_{\bm{\xi}}(\bm{\theta}^X)$ and listed below. $\bm{M}_j$ is a $p \times d_j$ matrix such that $[\bm{M}_{j}]_{a,b} = 1$ if the covariate corresponding to $[\bm{z}_{i,j}]_b$ is the same as the covariate for $[\bm{x}_i]_a$. All other elements of $\bm{M}_j$ are zero. For simplicity, assume that each element of $\bm{z}_i$ corresponds to some variable in $\bm{x}_i$, i.e. that each column of $\bm{M}_j$ has exactly one non-zero element.
	
	$$\left[\bm{\beta}, \bm{\alpha}, \{\bm{\Sigma}_j\}_{j=1}^J, \bm{\Omega}\right] = t_{\bm{\xi}}([\bm{\beta}^X,\bm{\alpha}^X, \{\bm{\Sigma}_j^X\}_{j=1}^J, \bm{\Omega}^X]) = \begin{cases}
	\bm{\beta} = \bm{\beta}^X + \sum_{j=1}^J \bm{M}_j \bm{R}_j \bm{\mu}_j \\
	\bm{\alpha}_{j,g} = \bm{R}_j\left(\bm{\alpha}^X_{j,g} - \bm{\mu}_j\right) \\
	\bm{\Sigma}_j = \bm{R}_j\bm{\Sigma}^X_j \bm{R}_j^T \\
	\bm{\Omega} = \bm{\Omega}^X
	\end{cases}$$
	
	The augmented model is listed below for an important special case treated in detail (``Mean Expansion'') in the empirical analysis.	The full expansion (``Translation Expansion'') is also listed.
	\begin{itemize}
		\item Mean Expansion: Assume all $\bm{R}_j = \bm{I}_{d_j}$.
		\begin{equation*}
		\begin{split}
		&\ln p(y_i | \omega_i, \bm{\beta}^X,\bm{\alpha}^X) \propto \bm{s}^T[\bm{X}\bm{\beta}^X + \bm{Z}\bm{\alpha}^X] - 1/2[\bm{X}\bm{\beta}^X + \bm{Z}\bm{\alpha}^X]^T\bm{\Omega}[\bm{X}\bm{\beta}^X + \bm{Z}\bm{\alpha}^X] \\
		&p(\bm{\beta}^X) \propto 1, \quad \bm{\alpha}_{j,g}^X | \bm{\Sigma}^X_j, \sim N\left(\bm{\mu}_j, \bm{\Sigma}^X_j\right), \quad p(\bm{\Sigma}_j^X) \sim IW(\nu_j, \bm{\Phi}_j)
		\end{split}
		\end{equation*}
		\item Translation Expansion:
		\begin{equation*}
		\begin{split}
		&\ln p(y_i | \omega_i, \bm{\beta}^X,\bm{\alpha}^X) \propto \bm{s}^T[\bm{X}\bm{\beta}^X + \bm{Z}\bm{R}\bm{\alpha}^X] - 1/2[\bm{X}\bm{\beta}^X + \bm{Z}\bm{R}\bm{\alpha}^X]^T\bm{\Omega}[\bm{X}\bm{\beta}^X + \bm{Z}\bm{R}\bm{\alpha}^X] \\
		&\bm{R} = \mathrm{blockdiag}\left(\{\bm{I}_{g_j} \otimes \bm{R}_j\}_{j=1}^J\right), \quad p(\bm{\beta}^X) \propto 1, \quad \bm{\alpha}_{j,g}^X | \bm{\Sigma}^X_j \sim N\left(\bm{\mu}_j, \bm{\Sigma}_j^X\right) \\
		&p(\bm{\Sigma}_j^X) \sim IW(\nu_j,\bm{R}_j^{-1} \bm{\Phi}_j\bm{R}_j^{-T})
		\end{split}
		\end{equation*}
	\end{itemize}
\end{definition}	

Given such an expanded version of the hierarchical model, there are two ways to improve the algorithms in this paper. First, drawing on \citet{jaakkola2007parameter}, it is possible to accelerate convergence of Algorithm~\ref{alg:CAVI} using ``parameter expanded variational Bayes'' (PX-VB). Appendix~\ref{section_app:acceleration} derives a new application of PX-VB to the models in Result~\ref{result:CAVI} and shows it can often improve the algorithm's convergence by decreasing the number of iterations required at effectively no computational cost as, functionally, it involves centering the random effects to be mean zero and adjusting the mean of $q(\bm{\beta})$ correspondingly.

The main use of parameter expansions in this paper, however, is to improve the \emph{quality} of the approximation by ``improving'' $q(\bm{\theta})$ by performing one step of marginal augmentation where the expansion parameters $\bm{\xi}$ are sampled and then the components of $\bm{\theta}$ are re-sampled. Definition~\ref{def:mavb} outlines the procedure in a general case. The notation and procedure mirrors that in \cite{liu1999parameter}.

\begin{definition}[Marginally Augmented Variational Bayes---MAVB]
	\label{def:mavb}
	Given an initial approximation $q(\bm{\theta})$, a proper prior on the expansion parameter $p_0(\bm{\xi})$, and a one-to-one and differentiable transformation such that $t_{\bm{\xi}}(\bm{\theta}^X) = \bm{\theta}$, create a new approximation $\tilde{q}(\bm{\theta})$ using the following procedure:
	\begin{enumerate}
		\item Sample $\bm{\theta} \sim q(\bm{\theta})$ and $\bm{\xi}_0 \sim p_0(\bm{\xi})$.
		\item Create $\bm{\theta}^X = t^{-1}_{\bm{\xi}_0}(\bm{\theta})$.
		\item Sample a new $\bm{\xi}_1$ as follows where $J_{\bm{\xi}}(\bm{\theta}^X)$ is the Jacobian of $t_{\bm{\xi}}$ with respect to $\bm{\theta}^X$ and $p(\bm{\theta}| \bm{y})$ denotes the \emph{true} posterior distribution.
		
		$$\bm{\xi}_1 \sim p\left(\bm{\xi} | \bm{\theta}^X, \bm{y}\right) \propto p(t_{\bm{\xi}}(\bm{\theta}^X) | \bm{y}) \cdot |J_{\bm{\xi}}(\bm{\theta}^X)| \cdot p_0(\bm{\xi})$$
		\item Define $\tilde{\bm{\theta}} = t_{\bm{\xi}_1}(\bm{\theta}^X) = t_{\bm{\xi}_1}\left(t^{-1}_{\bm{\xi}_0}(\bm{\theta})\right)$
	\end{enumerate}	
\end{definition}

Theorem~\ref{thm:MAVB} states a key result for MAVB.  

\begin{theorem}[Guaranteed Improvement with MAVB]
	\label{thm:MAVB}
	\vspace{1px}
	
	For any (proper) choice of prior $p_0(\bm{\xi})$, the MAVB approximation $\tilde{q}(\bm{\theta})$ has a better $\mathrm{ELBO}$ than the initial approximation:
	
	$$\mathrm{ELBO}_{\tilde{q}(\bm{\theta})} \geq \mathrm{ELBO}_{q(\bm{\theta})}$$
\end{theorem}

The proof is in Appendix~\ref{section_app:param_x} and uses two lemmas from existing results. First, Theorem 1 in \citet{liu1999parameter} demonstrates the transformation to generate MAVB maintains the stationarity of the posterior. Second, a data processing inequality noted by various authors (e.g. \citealt{ruiz2019contrastive}) showing that this transformation which keeps the \emph{true} posterior invariant results in a better approximating distribution. 

It is known from the data augmentation literature that an increasingly diffuse prior on the expansion parameters (``working prior'') allows for the parameters themselves to ``decide'' the best expansion parameter $\bm{\xi}$ rather than being weighed down by the prior (e.g. \citealt[p. 1268]{liu1999parameter}), and I conjecture that a similar intuition applies for MAVB. Thus, in all applications, I use an improper prior (i.e. $p_0(\bm{\xi}) \propto 1$); Appendix~\ref{section_app:param_x} discusses the validity of this prior using existing theory (\citealt{liu1999parameter,van2001art}), provides the result for a proper prior on $\bm{\xi}$, and notes Algorithm~\ref{alg:MAVB} can be found as the limit of a proper working prior $p_0(\bm{\mu}_j) \sim N(0, \tau^2 \bm{I})$ as $\tau \to \infty$. Algorithm~\ref{alg:MAVB} shows how MAVB is implemented using the mean expansion noted in Definition~\ref{def:expanded_hier}.\footnote{MAVB for ``Translation Expansion'' (i.e. $\bm{R}_j$ is not fixed) is more delicate and thus not explored here, as it requires a specific choice of prior on $\bm{\Sigma}_j$ and a specific choice of improper working prior to be tractable; see \citet{van2001art} for details. Examining whether this could be used with proper priors is an interesting area for future research.}

\begin{algorithm}[!ht]
	\caption{Applying MAVB to Non-Linear Hierarchical Models}
	\label{alg:MAVB}
	\begin{algorithmic}
		\State{\textbf{Set the Number of Samples Desired}: $M$}
		\State{\textbf{Estimate $q(\bm{\theta})$ using CAVI (e.g. Algorithm~\ref{alg:CAVI})}}
		\For{$m$ in $1, \cdots, M$}
		\State{1. Draw $\bm{\theta}^{(m)} \sim q(\bm{\theta})$}
		\State{2. Sample the expansion parameters $\bm{\mu}_j$ for each $j$}
		$$\tilde{\bm{\mu}}_j \sim N\left(\frac{1}{g_j} \sum_{g=1}^{g_j} \bm{\alpha}_{j,g}^{(m)}, \frac{1}{g_j} \bm{\Sigma}^{(m)}_j\right)$$
		\State{3. Adjust the initial draws to get the improved sample $\tilde{\bm{\theta}}^{(m)}$}
		$$\tilde{\bm{\alpha}}^{(m)}_{j,g} = \bm{\alpha}_{j,g}^{(m)} - \tilde{\bm{\mu}}_j, \quad \tilde{\bm{\beta}}^{(m)} = \bm{\beta}^{(m)} + \sum_{j=1}^J \bm{M}_j \tilde{\bm{\mu}}_j$$
		\EndFor
	\end{algorithmic}
\end{algorithm}

Thus, for this model and relying only on a location transformation, MAVB has a simple form that, as shown later, can result in considerable improvements in the performance of Scheme I. As noted in the earlier discussion, the presentation of MAVB in Algorithm~\ref{alg:MAVB} illustrates the close relationship to the location transformation noted earlier: It can be thought of a ``stochastic'' location transformation given that mean of the expansion parameter is the mean of the sampled $\bm{\alpha}_{j,g}$.

Some additional remarks are in order: First, if MAVB is applied to an approximation resulting from Scheme I (i.e. with independence assumed between $\bm{\beta}$ and $\bm{\alpha}$), the resulting approximation \emph{will not} imply such an assumption. Consider the correlation between $\bm{\alpha}_{j,g}$ and $\bm{\beta}$ in Algorithm~\ref{alg:MAVB}. Before applying MAVB, the two parameters are independent by assumption. After applying MAVB, they have a non-zero posterior correlation because of the shared dependence on $\tilde{\bm{\mu}}_j$. While not sufficient to restore \emph{all} missing dependencies (e.g. components of $\bm{\beta}$ that are not included in any random effect), this can at least address some of the shortcomings of Scheme I. MAVB can be applied to the outputs of Scheme II and III, although the expectation is that the improvement for these schemes should be less pronounced given that more of those dependencies are estimated directly.

Second, the cost of implementing MAVB is quite modest, unlike existing approaches that attempt to \emph{optimize} over the improved density (e.g. \citealt{ruiz2019contrastive}). After drawing a sample from $q(\bm{\theta})$, all that is needed to perform MAVB is drawing $\sum_j d_j$ univariate Gaussians (22 in the largest model considered in this paper [Model 9]), some summation of the sampled random effects and then subtracting off the sampled expansion parameter $\tilde{\bm{\mu}}_j$. Note that this MAVB procedures do not require sampling the Polya-Gammas as they are left un-transformed by the algorithm nor does the cost of MAVB depend on the size of the data ($N$) directly; even if $g_j$ is large, MAVB will still be fast.

Third, while MAVB is guaranteed to increase performance, the quality of MAVB is difficult to ascertain analytically in most complex models. However, insights come from simpler cases: In a stylized hierarchical model, \cite{liu1999parameter} show that marginal augmentation results in perfect sampling.  In the more realistic case where $\bm{\Sigma}_j$ is not fixed and $J = 1$, studies show that certain forms of marginal augmentation result vastly improved mixing of MCMC samplers  (\citealt{van2001art,gelman2008px}). Thus, there is reason to be optimistic about the ability of MAVB to improve initial approximations as the scale/location transformations in fully Bayesian marginal augmentation seem to provide quite considerable benefits over simple Gibbs samplers.

Overall, while MAVB is likely to be helpful in improving the variational schemes in this paper, it is not a panacea. Its major benefit appears to be in ``connecting'' blocks of parameters that were assumed to be independent in a way that is guaranteed to improve the approximation quality at a very limited computational cost. The key limitation is that its speed and scalability depends on it \emph{not} returning to the observed data ($\bm{y}$).  Interestingly, this suggests a ``stronger'' version of MAVB that could be performed by implementing one full sweep of the Gibbs Sampler, i.e. sampling Polya-Gammas and cycling through all full conditionals, and then performing marginal augmentation. If this were to be performed many times, the samples would converge to the true posterior by standard properties of MCMC. While this might raise its own computational concerns, exploring this is an interesting area of future research.

\section{Simulation Study}
\label{section:simulations}

I perform a simulation study to assess the accuracy of the proposed methods. I compare my variational algorithms against two gold standards (Laplace approximation using \texttt{blme} - \citealt{bates2015lmer,chung2015weakly}; HMC in \texttt{STAN} using \texttt{brms}; \citealt{burkner2017brms}) and Automatic Differentiation Variational Inference (ADVI; \citealt{kuckelbir2017advi}).\footnote{Using \texttt{blme} allows for an identical Inverse Wishart prior to be added to the Laplace approximation; models are fit using \texttt{optimx}'s \texttt{nlminb} algorithm (\citealt{nash2011unifying}) that returned noticeably better performance. \texttt{brms} generates a model that can be manually adapted to place an Inverse Wishart prior as this is not permitted in the default options in pre-written STAN models at the time of writing (e.g. \texttt{rstanarm} or \texttt{brms}).} The latter is a useful comparison as it is easily implemented in \texttt{STAN} and is a generic approach to approximate complex models. I show results using its mean-field approximation (MF) and full rank (FR). To begin, I conducted a simulation where the linear predictor $\psi_i$ was generated using the following scheme ($J = 2$).

\begin{framed}
	Draw the fixed effects $\bm{\beta} \sim N(\bm{0}, \left[0.2\right]^2 \bm{I}_{10})$
	
	For each group $g \in \{1, \cdots, 10\}$, draw the random intercept $\alpha_{1,g} \sim N(0, 1)$
	
	For each group $g' \in \{1, \cdots, 10\}$, draw the random intercept $\alpha_{2,g} \sim N(0, 1)$
	
	For each observation $i \in \{1, \cdots, 1000\}$, assign at random to groups $g, g'$. Draw its fixed effect $\bm{x}_i \sim N(\bm{0}, \bm{\Sigma})$ where $\bm{\Sigma}_{j,j'} = 0.5^{|j-j'|}; j,j' \in \{1, \cdots, 10\}$. Draw $y_i$ such that:
	
	\begin{equation*}
	y_i \sim \mathrm{Bern}(p_i), \quad p_i = \frac{\exp(\bm{x}_i^T\bm{\beta} + \alpha_{1,g[i]} + \alpha_{2,g'[i]})}{1 + \exp(\bm{x}_i^T\bm{\beta} + \alpha_{1,g[i]} + \alpha_{2,g'[i]})}
	\end{equation*}
\end{framed}

All models are fit with a standard Inverse Wishart prior of $\mathrm{IW}(d_j + 1; \bm{I}_{d_j})$ on $\bm{\Sigma}_j$. I run each variational algorithm until the change in the ELBO is less than $10^{-8}$ or the largest parameter changes by less than $10^{-5}$. For the HMC and MAVB methods, I draw 4,000 samples from the (approximate) posterior. 

Table~\ref{tab:sims} reports four measures of performance; the first two measures compare the point estimates (posterior mean) against HMC. The third measure compares the full posterior using a measure of ``accuracy'' that modifies the integrated absolute error (e.g. \citealt{faes2011variational}). Formally, this is defined as $1 - \frac{1}{2} \int_{-\infty}^\infty | q_k(\theta) - q_{\mathrm{HMC}}(\theta)|d\theta$. I use kernel density estimation with a range over the shared support of the samples (\texttt{bkde}, \texttt{KernSmooth}; \citealt{wand2020kernsmooth}) and then approximate the integral. Finally, to understand how the estimates of uncertainty fare against the unknown truth, I examine the ``frequentist coverage'': Does an interval of $\pm$ 1.96 times the standard deviation of the parameter contain the truth? A value of around 0.95 would indicate correct coverage at the expected frequentist level.

\begin{table}[!ht]
	\caption{Results from Simulations}
	\label{tab:sims}
	\begin{center}
		\begin{tabular}{rlrr|rr|rr|rr}
			\hline\hline
			& & \multicolumn{2}{c}{Bias} & \multicolumn{2}{c}{RMSE} & \multicolumn{2}{c}{Accuracy} & \multicolumn{2}{c}{Coverage} \\
			& & FE & RE & FE & RE & FE & RE & FE & RE \\
			\hline
			 & Laplace & -0.000 & 0.005 & 0.007 & 0.056 & 0.966 & 0.847 & 0.950 & 0.878 \\
 & HMC &  &  &  &  &  &  & 0.949 & 0.960 \\
 & ADVI (MF) & -0.002 & 0.003 & 0.041 & 0.074 & 0.775 & 0.781 & 0.796 & 0.830 \\
 & ADVI (FR) & -0.000 & 0.001 & 0.048 & 0.108 & 0.921 & 0.881 & 0.969 & 0.948 \\
 & Scheme I & -0.000 & 0.004 & 0.007 & 0.034 & 0.870 & 0.706 & 0.862 & 0.736 \\
 & Scheme II & -0.000 & 0.004 & 0.007 & 0.030 & 0.870 & 0.844 & 0.862 & 0.871 \\
 & Scheme III & -0.000 & 0.004 & 0.007 & 0.026 & 0.936 & 0.948 & 0.923 & 0.934 \\
\hline
\multirow{3}{*}{MAVB +} & Scheme I & -0.000 & 0.004 & 0.008 & 0.034 & 0.932 & 0.948 & 0.922 & 0.938 \\
 & Scheme II & -0.001 & 0.005 & 0.007 & 0.031 & 0.933 & 0.955 & 0.916 & 0.940 \\
 & Scheme III & -0.000 & 0.004 & 0.007 & 0.027 & 0.937 & 0.963 & 0.922 & 0.942 \\

			\hline\hline\\
			\multicolumn{10}{l}{
				\begin{minipage}{\textwidth}
					\footnotesize \emph{Note}: This reports the bias (Bias), root mean squared error (RMSE) of the estimated posterior means against those estimated from HMC. The distance between the distributions (Accuracy) and frequentist coverage (Coverage) are reported; see the main text for an explanation of these measures. The statistics are disaggregated by fixed (FE) and random effects (RE). All results are created using all relevant parameters in each simulation and then averaged across one hundred simulations. ADVI (MF) uses the mean-field approximation; ADVI (FR) uses the full rank approximation in \citet{kuckelbir2017advi}.
				\end{minipage}
			}
		\end{tabular}
	\end{center}
\end{table}

The results are promising; looking at the bias and RMSE, the variational methods perform well; they have very small bias against the means estimated from HMC and an RMSE that is quite small, comparable to the Laplace approximation, and out performs both ADVI implementations. 

Examining accuracy and frequentist coverage shows more separation across the methods. The accuracy and coverage of Scheme I are noticeably lower than the Laplace approximation. However, applying MAVB results in noticeable improvements in accuracy (around 6\% for fixed effects; and nearly 25\% for random effects) and increases coverage by similar amounts to near nominal levels. After this improvement, Scheme I is comparable to the best approximate method (Laplace approximation) having slightly lower accuracy for the fixed effects but noticeably better accuracy and coverage for the random effects. Scheme II has somewhat better initial performance but is also boosted considerably by applying MAVB.

Scheme III---the factorization that does not assume independence between $q(\bm{\alpha})$ and $q(\bm{\beta})$---performs nearly as well as the Laplace approximation (and better in terms of the random effects) before applying MAVB. Applying MAVB results in only slight improvements (e.g. a 1-2\% boost in accuracy and coverage for the random effects).

Appendix~\ref{section_app:simulations_extra} conducts additional simulations. First, I vary the magnitude of the true coefficients by changing the variance of the fixed and random effects. After applying MAVB, the coverage of the variational methods is near nominal (i.e. above 0.90) in all cases except when the variance of the true distribution of the fixed effects is larger where MAVB is insufficient to obtain nominal coverage on the fixed effects (0.80-0.85) although the coverage on the random effects remains good. While this is worthy of future exploration, I conjecture this occurs because of the large magnitudes of the linear predictors (with 5-95\% interval of around -5.9 to 5.5 vs. -2.7 and 2.3 in the simulations in Table~\ref{tab:sims}) and the highly bimodal distribution of $p_i$. It may be that a pass over the observed data and one full sweep of MCMC (discussed in Section~\ref{section:MAVB} as a ``stronger'' MAVB) could result in more significant improvements in coverage.

Second, to examine simulations in a more realistic case, I fit a simple MRP model on the data from \cite{ghitza2013mrp} with random effects for age, income, ethnicity and state (Model 1 from Table~\ref{tab:gg_models}, below) and take the parameter estimates from the Laplace approximation as ``ground truth'' to create simulated outcomes. It shows a similar pattern although with weaker performance across the board---Scheme I after applying MAVB outperforms ADVI (Mean Field) across all measures with noticeable improvements in accuracy (10\%) and is comparable to ADVI (Full Rank). Scheme III performs the best of all approximate methods, including beating the Laplace and ADVI (Full Rank). The values of the linear predictor are relatively modest in this case (90\% of the HMC posterior means are between -0.67 and 1.99) and more comparable to those in the simulations in Table~\ref{tab:sims}. This provides further evidence that for reasonably sized linear predictors, the variational approximations perform well and are improved by MAVB.

Finally, I examine the sensitivity of the algorithm to initial values to see if there is evidence of arriving at different local optima. I find little evidence of this for the models considered in this paper given reasonable random initializations, although researchers should check for this in their own applications.

\section{Application: Estimation for Complex MRP}
\label{section:gg_app}

This section re-analyses the results in \cite{ghitza2013mrp} where I compare my results against Hamiltonian Monte Carlo (HMC).  I then conduct 10-fold cross-validation using Scheme I to examine which model seems to be most appropriate to use for the final predictive task. I find that, contrary to the decision in \citet{ghitza2013mrp}, a model with intermediate complexity is preferred.

\subsection{Brief Explanation of MRP}

Before proceeding, I provide a brief explanation of MRP (see, e.g. \citealt{park2004mrp,lax2009estimation,ghitza2013mrp} for more detailed explanations). The key problem is that while it is easy to gather a representative survey at the national level, it is very expensive to gather a sufficiently large and representative survey at sub-national units (e.g. states) or sub-types of respondents (e.g. by race, education, income, their interactions, etc.). Further, the number of observations in any sub-group may be very small, rendering a direct analysis of their values unreliable (\citealt{lax2009estimation,warshaw2012district,buttice2013mrp}). However, the most substantively important questions exactly rely on drawing inferences about those sub-groups. MRP provides a model-based procedure to attempt to reliably estimate these sub-group effects by providing a principled way to extrapolate the nationally representative survey.

MRP is a two-step procedure. First, the researcher estimates a hierarchical model  (``multilevel regression'') on the initial survey including covariates such as demographic characteristics and indicators for the relevant geographic unit (e.g. state) to get estimates for various ``types'' of respondents (e.g. age-income-ethnicity by state). The hierarchical model usually has a binomial or binary outcome. The second step calculates the expected response for each demographic-state profile. These can be examined directly or aggregated to get a measure of opinion at the desired geographic level (e.g. state). The aggregation or ``post-stratification'' occurs by taking a weighted average of those sub-group predictions from the known joint distribution in the population from some ground truth such as the Census. This paper has focused on the first step (``multilevel regression'').

\citet{ghitza2013mrp} apply this method to explore the decision to turn out to vote and party choice by age-race-income-state sub-groups in the 2004 and 2008 American presidential elections. They note that traditional MRP includes the random effects linearly and thus may be failing to capture important complexities or interactions between demography and geography. They thus fit a highly complex model with eighteen random effects and nearly 4,000 parameters on a dataset with around 4,000 observations. After doing so, they draw a variety of subtle and nuanced conclusions about the behavior of particular demographic sub-groups. For example, they qualify the conventional wisdom to show that turnout increases were concentrated amongst \emph{non-white} younger voters instead of younger white voters (\citealt[p. 771-772]{ghitza2013mrp}).

\subsection{Estimating Complex Hierarchical Models}

I begin by performing a direct comparison of Schemes I, II, and III against the gold standard approaches applied to \citet{ghitza2013mrp}. To illustrate the many specifications available to the researcher, Table~\ref{tab:gg_models} shows nine possible specifications ranging from a simple MRP model with no interactions to the preferred model in \citet{ghitza2013mrp} (Model 9). I round $y_i$ and $n_i$ to the nearest integer to facilitate interpretation as a standard binomial regression. The intermediate models represent varying complexities that allow for some, but not all, interactions.

\begin{table}[!ht]
	\caption{Nine Possible Models for Predicting Turnout via MRP}
	\label{tab:gg_models}
	\resizebox{\textwidth}{!}{
		\begin{tabular}{l*{9}r}
			\hline\hline
			& \multicolumn{9}{c}{Model} \\
			& (1) & (2) & (3) & (4) & (5) & (6) & (7) & (8) & (9) \\
			\hline
			State & $\diamond$ & $\bullet$ & $\bullet$ & $\bullet$ & $\bullet$ & $\bullet$ & $\bullet$ & $\bullet$ & $\bullet$ \\
Age & $\diamond$ & $\bullet$ & $\bullet$ & $\bullet$ & $\bullet$ & $\bullet$ & $\bullet$ & $\bullet$ & $\bullet$ \\
Eth & $\diamond$ & $\bullet$ & $\bullet$ & $\bullet$ & $\bullet$ & $\bullet$ & $\bullet$ & $\bullet$ & $\bullet$ \\
Inc & $\diamond$ & $\diamond$ & $\diamond$ & $\diamond$ & $\diamond$ & $\diamond$ & $\diamond$ & $\diamond$ & $\diamond$ \\
Region &   &   &   &   & $\bullet$ & $\bullet$ & $\bullet$ & $\bullet$ & $\bullet$ \\
State * Age &   &   &   & $\diamond$ & $\diamond$ & $\diamond$ & $\diamond$ & $\diamond$ & $\diamond$ \\
State * Eth &   &   &   & $\diamond$ & $\diamond$ & $\diamond$ & $\diamond$ & $\diamond$ & $\diamond$ \\
State * Inc &   &   &   & $\diamond$ & $\diamond$ & $\diamond$ & $\diamond$ & $\diamond$ & $\diamond$ \\
Eth * Age &   &   & $\diamond$ & $\diamond$ & $\diamond$ & $\diamond$ & $\diamond$ & $\diamond$ & $\diamond$ \\
Eth * Inc &   &   & $\diamond$ & $\diamond$ & $\diamond$ & $\diamond$ & $\diamond$ & $\diamond$ & $\diamond$ \\
Inc * Age &   &   & $\diamond$ & $\diamond$ & $\diamond$ & $\diamond$ & $\diamond$ & $\diamond$ & $\diamond$ \\
Region * Age &   &   &   &   & $\diamond$ & $\diamond$ & $\diamond$ & $\diamond$ & $\diamond$ \\
Region * Eth &   &   &   &   & $\diamond$ & $\diamond$ & $\diamond$ & $\diamond$ & $\diamond$ \\
Region * Inc &   &   &   &   & $\diamond$ & $\diamond$ & $\diamond$ & $\diamond$ & $\diamond$ \\
State * Eth * Age &   &   &   &   &   &   &   & $\diamond$ & $\diamond$ \\
State * Eth * Inc &   &   &   &   &   &   & $\diamond$ & $\diamond$ & $\diamond$ \\
State * Inc * Age &   &   &   &   &   &   &   &   & $\diamond$ \\
Eth * Inc * Age &   &   &   &   &   & $\diamond$ & $\diamond$ & $\diamond$ & $\diamond$ \\

			\hline
			&\multicolumn{9}{c}{\emph{Number of Parameters}} \\
			 & 74 & 139 & 198 & 864 & 945 & 1026 & 2047 & 2864 & 3885 \\

			\hline
			&\multicolumn{9}{c}{\emph{Run Time of Model in Minutes}} \\
			Laplace - 2004 & 0.2 & 0.9 & 2.2 & 5.1 & 13.4 & 27.0 & 38.4 & 50.1 & 81.1 \\
Laplace - 2008 & 0.2 & 0.8 & 2.2 & 5.5 & 12.1 & 23.9 & 37.0 & 52.5 & 84.9 \\
HMC - 2004 & 113.8 & 131.5 & 166.9 & 199.9 & 326.6 & 288.6 & 375.4 & 430.0 & 469.0 \\
HMC - 2008 & 101.5 & 132.4 & 174.1 & 196.6 & 353.8 & 327.3 & 307.9 & 402.5 & 463.8 \\

			\hline\hline
	\end{tabular}}
	\bigskip
	\begin{tabular}{l}
		\begin{minipage}[b]{\textwidth}
			\footnotesize
			\emph{Note}: This table summarizes nine possible models to predict voter turnout. All models include six fixed effects: an intercept, (standardized) individual income, state-level income, state-level Republican vote share and the interaction between individual income and the latter two variables. \citet{ghitza2013mrp} use Model 9. The first panel indicates which random effects are included; a hollow diamond ($\diamond$) indicates that only a random intercept is used. A solid circle ($\bullet$) indicates that a random intercept and a random slope allowing for the effect of (standardized) individual income to vary by group are included. The number of parameters is the number of fixed effects, random effects, and variance components for the random effects. The run times are for a Laplace approximation using \texttt{blme} (\citealt{bates2015lmer,chung2015weakly}) and HMC in \texttt{STAN} (via \texttt{brms}; \citealt{burkner2017brms}). All models were run on an instance with 16 GB of memory and 4 cores. HMC was estimated using four chains distributed in parallel.
		\end{minipage}
	\end{tabular}
\end{table}

It clearly shows the scale of the difficulty for applied researchers: Fitting the published model (Model 9) takes hours using either specification on a machine similar to that available for many applied researchers (a Microsoft Azure instance; Ubuntu, 4 cores, 16 GB of RAM). Methods that require fitting the model repeatedly to facilitate common tasks as bootstrapping, model comparison via cross-validation, or ensemble analysis (\citealt{van2007super}, see \citealt{ornstein2019stacked} for an application to MRP) are clearly prohibitively expensive for all except the simplest models using the Laplace approximation.

\begin{figure}[!ht]
	\caption{Speed of Estimation}
	\label{fig:speed}
	\includegraphics[width=\textwidth]{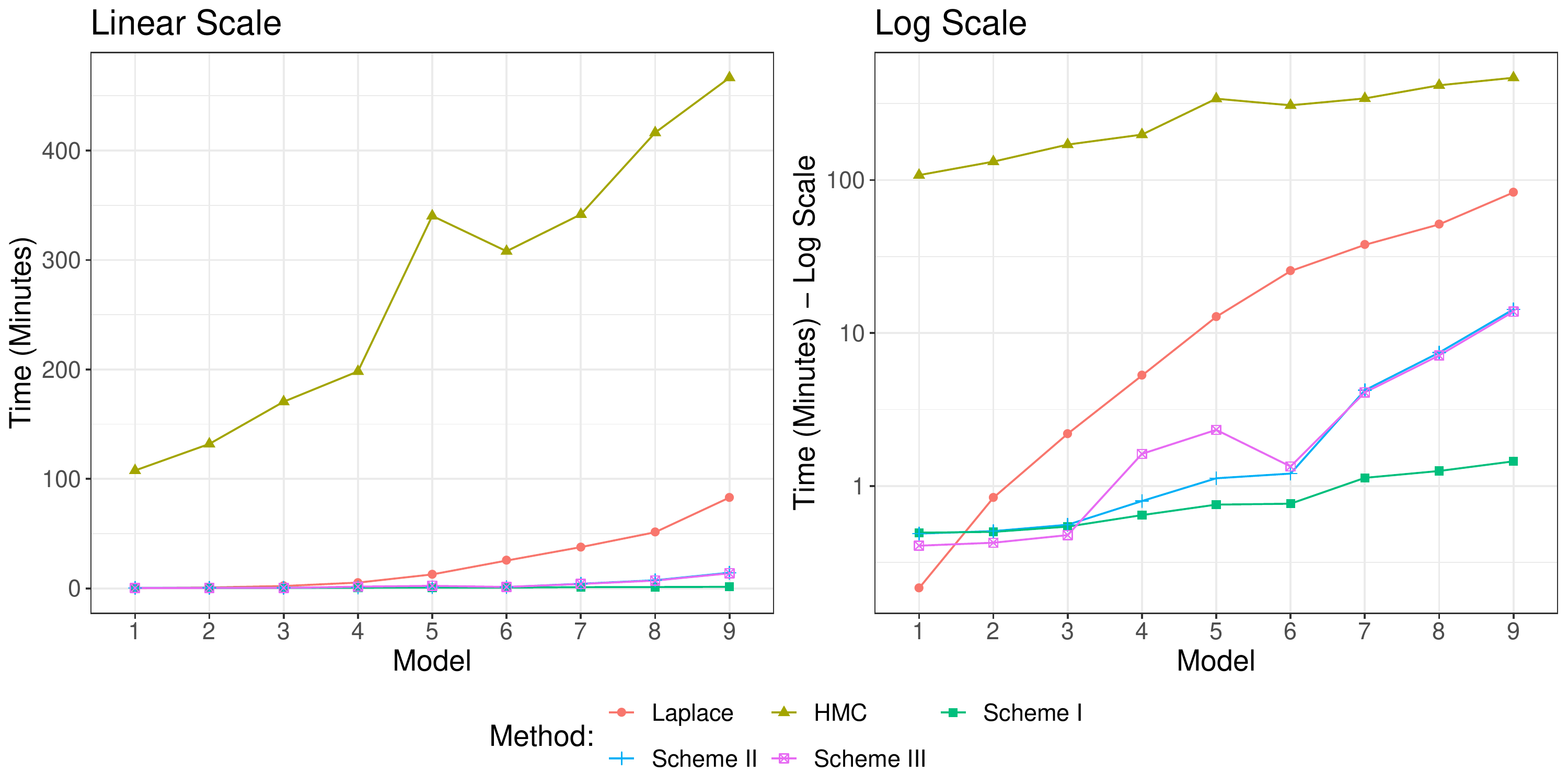}
	\caption*{\footnotesize \emph{Note}: Each figure plots the run-time of each of the five methods (Laplace approximation, Hamiltonian Monte Carlo [HMC], Schemes I-III with drawing 4,000 samples using MAVB). The reported times are averaged across the 2004 and 2008 elections. The left figure shows the time in minutes on a linear scale; the right figure reports the same information on a log-scale. Model 1-9 are described in Table~\ref{tab:gg_models}. All models are fit on a computer with 16 GB of RAM and 4 cores.}
\end{figure}

Figure~\ref{fig:speed} illustrates the improvement after applying the variational algorithms to each model in Table~\ref{tab:gg_models} and performing MAVB. All reported times include estimation of the variational algorithm and drawing 4,000 samples using MAVB. Appendix~\ref{section_app:gg_extra} shows the time for estimation and MAVB separately; it takes around thirty seconds for Scheme I on the most complex model.

As shown on a linear scale, the time to estimate either the Laplace approximation or Hamiltonian Monte Carlo dwarfs that of any of the variational schemes. The right panel shows the results on a log-scale to allow for clearer comparisons; it shows that Scheme I remains remarkably fast estimating even Model 9 in around one minute versus hour(s) for either gold standard method. The performance of Schemes II and III degrade somewhat---taking around fifteen minutes to fit. This is still very reasonable, but may still be onerous if repeated fitting is required as in cross-validation.

\begin{figure}[!ht]
	\caption{Comparing Posterior Means}
	\label{fig:gg_post_means}
	\includegraphics[width=\textwidth]{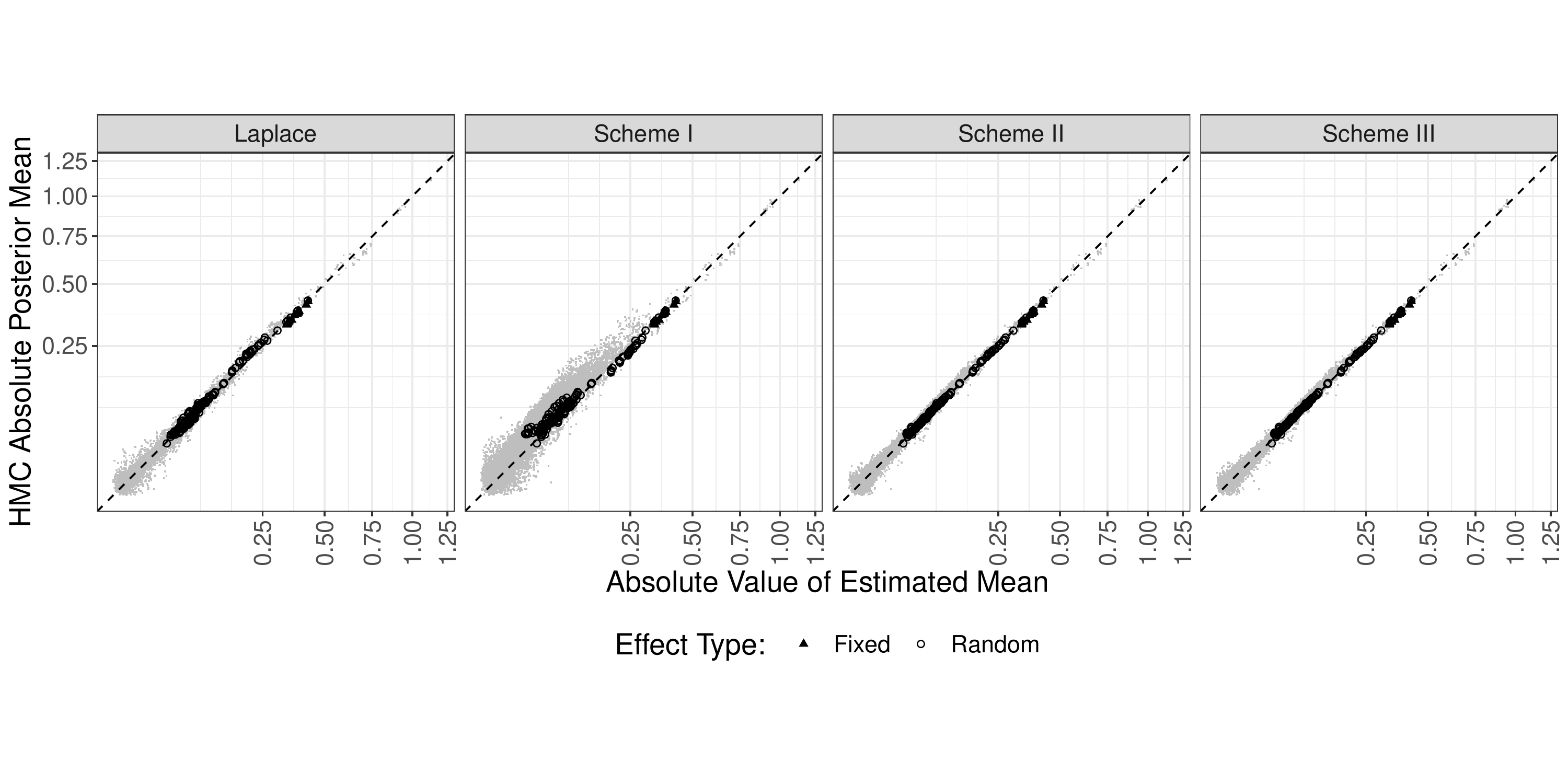}
	\caption*{\footnotesize \emph{Note}: This figure plots the absolute value of estimated mean value from Schemes I-III and the Laplace approximation on the horizontal axis against the absolute value of the posterior mean from Hamiltonian Monte Carlo [HMC] on the vertical axis. Each parameter is plotted as a thin grey point; the average of the values inside each random effect are shown as larger points. The axes are on a square-root scale.}
\end{figure}

The quality of the approximation is also crucial to assess. As the truth is unknown, I do this by comparing all methods against HMC as this seeks most directly to sample the posterior.\footnote{This method is, of course, itself approximate as it may fail to accurately sample the posterior. Experiments suggested that setting ``adapt delta'' to 0.99 was required to eliminate all divergent transitions (except for one in Model 3 in 2004).} Figure~\ref{fig:gg_post_means} begins by comparing the point estimates pooling across the 18 models. As there are thousands of parameters to plot, I simplify the picture in the following way; I plot the absolute magnitude of the estimates averaged across $j$: $\bar{\alpha}_j = \frac{1}{g_j} \sum_{g=1}^{g_j} |\alpha_{j,g}|$ in solid circles and shade the background of the plot based on the density of the individual $|\alpha_{j,g}|$. This prevents the domination of the $j$ with smaller numbers of groups (e.g. age, income, etc.) in the visualization. I also separately mark the fixed and random effects.

Consider first the Laplace approximation; it nearly exactly recovers the point estimates---its solid points and shading lie very near to the 45-degree line. For the variational methods, Scheme I is highly correlated with the posterior ($\rho = 0.996$ for $\bar{\alpha}_j$; $\rho = 0.964$ for the raw $|\alpha_{j,g}|$) although less so than the Laplace approximation. Schemes II and III show tight coupling with the estimates from HMC and are effectively equally accurate to the Laplace approximation. This matches the conventional wisdom that variational methods typically well-recover the posterior means.

\begin{figure}[!ht]
	\caption{Comparing Posterior Variability}
	\label{fig:gg_post_var}
	\includegraphics[width=\textwidth]{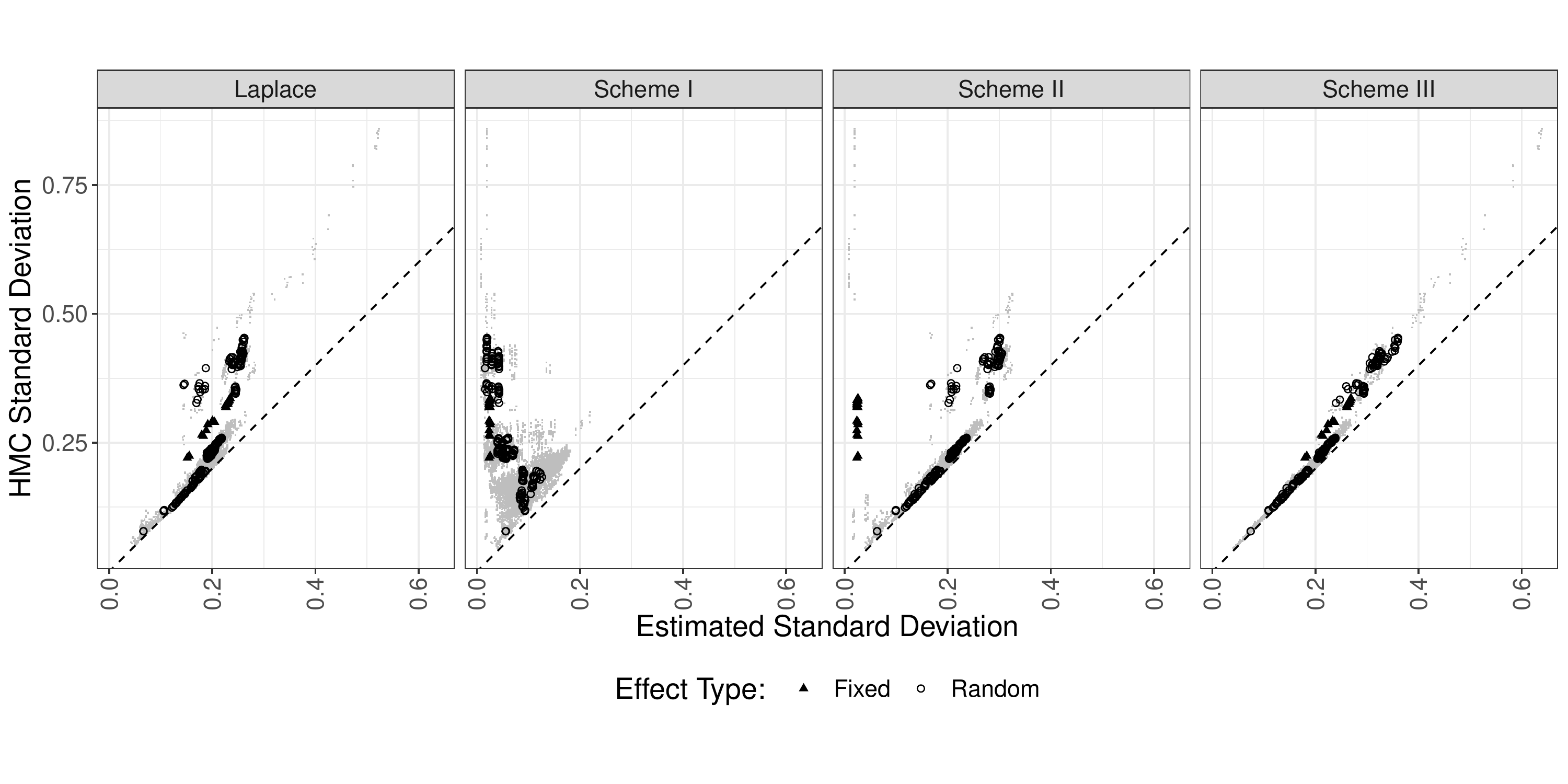}
	\caption*{\footnotesize \emph{Note}: This figure plots the estimated standard deviation from Schemes I-III and the Laplace approximation on the horizontal axis against the standard deviation of the posterior distribution from Hamiltonian Monte Carlo [HMC] on the vertical axis. Each parameter is plotted as a thin grey point; the average of the values inside each random effect are shown as larger points.}
\end{figure}

Figure~\ref{fig:gg_post_var} presents an analogous figure for the posterior variability, plotting the standard deviation of each parameter. It smooths across random effects in the same way as  Figure~\ref{fig:gg_post_means}. In interpreting this figure, note that points in the upper left quadrant (above the 45-degree line) indicate worrying performance as the posterior variability is below that coming from the HMC estimates.

Again consider first the Laplace approximation; the standard deviation of its point estimates are often tightly clustered near the 45-degree line but there are a number of random effects that are noticeably smaller (above the 45-degree line).

The performance for the variational algorithms is rather mixed, by comparison. Looking at Scheme I, almost all points show a too small standard deviation---with many random effects being considerably too small. Scheme III, however, improves the situation markedly. While slightly smaller---especially for points with large standard deviations--it tracks the 45-degree line closely and has better performance than the Laplace approximation. As expected, Scheme II is somewhat of an intermediate case; improving some parameters but still having significant problems.

Overall, therefore, Schemes I and II fall into the usual problem of understating posterior variance. By contrast, Scheme III appears to do rather well and lacks the obvious problems of lack of posterior variability versus a fully Bayesian baseline. This corroborates results from \cite{menictas2019streamlined} that estimating $q(\bm{\beta}, \bm{\alpha})$ jointly performs well for (linear) hierarchical models with $J > 1$.

\begin{figure}[!ht]
	\caption{Improvements from MAVB}
	\label{fig:gg_mavb}
	\includegraphics[width=\textwidth]{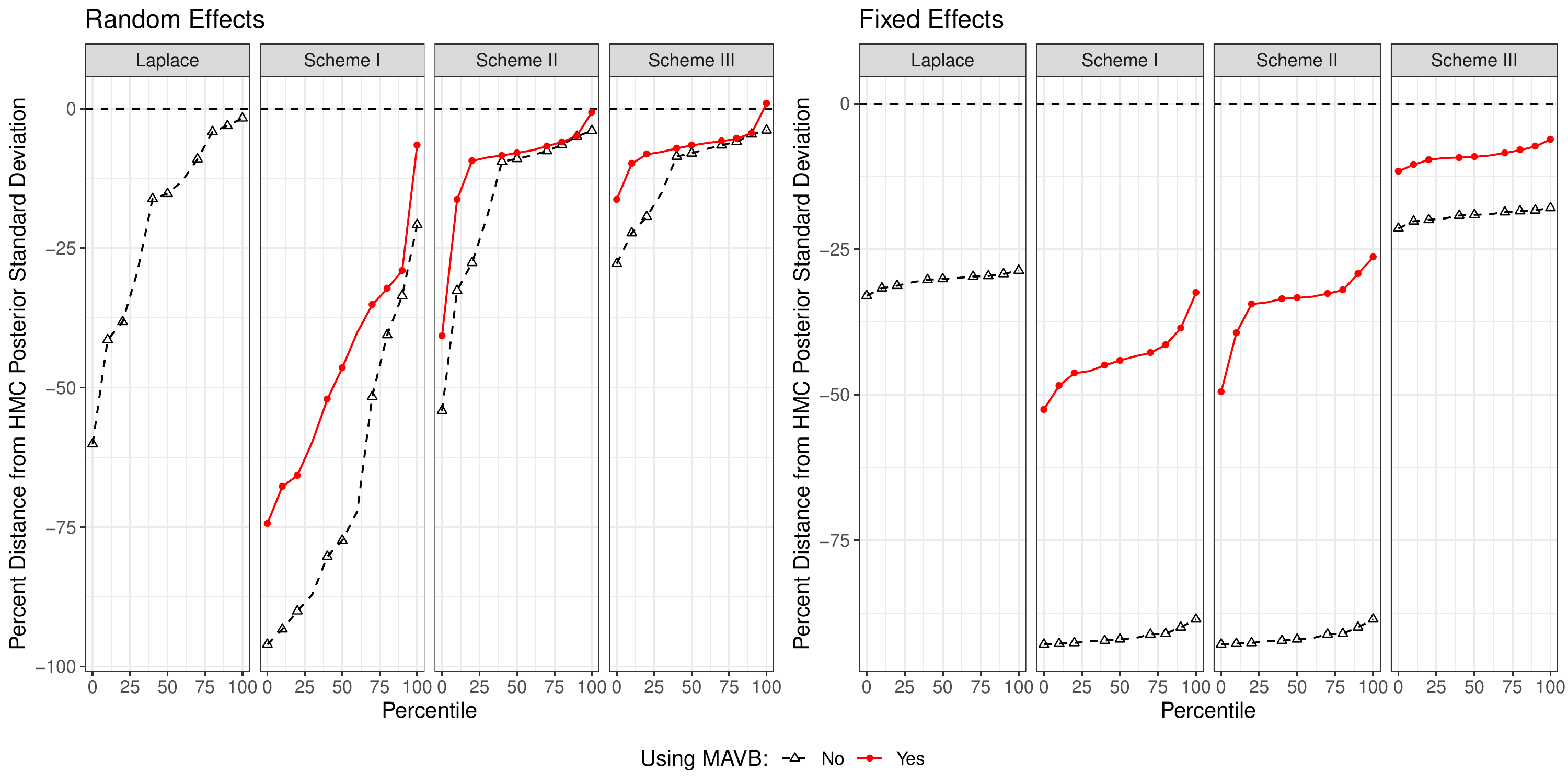}
	\caption*{\footnotesize \emph{Note}: This figure plots the percentile of the gap between the standard deviations estimated via Hamiltonian Monte Carlo [HMC] and the approximate methods. The percentage gap, i.e. $\left(\mathrm{sd}^{\mathrm{Laplace}}_k - \mathrm{sd}^{\mathrm{HMC}}_k\right)/\mathrm{sd}^{\mathrm{HMC}}_k \cdot 100$, is shown. A negative value on the vertical axis indicates that the corresponding percentile has a smaller variance than HMC. A vertical shift upward of the line indicates the variance of the parameters has increased. The solid markers indicate the deciles and extremes of the distribution. The dashed line with hollow triangles represents the estimates without using MAVB. The red line with solid circles represents the results after using MAVB.}
\end{figure}

Finally, I show how these estimates change when using MAVB. I focus on the effect on posterior variability as the means are not materially affected by MAVB; Appendix~\ref{section_app:gg_extra} shows the analogous figure. Figure~\ref{fig:gg_mavb} presents the distribution of the gap between the variability between the HMC estimates and the other methods where negative values indicates a smaller standard deviation for the competitor methods. Any point below the dotted line indicates that that percentile of effects has a smaller standard deviation than the HMC estimates. To make results interpretable, I report the percentage gap, e.g. $\left(\mathrm{sd}^{\mathrm{Laplace}}_k - \mathrm{sd}^{\mathrm{HMC}}_k\right)/\mathrm{sd}^{\mathrm{HMC}}_k \cdot 100$ for all parameters $k$ in $(\bm{\beta}, \bm{\alpha})$. To ensure that random effects with small $g_j$ are counted, it presents the averaged statistic across $g$ as in Figures~\ref{fig:gg_post_means} and~\ref{fig:gg_post_var}. 

The results provide clear evidence for the important role of MAVB. Considering first random effects in the left panel, it is first worth noting that the Laplace approximation---commonly used by researchers---has poor performance for a number of parameter blocks (e.g. the lower percentiles). Scheme I shows a clear lack of variability in the posterior estimates with all estimates being estimated at least 20\% too precisely and around \emph{half} of all estimates having less than 75\% of the variability estimated in the fully Bayesian setting. After applying MAVB, the (red) solid line shows a considerable improvement although still markedly below the HMC estimates and performing worse than the Laplace approximation. Large improvements are seen for the fixed effects ($\bm{\beta}$) where the estimates of the variability go from extremely poor to being much closer to the Laplace approximation which, itself, is markedly below the HMC coverage.

Scheme III is worth also considering in detail; even before applying MAVB, it has stronger performance than the Laplace approximation in that its curve has a much less poor ``tail'' (i.e. its worst blocks are around 25\% too small vs around 60\% for the Laplace approximation). MAVB provides some additional gains ensuring that most parameter blocks are only around 10\% too small in terms of their variability. Scheme II is again somewhat intermediate; after applying MAVB, it is broadly comparable to the Laplace approximation.

To provide another interpretation of the role of MAVB, consider the accuracy measure in Section~\ref{section:simulations} that measures similarity between two distributions, averaged within and then across parameter blocks: The Laplace approximation performs relatively well (90\%). Scheme I performs poorly (43\%) because it clearly fails to capture the posterior variance. MAVB increases this considerably (68\%) although it still falls below the Laplace approximation. Scheme III, however, out-performs the Laplace approximation (95\%) with a slight improvement from MAVB (97\%).

Appendix~\ref{section_app:gg_extra} provides some additional results. First, it breaks apart Figure~\ref{fig:gg_post_var} by the type of random effect; the main implication is that the initial lack of variability from Scheme I is most pronounced for the fixed effects and random effects with small $g_j$ (age, ethnicity, income). The improvements for MAVB for those random effects are large and resolve the much of the negative gap.

Second, it examines the linear predictor (i.e. $\bm{x}_i^T\bm{\beta} + \bm{z}_i^T\bm{\alpha}$). It shows that MAVB has little effect, although all schemes perform well. In addition to closely estimating the posterior mean (Scheme I has a bias of -0.002 vs HMC), the standard deviation is also fairly close (bias of -0.013 or about -2\%), especially compared to the gaps seen in Figure~\ref{fig:gg_mavb}. A conjecture would be that MAVB as implemented here has little impact on the linear predictor as it is more about building correlations between parameter blocks, but the ``stronger'' MAVB noted above might address such limitations.

\subsection{Choosing an Optimal Model}

Finally, I return to the substantive analysis in \citet{ghitza2013mrp}. A key question when performing MRP is the complexity of the accompanying model. Even with the regularization implied by the hierarchical effects, it is still possible to over-fit to the survey sample (\citealt{goplerud2018sparse}). The reported analysis relies on Model 9 without exploring this possibility. The computational burden needed to estimate multiple models and thereby engage in model testing and checking is often onerous for the applied researcher. I thus use the ability to rapidly fit variational approximations to deploy a standard model comparison technique (cross-validation) and examine whether a model of intermediate complexity should be preferred. Table~\ref{tab:gg_cv} reports a number of statistics on model fit.

\begin{table}[!ht]
	\caption{Cross-Validation to Choose Optimal Model}
	\label{tab:gg_cv}
	\resizebox{\textwidth}{!}{
		\begin{tabular}{l*{9}cr}
			\hline\hline
			\multicolumn{1}{l}{\multirow{2}{*}{Method}} 
			& \multicolumn{9}{c}{Models} & \multirow{2}{*}{Time} \\
			& 1 & 2 & 3 & 4 & 5 & 6 & 7 & 8 & 9 &  \\
			\hline
			& \multicolumn{9}{c}{2004 Election} & \\
			LOO & 11659 & 11339 & 11296 & 11150 & 11163 & 11164 & 11155 & 11170 & \textbf{11129} & 2504 \\
WAIC & 11658 & 11336 & 11292 & 11107 & 11115 & 11113 & 11083 & 11074 & \textbf{10968} & 2503 \\
VI-CV & 28.617 & 28.518 & 28.508 & \textbf{28.488} & 28.498 & 28.499 & 28.499 & 28.508 & 28.508 &   21 \\
\hline
 & \multicolumn{9}{c}{2008 Election} & \\
LOO & 11651 & 11270 & 11236 & 11095 & 11112 & 11118 & 11113 & 11111 & \textbf{11094} & 2463 \\
WAIC & 11651 & 11268 & 11233 & 11060 & 11073 & 11075 & 11053 & 11030 & \textbf{10957} & 2462 \\
VI-CV & 26.963 & 26.846 & 26.839 & \textbf{26.813} & 26.817 & 26.821 & 26.824 & 26.829 & 26.820 &   20 \\

			\hline\hline\\
	\end{tabular}}
	\bigskip
	\begin{tabular}{l}
		\begin{minipage}{\textwidth}%
			\footnotesize \emph{Note} This table reports statistics for model fit. The first two rows for each election report fit statistics on the model estimated via Hamiltonian Monte Carlo that approximate cross-validation; the ``LOO'' information criterion and the WAIC information criterion (\citealt{gelman2013understanding,vehtari2017practical}). The third row reports the average out-of-sample deviance from a model fit using Scheme I. For all statistics, smaller is better and the best value is bolded. The time in minutes for each row to be estimated is shown in the final column; this includes estimation time and the time needed to estimate the relevant fit statistic.
		\end{minipage}
	\end{tabular}
\end{table}

The first two rows (LOO and WAIC) are popular tools for deciding between non-nested Bayesian models (\citealt{vehtari2017practical}). Details on their exact calculation can be found in the relevant articles (\citealt{gelman2013understanding,vehtari2017practical}), but both are designed to be 
approximations to cross-validation that do not require fitting the Bayesian model repeatedly. 

Fortunately, both have diagnostics to assess whether the underlying approximations are reliable; unfortunately, the diagnostics tests fail in this setting. Almost all models report unacceptable violations of the underlying assumptions for both the LOO and WAIC, and the associated software explicitly encourages the user to resort to $K$-fold cross-validation. On the other hand, variational inference provides a fast approximate method. The final row of the table (VI-CV) reports the average deviance (twice the negative log-likelihood) of the held-out predictions after conducting 10-fold cross-validation where observations are allocated to each fold with equal probability using Scheme I. Formally, if observation $i$ has a prediction $\hat{p}_i$, the individual deviance is $-2 \left[y_i \ln(\hat{p}_i) + (n_i - y_i) \ln(1-\hat{p}_i)\right]$. Observations with $n_i = 0$ are excluded from the reported average. Model 4 is also selected if Schemes II or III are used.

The results are interesting and push against the decision to use Model 9; it finds that while Model 1 performs noticeably worse than all other models, it is not necessarily best to use the most complex model. Indeed, an intermediate model---Model 4---performs the best although the differences in the error are quantitatively small between Models 4 and 9. 

Appendix~\ref{section_app:gg_extra} provides a more detailed exploration against a ``Bayesian gold standard.'' Using a new set of folds, it fits Models 1, 4, and 9 using ten-fold cross-validation and performing HMC and the Laplace approximation on each fold. This is extremely time intensive---taking around \emph{ten} days to complete the whole process. It confirms that cross-validated HMC, Laplace approximation, and Scheme I all select Model 4. The out-of-sample predictions between Scheme I and HMC are highly correlated (0.998). This gives some confidence that the results of the variational method can be used in lieu of prohibitively expensive classical cross-validation. When it is too expensive to conduct such an analysis, relying on methods such as simulation-based calibration (e.g. \citealt{yao2018work}) may be a feasible way to assess whether the variational approximation ``successfully'' approximated the posterior.

Returning to Table~\ref{tab:gg_models}, the major feature that distinguishes Model 4 from less complex models is interactions between the core random effects (age, ethnicity, income) and state. This matches a reasonable expectation from political science that demographics are likely to vary across state but the complex higher-order interactions between region and three-way-interactions do not seem to add much predictive power. 

These results are useful to practitioners of MRP in three ways; first, complex hierarchical models can now be compared against other state-of-the-art machine learning methods versus relying on a very simple model (analogous to Model 1) due to computational costs (\citealt{bisbee2019barp,ornstein2019stacked}). Thus, it is an interesting and open question whether methods such as BART are actually superior for MRP tasks (\citealt{bisbee2019barp}) or whether properly specified complex hierarchical models can be competitive. Second, it suggests that interactions between demographics and state characteristics are important to include although the evidence for going extremely ``deep'' and adding many higher-order interactions appears more limited. Finally, even if one prefers to fit a Bayesian model for the final regression, the ability to quickly search between models allows the researcher to narrow down a set of plausible candidate models for final exploration and model testing.

\section{Conclusion}
\label{section:conclusion}

This paper provided a new set of variational algorithms that, leveraging Polya-Gamma data augmentation (\citealt{polson2013polyagamma}), require only a mean-field assumption to estimate a logistic hierarchical regression with an arbitrary number and size of random effects. It provided multiple factorization assumptions; Scheme I required the independence of the fixed effects and each block of random effects whereas Scheme III relaxed that assumption at the expense of increased computational cost. All methods seemed to quite accurately capture the posterior means in even complex models. As expected in both simulations and real data, Scheme I performed worse--especially in terms of understating posterior variance for many random effects.

The paper also provided a generic way to improve the performance of Scheme I, and Schemes II and III to a lesser extent. By leveraging the existence of a parameter expansion of the underlying model either by allowing the means of the random effects to be non-zero or by imposing some translation, one can use a marginal augmentation sampler to improve the posterior approximation. This procedure (``marginally augmented variational Bayes''; MAVB) showed promising performance when applied to Scheme I: It increased the variance of the estimated approximations to be closer to the samples drawn using a fully Bayesian procedure, although still remaining too small on real data. However, given its speed even on complex models, MAVB provides a cheap way to make Scheme I a more viable approximation to the true posterior. It is also worth noting that Scheme III performed very well---often beating the very popular Laplace approximation on both real and simulated data.

Future work could proceed in at least two directions. First, the algorithms here can be naturally extended to count and multinomial outcomes providing a more unified approach to variational estimation of non-linear hierarchical models. Extending the model to include a weakly informative prior such as \cite{huang2013simple} is also an important extension.

Second, the usefulness of MAVB should be explored both theoretically and in the context of other models. As noted earlier, there is nothing about using MAVB that is specific to logistic hierarchical models \emph{per se}. Indeed, this idea of ``improving'' an approximation by pushing it through a Markov transition kernel can be generalized to a wide variety of MCMC samplers and models. It thus opens a question of which Markov transition density to use for other models that do not admit marginal augmentation. A reasonable conjecture is that as the mixing of the sampler improves, the transformed sample will be closer to the true posterior.

\section*{Supplemental Materials}
The supporting information contains derivations of the variational algorithm (Appendix~\ref{section_app:pg}), formal definitions and proofs of MAVB (Appendix~\ref{section_app:param_x}), results on accelerating CAVI using PX-VB and joint updates of certain parameters (Appendix~\ref{section_app:acceleration}), additional simulations (Appendix~\ref{section_app:simulations_extra}), and additional analyses on \citet{ghitza2013mrp} (Appendix~\ref{section_app:gg_extra}).

Open-source statistical software to implement the algorithms in this paper is available on GitHub as noted in the acknowledgements. Materials to replicate the analyses in the paper can be found at the following link: \url{https://doi.org/10.7910/DVN/DI19IB}.

\bibliographystyle{ba}
\bibliography{MAVB_bib}

\begin{thebibliography}{56}
\newcommand{\enquote}[1]{``#1''}
\expandafter\ifx\csname natexlab\endcsname\relax\def\natexlab#1{#1}\fi
\expandafter\ifx\csname url\endcsname\relax
  \def\url#1{{\tt #1}}\fi
\expandafter\ifx\csname urlprefix\endcsname\relax\def\urlprefix{URL }\fi
\ifx\endbibitem\undefined \let\endbibitem\relax\fi

\bibitem[{Bates et~al.(2015)Bates, M{ä}chler, Bolker, and
  Walker}]{bates2015lmer}
Bates, D., M{ä}chler, M., Bolker, B., and Walker, S. (2015).
\newblock \enquote{Fitting Linear Mixed-Effects Models Using lme4.}
\newblock {\em Journal of Statistical Software\/}, 67(1): 1--48.
\endbibitem

\bibitem[{Bell and Jones(2015)}]{bell2015explaining}
Bell, A. and Jones, K. (2015).
\newblock \enquote{Explaining Fixed Effects: Random Effects Modeling of
  Time-Series Cross-Sectional and Panel Data.}
\newblock {\em Political Science Research and Methods\/}, 3(1): 133--153.
\endbibitem

\bibitem[{Bisbee(2019)}]{bisbee2019barp}
Bisbee, J. (2019).
\newblock \enquote{{BARP}: Improving {Mister} {P} Using Bayesian Additive
  Regression Trees.}
\newblock {\em American Political Science Review\/}, 113(4): 1060--1065.
\endbibitem

\bibitem[{Bishop(2006)}]{bishop2006ml}
Bishop, C. (2006).
\newblock {\em Pattern Recognition and Machine Learning\/}.
\newblock Springer.
\endbibitem

\bibitem[{Blei et~al.(2017)Blei, Kucukelbir, and McAuliffe}]{blei2017vi}
Blei, D.~M., Kucukelbir, A., and McAuliffe, J.~D. (2017).
\newblock \enquote{Variational Inference: A Review for Statisticians.}
\newblock {\em Journal of the American Statistical Association\/}, 112(518):
  859--877.
\endbibitem

\bibitem[{B{ü}rkner(2017)}]{burkner2017brms}
B{ü}rkner, P.-C. (2017).
\newblock \enquote{brms: An R Package for Bayesian Multilevel Models Using
  Stan.}
\newblock {\em Journal of Statistical Software\/}, 80(1): 1--28.
\endbibitem

\bibitem[{Buttice and Highton(2013)}]{buttice2013mrp}
Buttice, M.~K. and Highton, B. (2013).
\newblock \enquote{How Does Multilevel Regression and Poststratification
  Perform with Conventional National Surveys?}
\newblock {\em Political Analysis\/}, 21(4): 449--467.
\endbibitem

\bibitem[{Carpenter et~al.(2017)Carpenter, Gelman, Hoffman, Lee, Goodrich,
  Betancourt, Brubaker, Guo, Li, and Riddell}]{carpenter2017stan}
Carpenter, B., Gelman, A., Hoffman, M.~D., Lee, D., Goodrich, B., Betancourt,
  M., Brubaker, M., Guo, J., Li, P., and Riddell, A. (2017).
\newblock \enquote{Stan: A Probabilistic Programming Language.}
\newblock {\em Journal of Statistical Software\/}, 76(1): 1--32.
\endbibitem

\bibitem[{Chung et~al.(2015)Chung, Gelman, Rabe-Hesketh, Liu, and
  Dorie}]{chung2015weakly}
Chung, Y., Gelman, A., Rabe-Hesketh, S., Liu, J., and Dorie, V. (2015).
\newblock \enquote{Weakly Informative Prior for Point Estimation of Covariance
  Matrices in Hierarchical Models.}
\newblock {\em Journal of Educational and Behavioral Statistics\/}, 40(2):
  136--157.
\endbibitem

\bibitem[{Clark and Linzer(2015)}]{clark2015fe}
Clark, T.~S. and Linzer, D.~A. (2015).
\newblock \enquote{Should {I} Use Fixed or Random Effects?}
\newblock {\em Political Science Research and Methods\/}, 3(2): 399--408.
\endbibitem

\bibitem[{Cover and Thomas(2006)}]{cover2006elements}
Cover, T.~M. and Thomas, J.~A. (2006).
\newblock {\em Elements of Information Theory\/}.
\newblock John Wiley \& Sons.
\endbibitem

\bibitem[{Faes et~al.(2011)Faes, Ormerod, and Wand}]{faes2011variational}
Faes, C., Ormerod, J.~T., and Wand, M.~P. (2011).
\newblock \enquote{Variational Bayesian Inference for Parametric and
  Nonparametric Regression with Missing Data.}
\newblock {\em Journal of the American Statistical Association\/}, 106(495):
  959--971.
\endbibitem

\bibitem[{Gao et~al.(2020)Gao, Kennedy, Simpson, and Gelman}]{gao2019improving}
Gao, Y., Kennedy, L., Simpson, D., and Gelman, A. (2020).
\newblock \enquote{Improving Multilevel Regression and Poststratification with
  Structured Priors.}
\newblock {\em Bayesian Analysis\/}, Advanced Access: doi 10.1214/20-BA1223.
\endbibitem

\bibitem[{Gelman and Hill(2006)}]{gelman2006multi}
Gelman, A. and Hill, J. (2006).
\newblock {\em Data Analysis Using Regression and Multilevel/Hierarchical
  Models\/}.
\newblock Cambridge University Press.
\endbibitem

\bibitem[{Gelman et~al.(2014)Gelman, Hwang, and
  Vehtari}]{gelman2013understanding}
Gelman, A., Hwang, J., and Vehtari, A. (2014).
\newblock \enquote{Understanding Predictive Information Criteria for {Bayesian}
  Models.}
\newblock {\em Statistics and Computing\/}, 24(6): 997--1016.
\endbibitem

\bibitem[{Gelman et~al.(2016)Gelman, Lax, Phillips, Gabry, and
  Trangucci}]{gelman2016using}
Gelman, A., Lax, J., Phillips, J., Gabry, J., and Trangucci, R. (2016).
\newblock \enquote{Using Multilevel Regression and Poststratification to
  Estimate Dynamic Public Opinion.}
\newblock {\em Unpublished manuscript\/}.
\endbibitem

\bibitem[{Gelman and Little(1997)}]{gelman1997mrp}
Gelman, A. and Little, T.~C. (1997).
\newblock \enquote{Poststratification Into Many Categories Using Hierarchical
  Logistic Regression.}
\newblock {\em Survey Methodology\/}, 23: 127–--135.
\endbibitem

\bibitem[{Gelman et~al.(2008)Gelman, van Dyk, Huang, and
  Boscardin}]{gelman2008px}
Gelman, A., van Dyk, D.~A., Huang, Z., and Boscardin, J.~W. (2008).
\newblock \enquote{Using Redundant Parameterizations to Fit Hierarchical
  Models.}
\newblock {\em Journal of Computational and Graphical Statistics\/}, 17(1):
  95--122.
\endbibitem

\bibitem[{Gerchinovitz et~al.(2020)Gerchinovitz, M{\'e}nard, and
  Stoltz}]{gerchinovitz2020fanno}
Gerchinovitz, S., M{\'e}nard, P., and Stoltz, G. (2020).
\newblock \enquote{Fano's Inequality for Random Variables.}
\newblock {\em Statistical Science\/}, 35: 178--201.
\endbibitem

\bibitem[{Ghitza and Gelman(2013)}]{ghitza2013mrp}
Ghitza, Y. and Gelman, A. (2013).
\newblock \enquote{Deep Interactions with {MRP}: Election Turnout and Voting
  Patterns Among Small Electoral Subgroups.}
\newblock {\em American Journal of Political Science\/}, 57(3): 762--776.
\endbibitem

\bibitem[{Giordano et~al.(2015)Giordano, Broderick, and
  Jordan}]{giordano2015lrvb}
Giordano, R.~J., Broderick, T., and Jordan, M.~I. (2015).
\newblock \enquote{Linear Response Methods for Accurate Covariance Estimates
  from Mean Field Variational Bayes.}
\newblock In {\em Neural {Information} {Processing} {Systems} 2015\/}.
\endbibitem

\bibitem[{Goplerud et~al.(2018)Goplerud, Kuriwaki, Ratkovic, and
  Tingley}]{goplerud2018sparse}
Goplerud, M., Kuriwaki, S., Ratkovic, M., and Tingley, D. (2018).
\newblock \enquote{Sparse Multilevel Regression (and Poststratification
  [sMRP]).}
\newblock {\em Unpublished manuscript\/}.
\endbibitem

\bibitem[{Hall et~al.(2019)Hall, Johnstone, Ormerod, Wand, and
  Yu}]{hall2019fast}
Hall, P., Johnstone, I.~M., Ormerod, J.~T., Wand, M.~P., and Yu, J.~C. (2019).
\newblock \enquote{Fast and Accurate Binary Response Mixed Model Analysis via
  Expectation Propagation.}
\newblock {\em Journal of the American Statistical Association\/}, 1--15.
\endbibitem

\bibitem[{Hall et~al.(2011)Hall, Ormerod, and Wand}]{hall2011poisson}
Hall, P., Ormerod, J.~T., and Wand, M.~P. (2011).
\newblock \enquote{Theory of Gaussian Variational Approximation for a Poisson
  Mixed Model.}
\newblock {\em Statistica Sinica\/}, 21(1): 369--389.
\endbibitem

\bibitem[{Huang and Wand(2013)}]{huang2013simple}
Huang, A. and Wand, M.~P. (2013).
\newblock \enquote{Simple Marginally Noninformative Prior Distributions for
  Covariance Matrices.}
\newblock {\em Bayesian Analysis\/}, 8(2): 439--452.
\endbibitem

\bibitem[{Jaakkola and Qi(2007)}]{jaakkola2007parameter}
Jaakkola, T.~S. and Qi, Y. (2007).
\newblock \enquote{Parameter Expanded Variational Bayesian Methods.}
\newblock In {\em Neural {Information} {Processing} {Systems} 2007\/}.
\endbibitem

\bibitem[{Jeon et~al.(2017)Jeon, Rijmen, and
  Rabe-Hesketh}]{jeon2017variational}
Jeon, M., Rijmen, F., and Rabe-Hesketh, S. (2017).
\newblock \enquote{A Variational Maximization--Maximization Algorithm for
  Generalized Linear Mixed Models with Crossed Random Effects.}
\newblock {\em Psychometrika\/}, 82(3): 693--716.
\endbibitem

\bibitem[{Kucukelbir et~al.(2017)Kucukelbir, Tran, Ranganath, Gelman, and
  Blei}]{kuckelbir2017advi}
Kucukelbir, A., Tran, D., Ranganath, R., Gelman, A., and Blei, D.~M. (2017).
\newblock \enquote{Automatic Differentiation Variational Inference.}
\newblock {\em Journal of Machine Learning Research\/}, 18(14): 1--45.
\endbibitem

\bibitem[{Lax and Phillips(2009{\natexlab{a}})}]{lax2009gay}
Lax, J.~R. and Phillips, J.~H. (2009{\natexlab{a}}).
\newblock \enquote{Gay Rights in the States: Public Opinion and Policy
  Responsiveness.}
\newblock {\em American Political Science Review\/}, 103(3): 367--386.
\endbibitem

\bibitem[{Lax and Phillips(2009{\natexlab{b}})}]{lax2009estimation}
--- (2009{\natexlab{b}}).
\newblock \enquote{How Should We Estimate Public Opinion in The States?}
\newblock {\em American Journal of Political Science\/}, 53(1): 107--121.
\endbibitem

\bibitem[{Lax and Phillips(2012)}]{lax2012deficit}
--- (2012).
\newblock \enquote{The Democratic Deficit in the States.}
\newblock {\em American Journal of Political Science\/}, 56(1): 148--166.
\endbibitem

\bibitem[{Liu et~al.(1998)Liu, Rubin, and Wu}]{liu1998pxem}
Liu, C., Rubin, D.~B., and Wu, Y.~N. (1998).
\newblock \enquote{Parameter Expansion to Accelerate {EM}: The {PX-EM}
  Algorithm.}
\newblock {\em Biometrika\/}, 85(4): 755--770.
\endbibitem

\bibitem[{Liu and Wu(1999)}]{liu1999parameter}
Liu, J.~S. and Wu, Y.~N. (1999).
\newblock \enquote{Parameter Expansion for Data Augmentation.}
\newblock {\em Journal of the American Statistical Association\/}, 94(448):
  1264--1274.
\endbibitem

\bibitem[{Menictas et~al.(2019)Menictas, Di~Credico, and
  Wand}]{menictas2019streamlined}
Menictas, M., Di~Credico, G., and Wand, M.~P. (2019).
\newblock \enquote{Streamlined Variational Inference for Linear Mixed Models
  with Crossed Random Effects.}
\newblock {\em arxiv preprint: 1910.01799\/}.
\endbibitem

\bibitem[{Nash and Varadhan(2011)}]{nash2011unifying}
Nash, J.~C. and Varadhan, R. (2011).
\newblock \enquote{Unifying Optimization Algorithms to Aid Software System
  Users: optimx for R.}
\newblock {\em Journal of Statistical Software\/}, 43(9): 1--14.
\endbibitem

\bibitem[{Ormerod and Wand(2012)}]{ormerod2012gva}
Ormerod, J.~T. and Wand, M.~P. (2012).
\newblock \enquote{Gaussian Variational Approximate Inference for Generalized
  Linear Mixed Models.}
\newblock {\em Journal of Computational and Graphical Statistics\/}, 21(1):
  2--17.
\endbibitem

\bibitem[{Ornstein(2020)}]{ornstein2019stacked}
Ornstein, J.~T. (2020).
\newblock \enquote{Stacked Regression and Poststratification.}
\newblock {\em Political Analysis\/}, 28(2): 293--301.
\endbibitem

\bibitem[{Park et~al.(2004)Park, Gelman, and Bafumi}]{park2004mrp}
Park, D.~K., Gelman, A., and Bafumi, J. (2004).
\newblock \enquote{Bayesian Multilevel Estimation with Poststratification:
  State-Level Estimates from National Polls.}
\newblock {\em Political Analysis\/}, 12(4): 375--385.
\endbibitem

\bibitem[{Polson et~al.(2013)Polson, Scott, and Windle}]{polson2013polyagamma}
Polson, N.~G., Scott, J.~G., and Windle, J. (2013).
\newblock \enquote{Bayesian Inference for Logistic Models Using
  P{\'o}lya–Gamma Latent Variables.}
\newblock {\em Journal of the American Statistical Association\/}, 108(504):
  1339--1349.
\endbibitem

\bibitem[{Rabe-Hesketh and Skrondal(2008)}]{skrondal2008multilevel}
Rabe-Hesketh, S. and Skrondal, A. (2008).
\newblock {\em Multilevel and Longitudinal Modeling using {STATA}\/}.
\newblock {STATA} Press.
\endbibitem

\bibitem[{Rabe-Hesketh et~al.(2004)Rabe-Hesketh, Skrondal, and
  Pickles}]{rabehesketh2004gllamm}
Rabe-Hesketh, S., Skrondal, A., and Pickles, A. (2004).
\newblock \enquote{Generalized Multilevel Structural Equation Modeling.}
\newblock {\em Psychometrika\/}, 69(2): 167--190.
\endbibitem

\bibitem[{Ruiz and Titsias(2019)}]{ruiz2019contrastive}
Ruiz, F.~J. and Titsias, M.~K. (2019).
\newblock \enquote{A Contrastive Divergence for Combining Variational Inference
  and MCMC.}
\newblock In {\em International Conference on Machine Learning\/}.
\endbibitem

\bibitem[{Salimans et~al.(2015)Salimans, Kingma, and
  Welling}]{salimans2015markov}
Salimans, T., Kingma, D., and Welling, M. (2015).
\newblock \enquote{Markov Chain Monte Carlo and Variational Inference: Bridging
  the Gap.}
\newblock In {\em International Conference on Machine Learning\/}.
\endbibitem

\bibitem[{Steenbergen and Jones(2002)}]{steenbergen2002multilevel}
Steenbergen, M.~R. and Jones, B.~S. (2002).
\newblock \enquote{Modeling Multilevel Data Structures.}
\newblock {\em American Journal of Political Science\/}, 46(1): 218--237.
\endbibitem

\bibitem[{Stegmueller(2013)}]{stegmueller2013multilevel}
Stegmueller, D. (2013).
\newblock \enquote{How Many Countries for Multilevel Modeling? A Comparison of
  Frequentist and Bayesian Approaches.}
\newblock {\em American Journal of Political Science\/}, 57(3): 748--761.
\endbibitem

\bibitem[{Tan and Nott(2013)}]{tan2013variational}
Tan, L.~S. and Nott, D.~J. (2013).
\newblock \enquote{Variational Inference for Generalized Linear Mixed Models
  using Partially Noncentered Parametrizations.}
\newblock {\em Statistical Science\/}, 28(2): 168--188.
\endbibitem

\bibitem[{Tan(2021)}]{tan2021rvb}
Tan, L. S.~L. (2021).
\newblock \enquote{Use of Model Reparametrization to Improve Variational
  Bayes.}
\newblock {\em Journal of the Royal Statistical Society: Series B (Statistical
  Methodology)\/}, 83(1): 30--57.
\endbibitem

\bibitem[{Tausanovitch and Warshaw(2014)}]{tausanovitch2014representation}
Tausanovitch, C. and Warshaw, C. (2014).
\newblock \enquote{Representation in Municipal Government.}
\newblock {\em American Political Science Review\/}, 108: 605--641.
\endbibitem

\bibitem[{Van~der Laan et~al.(2007)Van~der Laan, Polley, and
  Hubbard}]{van2007super}
Van~der Laan, M.~J., Polley, E.~C., and Hubbard, A.~E. (2007).
\newblock \enquote{Super Learner.}
\newblock {\em Statistical Applications in Genetics and Molecular Biology\/},
  6(1).
\endbibitem

\bibitem[{Van~Dyk and Meng(2001)}]{van2001art}
Van~Dyk, D.~A. and Meng, X.-L. (2001).
\newblock \enquote{The Art of Data Augmentation.}
\newblock {\em Journal of Computational and Graphical Statistics\/}, 10(1):
  1--50.
\endbibitem

\bibitem[{Vehtari et~al.(2017)Vehtari, Gelman, and
  Gabry}]{vehtari2017practical}
Vehtari, A., Gelman, A., and Gabry, J. (2017).
\newblock \enquote{Practical Bayesian Model Evaluation using Leave-One-Out
  Cross-Validation and {WAIC}.}
\newblock {\em Statistics and Computing\/}, 27(5): 1413--1432.
\endbibitem

\bibitem[{Wand and Ripley(2020)}]{wand2020kernsmooth}
Wand, M. and Ripley, B. (2020).
\newblock \enquote{KernSmooth: Functions for kernel smoothing for Wand \& Jones
  (1995).}
\newblock {\em R package version\/}, 2.23-18.
\endbibitem

\bibitem[{Warshaw and Rodden(2012)}]{warshaw2012district}
Warshaw, C. and Rodden, J. (2012).
\newblock \enquote{How Should We Measure District-Level Public Opinion on
  Individual Issues?}
\newblock {\em The Journal of Politics\/}, 74(1): 203--219.
\endbibitem

\bibitem[{Yao et~al.(2018)Yao, Vehtair, Simpson, and Gelman}]{yao2018work}
Yao, Y., Vehtair, A., Simpson, D., and Gelman, A. (2018).
\newblock \enquote{Yes, but Did It Work? Evaluating Variational Inference.}
\newblock In {\em International Conference on Machine Learning\/}.
\endbibitem

\bibitem[{Yin and Zhou(2018)}]{yin2018semi}
Yin, M. and Zhou, M. (2018).
\newblock \enquote{Semi-Implicit Variational Inference.}
\newblock In {\em International Conference on Machine Learning\/}.
\endbibitem

\bibitem[{Zhao et~al.(2006)Zhao, Staudenmayer, Coull, and Wand}]{zhao2006mlm}
Zhao, Y., Staudenmayer, J., Coull, B.~A., and Wand, M.~P. (2006).
\newblock \enquote{General Design {Bayesian} Generalized Linear Mixed Models.}
\newblock {\em Statistical Science\/}, 21(1): 35--51.
\endbibitem

\end{thebibliography}

\clearpage

\renewcommand{\theequation}{A.\arabic{equation}}
\renewcommand\thefigure{A.\arabic{figure}}    
\renewcommand\thetable{A.\arabic{table}}    

\appendix

\section{Derivation of Variational Algorithms}
\label{section_app:pg}

\subsection{Alternative Notations: Gelman and Hill and Plate Diagram}

Equation~\ref{eq:model_gh} expresses the generative model in Equation~\ref{eq:model_gendesign} using the popular notation in \citet{gelman2006multi} where $\bm{\alpha}_{j,g[i]}$ denotes the random effect $j$ for group $g$ of which $i$ is a member.
\begin{align}
\label{eq:model_gh}
\begin{split}
y_i | \bm{\beta}, \{\{\bm{\alpha}_{j,g}\}_{g=1}^{g_j}\}_{j=1}^J &\sim \mathrm{Binom}(n_i, p_i), \quad p_i = \frac{\exp(\psi_i)}{1+\exp(\psi_i)}, \quad \psi_i = \bm{x}_i^T\bm{\beta} + \sum_{j=1}^J \bm{z}^b_{i,j} \bm{\alpha}_{j, g[i]} \\
&\bm{\alpha}_{j,g} | \bm{\Sigma}_j \sim N(\bm{0}_{d_j}, \bm{\Sigma}_j), \quad \bm{\Sigma}_j \sim \mathrm{IW}(\nu_j, \bm{\Phi}_j) \quad \forall (j,g), \quad p(\bm{\beta}) \propto 1
\end{split}
\end{align}

A plate diagram is shown below: Three arrows from $\bm{z}^{b}_{i,j}$, $\bm{\alpha}_{j,g}$ and $\bm{m}_{i,j}$ intersecting to the right of $y_i$ denote picking out the specific random effect corresponding to observation $i$, i.e. $\bm{\alpha}_{j,g[i]}$ in \cite{gelman2006multi} notation.

\begin{figure}[!ht]
	\caption{Plate Diagram of Hierarchical Model}
	\begin{tikzpicture}
	\node[obs] (y) {$y_i$};
	\node[obs, left=of y, xshift=-0.5cm] (x) {$\bm{x}_i$};
	\node[latent, above=of x] (beta) {$\bm{\beta}$}; 
	\factor[left=of y] {fe} {} {x,beta} {y};
	\node[obs, right=of y, xshift=0.5cm] (z) {$\bm{z}^b_{i,j}$};
	\node[obs, below=of z] (m) {$\bm{m}_{i,j}$};
	\node[latent, above=of z, xshift=2cm] (a) {$\bm{\alpha}_{j,g}$};
	\factor[right=of y] {re} {} {z,a,m} {y};
	\node[latent, right=of a] (sigma) {$\bm{\Sigma}_j$};
	\node[const, above=of sigma, yshift=-0.5cm] (nu) {$\nu_j$};
	\node[const, below=of sigma, yshift=0.5cm] (phi) {$\bm{\Phi}_j$};
	{\tikzset{plate caption/.append style={below=10pt of #1.south west}}
		\plate {repg} {(a)} {$g \in \{1, \cdots, g_j\}$} ;}
	
	\plate {repj} {(nu) (m) (z.west)(sigma.east)} {$j \in \{1, \cdots, J\}$} ;
	
	{\tikzset{plate caption/.append style={above=5pt of #1.south west}}
		\plate[] {repi} {(x) (y) (z) (m)} {$i \in \{1, \cdots N\}$};}
	
	\edge {sigma} {a};
	\edge {nu,phi} {sigma};
	\end{tikzpicture}
\end{figure}
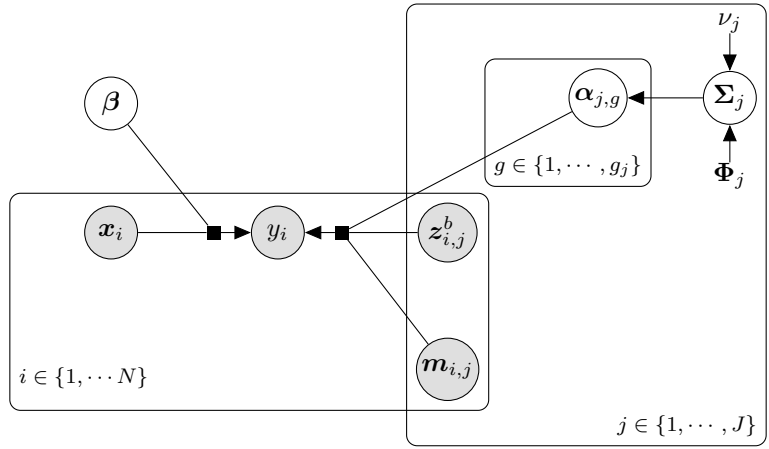

\subsection{Derivation of Variational Algorithms}

Equation~\ref{eq_app:log_complete} provides the log-complete joint density of the model. I assume a flat prior on $\bm{\beta}$ in all of the subsequent derivations where $\Gamma_d$ is the multivariate Gamma function.
\begin{equation}
\label{eq_app:log_complete}
\begin{split}
&\ln p\left(\bm{y}, \bm{\Omega}, \bm{\beta}, \bm{\alpha},\{\bm{\Sigma}_j\}_{j=1}^J\right) = \bm{s}^T\left[\bm{X}\bm{\beta} + \bm{Z}\bm{\alpha}\right] - \frac{1}{2} \left[\bm{X}\bm{\beta} + \bm{Z}\bm{\alpha}\right]^T \bm{\Omega} \left[\bm{X}\bm{\beta} + \bm{Z}\bm{\alpha}\right] + \\
&\left(\sum_i \ln f_{PG}(\omega_i | n_i, 0)\right) + \sum_{j=1}^J \left[-(d_j g_j)/2 \ln(2\pi) - g_j/2 \ln(|\bm{\Sigma}_j|) + \sum_{g=1}^{g_j} -\frac{1}{2} \bm{\alpha}_{jg}^T \bm{\Sigma}_j^{-1} \bm{\alpha}_{jg}\right] +\\
&-\left(\sum_i n_i\right) \ln(2) + \sum_{j=1}^J \ln c(\nu_j , \bm{\Phi}_j) - (\nu_j + d_j + 1)/2 \ln(|\bm{\Sigma}_j|) - \frac{1}{2} \mathrm{tr}\left(\bm{\Phi}_j \bm{\Sigma}_j^{-1} \right) \\
& \mathrm{where}\quad\ln c(\nu_j, \bm{\Phi}_j) = \nu_j/2 \ln |\bm{\Phi}_j| - (\nu_j d)/2 \ln(2) - \ln(\Gamma_d(\nu_j/2))
\end{split} 
\end{equation}

A standard result in variational inference is that, under a mean-field assumption as imposed in Schemes I, II, and III, the optimal approximating distribution is proportional to the exponential of the expectation of Equation~\ref{eq_app:log_complete} over all other parameters (e.g. \citealt{bishop2006ml,blei2017vi}). From this, the updates can be derived as follows.\footnote{Note that a corresponding fully Bayesian Gibbs Sampler can be mostly read-off from these algorithms if one instead samples from the noted variational distribution \emph{and} plugs in the sampled parameters instead of their expectations/variances.}

\begin{itemize}
	\item $q(\bm{\Omega})$. This factorizes into $N$ independent Polya-Gamma variables:
	
	$$q(\omega_i) \sim PG(\tilde{b}_i, \tilde{c}_i)$$
	
	Under Scheme I and Scheme II, the approximating distribution is as follows.
	$$\tilde{b}_i = n_i, \quad \tilde{c}_i = \sqrt{\left[\bm{x}_i^T\tilde{\bm{\mu}}_\beta + \bm{z}_i^T\tilde{\bm{\mu}}_\alpha\right]^2 + \bm{x}_i^T \tilde{\bm{\Lambda}}_\beta \bm{x}_i + \bm{z}_i^T\tilde{\bm{\Lambda}}_\alpha \bm{z}_i}$$
	
	Under Scheme III, this depends on the covariance between $\bm{\alpha}$ and $\bm{\beta}$. I use $\bm{\Lambda}_{\beta-\alpha}$ to denote the estimated covariance matrix from the variational approximation on the stacked vector $[\bm{\beta}^T, \bm{\alpha}^T]$. This is block diagonal under Schemes I and II.
	
	$$\tilde{b}_i = n_i, \quad \tilde{c}_i = \sqrt{\left[\bm{x}_i^T\tilde{\bm{\mu}}_\beta + \bm{z}_i^T\tilde{\bm{\mu}}_\alpha\right]^2 + [\bm{x}_i^T, \bm{z}_i^T] \tilde{\bm{\Lambda}}_{\beta-\alpha} \left[\begin{array}{l} \bm{x}_i \\\bm{z}_i\end{array}\right]}$$
	
	I define the stacked expectation of the Polya-Gammas into a diagonal matrix as follows:
	
	$$\tilde{\bm{\Lambda}}_{\Omega} = \mathrm{diag}\left(\frac{\tilde{b}_i}{2\tilde{c}_i}\tanh(\tilde{c}_i/2)\right)$$
	\item $q(\bm{\Sigma})$. This factorizes into $J$ independent Inverse Wishart distributions whose parameters are as follows where the subscript $jg$ denotes taking the sub vector or matrix corresponding to $\bm{\alpha}_{j,g}$ that is stacked into $\bm{\alpha}$.
	
	$$q(\bm{\Sigma}_j) \sim \mathrm{IW}\left(\tilde{\nu}_j, \tilde{\bm{\Phi}}_j\right)$$
	$$\tilde{\nu}_j = \nu_j + g_j, \quad \tilde{\bm{\Phi}}_j = \bm{\Phi}_j + \sum_{g=1}^{g_j} \left([\tilde{\bm{\mu}}_\alpha]_{jg}\right)\left([\tilde{\bm{\mu}}_\alpha]_{jg}\right)^T + [\tilde{\bm{\Lambda}}_\alpha]_{jg, jg}$$
	
	Note that $E_{q(\bm{\Sigma}_j)}[\bm{\Sigma}_j^{-1}] = \tilde{\nu}_j \left[\tilde{\bm{\Phi}}_j\right]^{-1}$
	\item $q(\bm{\beta}, \bm{\alpha})$. The optimal approximating distribution under Scheme I, II, or III is multivariate normal. The exact details depend on the factorization scheme and are each enumerated below. The stability and convergence of the algorithm improves when all of the mean parameters ($\tilde{\bm{\mu}}_\beta, \tilde{\bm{\mu}}_{\alpha,j}$) are updated jointly.
	\begin{itemize}
		\item Scheme I: 
		$$q(\bm{\beta}) \sim N\left(\tilde{\bm{\mu}}_\beta, \tilde{\bm{\Lambda}}_\beta\right), \quad \tilde{\bm{\Lambda}}_\beta = \left(\bm{X}^T \tilde{\bm{\Lambda}}_{\Omega}\bm{X}\right)^{-1}, \quad \tilde{\bm{\mu}}_\beta = \tilde{\bm{\Lambda}}_\beta \bm{X}^T \left(\bm{s} - \tilde{\bm{\Lambda}}_{\Omega}\bm{Z} \tilde{\bm{\mu}}_\alpha\right)$$
		
		Recall that $\bm{Z} \bm{\alpha} = \sum_{j=1}^J \bm{Z}_j \bm{\alpha}_j$. The approximation distributions $q(\bm{\alpha}_j)$ can be updated cyclically by iterating through $j \in \{1, \cdots, J\}$ with each update as follows
		
		$$q(\bm{\alpha}_j) \sim N\left(\tilde{\bm{\mu}}_{\alpha,j}, \tilde{\bm{\Lambda}}_{\alpha,j}\right), \quad \tilde{\bm{\Lambda}}_{\alpha,j} = \left(\bm{Z}_j^T \tilde{\bm{\Lambda}}_\Omega \bm{Z}_j + \left[\bm{I}_{g_j} \otimes E_{q(\bm{\Sigma}_j)}[\bm{\Sigma}_j^{-1}]\right]\right)^{-1} $$ $$\tilde{\bm{\mu}}_{\alpha,j} = \tilde{\bm{\Lambda}}_{\alpha,j} \bm{Z}_j^T \left(\bm{s} - \tilde{\bm{\Lambda}}_{\Omega}\bm{X}\tilde{\bm{\mu}}_\beta - \sum_{\ell \in \{1, \cdots, J\} \setminus j } \tilde{\bm{\Lambda}}_{\Omega}\bm{Z}_{\ell} \tilde{\bm{\mu}}_{\alpha,\ell}\right)$$
		
		Note that $\tilde{\bm{\Lambda}}_\alpha$ can be made by stacking $\tilde{\bm{\Lambda}}_{\alpha,j}$ block diagonally. $\tilde{\bm{\mu}}_\alpha$ is formed by stacking the $\tilde{\bm{\mu}}_{\alpha,j}$ vertically.
		
		\item Scheme II: The update for $q(\bm{\beta})$ is unchanged from Scheme I. The update for $q(\bm{\alpha})$ is as follows.
		
		\begin{align*}
		q(\bm{\alpha}) &\sim N\left(\tilde{\bm{\mu}}_{\alpha}, \tilde{\bm{\Lambda}}_{\alpha}\right) \\
		\tilde{\bm{\Lambda}}_{\alpha} &= \left(\bm{Z}^T \tilde{\bm{\Lambda}}_\Omega \bm{Z} + \mathrm{blockdiag}\left(\{\bm{I}_{g_j} \otimes E_{q(\bm{\Sigma}_j)}[\bm{\Sigma}_j^{-1}]\}_{j=1}^J\right)\right)^{-1} \\
		\tilde{\bm{\mu}}_{\alpha} &= \tilde{\bm{\Lambda}}_{\alpha} \bm{Z}^T \left(\bm{s} - \tilde{\bm{\Lambda}}_{\Omega}\bm{X}\tilde{\bm{\mu}}_\beta\right)
		\end{align*}
		
		\item Scheme III: $q(\bm{\beta}, \bm{\alpha})$ is no longer assumed to contain any independent components and is updated jointly. 
		
		\begin{align*}
		q(\bm{\beta}, \bm{\alpha}) &\sim N\left(\left[\begin{array}{l} \tilde{\bm{\mu}}_\beta \\ \tilde{\bm{\mu}}_\alpha \end{array}\right], \quad \tilde{\bm{\Lambda}}_{\beta-\alpha}\right), \quad \tilde{\bm{\Lambda}}_{\beta-\alpha} = \left([\bm{X}, \bm{Z}]^T \tilde{\bm{\Lambda}}_\Omega \left[\bm{X}, \bm{Z}\right] + \bm{T}\right)^{-1} \\
		\bm{T} &= \left(\begin{array}{ll} \bm{0}_{p \times p} & \bm{0}_{p \times \sum_j d_j g_j} \\ \bm{0}_{\sum_j d_j g_j \times p} & \mathrm{blockdiag}\left(\{\bm{I}_{g_j} \otimes E_{q(\bm{\Sigma}_j)}[\bm{\Sigma}_j^{-1}]\}_{j=1}^J \right)\end{array}\right) \\
		\left[\begin{array}{l} \tilde{\bm{\mu}}_\beta \\ \tilde{\bm{\mu}}_\alpha \end{array}\right] &= \tilde{\bm{\Lambda}}_{\beta-\alpha} [\bm{X},\bm{Z}]^T \bm{s}
		\end{align*}
	\end{itemize}
\end{itemize}

The ELBO can be expressed as in Equation~\ref{eq_app:elbo_lc} and~\ref{eq_app:elbo_entropy}. It can be decomposed into two parts; the expectation of Equation~\ref{eq_app:log_complete} and the differential entropy of the approximating distribution. The sum of the two terms is the ELBO, i.e. $\mathrm{ELBO} = \mathrm{LogComplete} + \mathrm{Entropy}$.

\begin{equation}
\label{eq_app:elbo_lc}
\begin{split}
&\mathrm{LogComplete} = -\left(\sum_i n_i\right) \ln(2) +\bm{s}^T\left[\bm{X}\tilde{\bm{\mu}}_\beta + \bm{Z}\tilde{\bm{\mu}}_\alpha\right] + \\
&- \frac{1}{2} \left[\bm{X}\tilde{\bm{\mu}}_\beta + \bm{Z}\tilde{\bm{\mu}}_\alpha\right]^T \tilde{\bm{\Lambda}}_\Omega \left[\bm{X}\tilde{\bm{\mu}}_\beta + \bm{Z}\tilde{\bm{\mu}}_\alpha\right] + -\frac{1}{2}\mathrm{tr}\left(\tilde{\bm{\Lambda}}_\Omega \mathrm{Var}(\bm{X}\bm{\beta} + \bm{Z} \bm{\alpha})\right) + \\
& \sum_{i=1}^N E_{q(\omega_i)}\left[\ln f_{PG}(\omega_i | n_i, 0)\right] + \\
&\sum_{j=1}^J \left[\begin{split}-(d_j g_j)/2 \ln(2\pi) -\frac{g_j}{2} E_{q(\bm{\Sigma}_j)}\left[\ln(|\bm{\Sigma}_j|)\right] + \\ -\frac{1}{2} \left[\sum_{g=1}^{g_j}  [\tilde{\bm{\mu}}_{jg}]^T \left(E_{q(\bm{\Sigma}_j)}\bm{\Sigma}_j^{-1}\right) \tilde{\bm{\mu}}_{jg} + \mathrm{tr}\left(E_{q(\bm{\Sigma}_j)}\left[\bm{\Sigma}_j^{-1}\right] \left[\tilde{\bm{\Lambda}}_{\alpha}\right]_{jg}\right)\right]\end{split}\right] +\\
&\sum_{j=1}^J \ln c(\nu_j , \bm{\Phi}_j) - \frac{(\nu_j + d_j + 1)}{2} E_{q(\bm{\Sigma}_j)}\left[\ln(|\bm{\Sigma}_j|)\right] - \frac{1}{2} \mathrm{tr}\left(\bm{\Phi}_j E_{q(\bm{\Sigma}_j)}\left[\bm{\Sigma}_j^{-1}\right] \right)
\end{split}
\end{equation}
\begin{equation}
\label{eq_app:elbo_entropy}
\begin{split}
\mathrm{Entropy} &= \frac{1}{2} \ln\left[2\pi e |\tilde{\bm{\Lambda}}_{\alpha-\beta}|\right] + \\
&\sum_{i=1}^N \frac{\tilde{b}_i \tilde{c}_i}{4}\tanh(\tilde{c}_i/2) - E_{q(\omega_i)}\left[\ln f_{PG}(\omega_i | \tilde{b}_i, 0)\right] - \tilde{b}_i \ln\left[\cosh(\tilde{c}_i/2)\right] + \\
&\sum_{j=1}^J -\ln c(\tilde{\nu}_j, \tilde{\bm{\Phi}}_j) + \frac{\tilde{\nu}_j + d_j + 1}{2} E_{q(\bm{\Sigma}_j)}\left[\ln |\bm{\Sigma}_j|\right] + \frac{1}{2}\mathrm{tr}\left(\tilde{\bm{\Phi}}_j E_{q(\bm{\Sigma}_j)}\left[\bm{\Sigma}_j^{-1}\right]\right)
\end{split}
\end{equation}

Note that neither Equation~\ref{eq_app:elbo_entropy} nor Equation~\ref{eq_app:elbo_lc} are individually computable in closed form as each contains an intractable expectation of the log-density of the Polya-Gamma density as $\tilde{b}_i = n_i$. Fortunately, the terms cancel and thus the ELBO is tractable. Some final intermediate results are necessary to calculate the ELBO.

\begin{itemize}
	\item The entropy of a Polya-Gamma random variable. The key identity follows from Equation 5 in \citet{polson2013polyagamma}.
	
	$$E_{q(\omega | b, c)}[-\ln f_{PG}(\omega | b, c)] = E_{q(\omega | b,c)}\left[\frac{c^2 \omega}{2} - \ln f_{PG}(\omega | b, 0) - b \ln\cosh(c/2)\right] $$
	\item Some expectations over an Inverse Wishart distribution. The results are stated in \citet{tan2013variational} and follow from noting that if $\bm{\Sigma}_j \sim \mathrm{IW}(\nu, \bm{\Phi})$ then $\bm{\Sigma}_j^{-1} \sim \mathrm{Wishart}(\nu, \bm{\Phi}^{-1})$. For $\bm{\Sigma}_j \sim \mathrm{IW}(\nu_0, \bm{\Phi}_0)$ where $\bm{\Sigma}_j$ is $d \times d$:
	
	$$E_{q(\bm{\Sigma}_j)}[\bm{\Sigma}_j^{-1}] = \nu_0 [\bm{\Phi}_0]^{-1}, \quad E_{q(\bm{\Sigma}_j)}[\ln |\bm{\Sigma}_j|] = \ln|\bm{\Phi}_0| - \sum_{k=1}^d \psi((\nu_0-k+1)/2) - d \ln(2)$$

\end{itemize}

\section{Derivation of Parameter Expansions and MAVB}
\label{section_app:param_x}

This section contains a number of results on using MAVB. First, I prove Theorem~\ref{thm:MAVB}. Second, I derive the implementation for a proper working prior ($\bm{\mu}_j \sim N(\bm{0}, \tau^2 \bm{I})$) and note that the limiting case as $\tau^2 \to \infty$ corresponds to the result in the main text.

\subsection{Definitions}

First, I explicitly characterize the definition of parameter expansion more generally below.

\begin{definition}[Parameter Expansion]
	\label{def:px}
	Drawing on the results from \citet{liu1999parameter} and \citet{van2001art}, define a parameter expansion as follows: $\bm{\xi}$ is a $d_\xi$-dimensional parameter that is defined on some space $\mathcal{S}_\xi \subseteq \mathbb{R}^{d_\xi}$. A parameter expansion of the original model $p(\bm{y}, \bm{\theta})$ is defined via a transformation function $t_{\bm{\xi}}$ generating expanded parameters $\bm{\theta}^X$ satisfying the following conditions:
	
	\begin{itemize}
		\item Equivalence: The transformation preserves the likelihood of the observed data
		\begin{equation*}
		p(\bm{y}) = \int p(\bm{y}, \bm{\theta}) d\bm{\theta} = \int p^X(\bm{y}, \bm{\theta}^X | \bm{\xi}) d\bm{\theta}^X \quad \forall \bm{\xi} \in \mathcal{S}_\xi
		\end{equation*}
		\item Reduction: $t_{\bm{\xi}}$ is a one-to-one and differentiable function for $\bm{\xi} \in \mathcal{S}_\xi$ where $t_{\bm{\xi}}(\bm{\theta}^X) = \bm{\theta}$
		
		\item Null Value: There exists some $\bm{\xi}_{\mathrm{Null}} \in \mathcal{S}_\xi$ such that $t_{\bm{\xi}_\mathrm{Null}}(\bm{\theta}^X) = \bm{\theta}^X$
	\end{itemize}
	
	The expanded model has an associated evidence lower bound conditional on $\bm{\xi}$ as follows:
	
	$$\mathrm{ELBO}^{X-\bm{\xi}}_{q(\bm{\theta}^X)} = E_{q(\bm{\theta}^X)}\left[\ln p^X(\bm{y}, \bm{\theta}^X | \bm{\xi})\right] - E_{q(\bm{\theta}^X)}\left[\ln q(\bm{\theta}^X)\right]$$
\end{definition}

\subsection{Proof of Theorem~\ref{thm:MAVB}}

Theorem~\ref{thm:MAVB} can be proved directly following results from \citet{liu1999parameter} and a data processing inequality stated by a number of authors. I explicitly restate those here with notation adapted to this paper.

\begin{lemma}[Liu and Wu -- Theorem 1]
	\label{lemma_app:liuwu}
	For a transformation function $t_{\bm{\xi}}(\bm{\theta}^X)$ satisfying the conditions in Definition~\ref{def:px}, i.e. one-to-one and differentiable for a fixed $\bm{\xi}$, the following result holds: Assume that $\bm{z} \sim p(\bm{z})$ and $\bm{\xi} \sim p_0(\bm{\xi})$. Let $\bm{\xi}_0$ be a random draw from the prior $p_0(\bm{\xi})$ and define $\bm{w} = t_{\bm{\xi}_0}^{-1}(\bm{z})$.
	
	If $\bm{\xi}_1 \sim p(\bm{\xi} | \bm{w}) \propto p(t_{\bm{\xi}}(\bm{w})) |J_{\bm{\xi}}(\bm{w})|p_0(\bm{\xi})$, then $\bm{z}' = t_{\bm{\xi}_1}(\bm{w})$ has the same distribution as $\bm{z}$.
\end{lemma}

\begin{lemma}[Data Processing Inequality - Various]
	\label{lemma_app:dpi}
	
	If $\pi(\bm{y} | \bm{x})$ is a conditional distribution to generate $\bm{y}$ given $\bm{x}$, the following identity holds for any choice of $\pi(\bm{y} | \bm{x})$ assuming that (i) $p(\bm{x})$, $q(\bm{x})$, $p(\bm{y})$, $q(\bm{y})$ are all proper densities and (ii) the KL-divergence between $q(\bm{x})$ and $p(\bm{x})$ is finite.	
	
	\begin{subequations}
		\begin{alignat*}{2}
		&\kl{q(\bm{x})}{p(\bm{x})} \geq \kl{q(\bm{y})}{p(\bm{y})} \\ 
		&\mathrm{where} \quad q(\bm{y}) = \int \pi(\bm{y} | \bm{x}) q(\bm{x}) d\bm{x}, \quad p(\bm{y}) = \int \pi(\bm{y} | \bm{x}) p(\bm{x}) d\bm{x}
		\end{alignat*}
	\end{subequations}
\end{lemma}

The proof of the data processing inequality has been noted by a variety of authors (e.g. \citealt{ruiz2019contrastive} citing \citealt{cover2006elements}). The result also appears to follow from \citet{gerchinovitz2020fanno}'s result on $f$-divergences (Lemma 2.1). For completeness, I show a proof via a manipulation of the KL-divergence below where I assume that $\pi(\bm{y}|\bm{x})$ is a conditional density for sampling $y$ given $x$. This assumes that $p(\bm{x}), p(\bm{y}), q(\bm{x}), q(\bm{y})$ are all well-defined and proper. I also assume the initial KL divergence between $p(\bm{x})$ and $q(\bm{x})$ is finite.

\begin{subequations}
	\begin{alignat}{2}
	\kl{p(\bm{x})}{q(\bm{x})} &= \int \ln\left(\frac{p(\bm{x})}{q(\bm{x})}\right)p(\bm{x})d\bm{x}\\
	&= \int\int \ln\left(\frac{p(\bm{x})\pi(\bm{y}|\bm{x})}{q(\bm{x})\pi(\bm{y}|\bm{x})}\right)p(\bm{x})\pi(\bm{y}|\bm{x})d\bm{x}d\bm{y}\\
	&=E_{p(\bm{y})}\left[E_{p(\bm{x}|\bm{y})}\left[\ln\left(\frac{p(\bm{x})\pi(\bm{y}|\bm{x})}{q(\bm{x})\pi(\bm{y}|\bm{x})}\right)\right]\right]\\
	&=E_{p(\bm{y})}\left[E_{p(\bm{x}|\bm{y})}\left[-\ln\left(\frac{q(\bm{x})\pi(\bm{y}|\bm{x})}{p(\bm{x})\pi(\bm{y}|\bm{x})}\right)\right]\right]\\
	&\geq E_{p(\bm{y})}\left[-\ln\left[E_{p(\bm{x}|\bm{y})}\left(\frac{q(\bm{x})\pi(\bm{y}|\bm{x})}{p(\bm{x})\pi(\bm{y}|\bm{x})}\right)\right]\right] \\
	&= E_{p(\bm{y})}\left[-\ln\left[\int \frac{q(\bm{x})\pi(\bm{y}|\bm{x})}{p(\bm{x})\pi(\bm{y}|\bm{x})} p(\bm{x}|\bm{y}) d\bm{x}\right]\right]\\
	&=E_{p(\bm{y})}\left[-\ln\left[\int \frac{q(\bm{y}) q(\bm{x}|\bm{y})}{p(\bm{y})p(\bm{x} | \bm{y})} p(\bm{x} | \bm{y}) d\bm{x}\right]\right]\\
	&= E_{p(\bm{y})}\left[-\ln\left[\int \frac{q(\bm{y})}{p(\bm{y})} q(\bm{x}|\bm{y})d\bm{x}\right]\right] = E_{p(\bm{y})}\left[-\ln\left[ \frac{q(\bm{y})}{p(\bm{y})}\right]\right]\\
	&= E_{p(\bm{y})}\left[\ln\left(\frac{p(\bm{y})}{q(\bm{y})}\right)\right] = \kl{p(\bm{y})}{q(\bm{y})}
	\end{alignat}
\end{subequations}

With these in mind, and assuming the posterior is proper, applying the procedure in Lemma~\ref{lemma_app:liuwu} maintains the stationarity of the posterior. Thus, it satisfies the conditions in Lemma~\ref{lemma_app:dpi} and reduces the KL divergence between the new approximation $\tilde{q}(\bm{\theta})$ and the posterior target. Noting the following identity, Theorem~\ref{thm:MAVB} follows as $\ln p(\bm{y})$ is unchanged by the transformation as the posterior remains invariant. Note this proof works for other transition kernels that maintain the stationary of the true posterior.

\begin{equation}
\kl{q(\bm{\theta})}{p(\bm{\theta} | \bm{y})} = E_{q(\bm{\theta})}[\ln q(\bm{\theta})] - E_{q(\bm{\theta})}\left[\ln p(\bm{\theta} | \bm{y})\right] = -\mathrm{ELBO}_{q(\bm{\theta})} + \ln p(\bm{y}) 
\end{equation}

\subsection{Applying MAVB}

To apply the above results to the specific case in the paper, I assume a conditionally conjugate prior on the expansion parameter, i.e. $p_0(\bm{\mu}_j) \sim N(\bm{0}, \tau^2 \bm{I}_{d_j})$. Applying Definition~\ref{def:mavb} of MAVB (i.e. following Theorem 1 in \citealt{liu1999parameter}) gives the following Algorithm~\ref{alg:MAVB_working}.

\begin{algorithm}[!ht]
	\caption{MAVB with a Proper Working Prior}
	\label{alg:MAVB_working}
	\begin{algorithmic}
		\State{\textbf{Set the Number of Samples Desired}: $M$}
		\State{\textbf{Estimate $q(\bm{\theta})$ using CAVI (Algorithm~\ref{alg:CAVI})}}
		\For{$m$ in $1, \cdots, M$}
		\State{1. Draw $\bm{\theta}^{(m)} \sim q(\bm{\theta})$}
		\State{2. Draw an expansion parameter from its working prior: $\bm{\mu}^{(m)}_j \sim p_0(\bm{\mu}_j)$}
		\State{3. Create the transformed versions of the parameters as follows}
		$$\bm{\alpha}^X_{j,g} = \bm{\alpha}^{(m)}_{j,g} + \bm{\mu}^{(m)}_j$$
		$$\bm{\beta}^X = \bm{\beta}^{(m)} - \sum_{j=1}^J \bm{M}_j \bm{\mu}^{(m)}_j$$
		
		\State{4. Sample a second expansion parameter $\bm{\mu}_j$ for each $j$}
		$$\tilde{\bm{\mu}}^{(m)}_j | \bm{\alpha}^X_{j,g}, \bm{\Sigma}^{(m)}_j \sim N\left( \left[\left(\bm{\Sigma}^{(m)}_j\right)^{-1} g_j + 1/\tau^2 \bm{I}_{d_j}\right]^{-1}\left[\bm{\Sigma}^{(m)}_j\right]^{-1}\left(\sum_{g=1}^{g_j} \bm{\alpha}^X_{j,g}\right), \left[\left(\bm{\Sigma}^{(m)}_j\right)^{-1} g_j + 1/\tau^2 \bm{I}_{d_j}\right]^{-1}\right)$$
		
		\State{5. Adjust the draws to get the improved sample $\tilde{\bm{\theta}}^{(m)}$}
		$$\tilde{\bm{\alpha}}^{(m)}_{j,g} = \bm{\alpha}^X_{j,g} - \tilde{\bm{\mu}}^{(m)}_j, \quad \tilde{\bm{\beta}}^{(m)} = \bm{\beta}^X + \sum_{j=1}^J \bm{M}_j \tilde{\bm{\mu}}^{(m)}_j$$
		\EndFor
	\end{algorithmic}
\end{algorithm}

The limiting case can be found as $\tau^2 \to \infty$ or by applying the logic of Scheme 2.1 in \cite{liu1999parameter}. A MAVB corresponding to the optimal marginal augmentation scheme and a flat improper prior on $\bm{\mu}_j$ is shown in the main text's Algorithm~\ref{alg:MAVB}. Putting a highly diffuse working prior, e.g. $\tau^2 = 10^4$, should lead to nearly identical results. Note that a highly \emph{informative} working prior, e.g. $\tau^2 = 10^{-6}$, results in effectively no change from applying MAVB. 
\clearpage
\section{Accelerating CAVI using PX-VB and Joint Updates}
\label{section_app:acceleration}

This section notes two simple ways to accelerate convergence of CAVI at limited computational cost. I first derive the methods and show their impact.

\subsection{Derivation of Acceleration Techniques}

First, I derive a new application of PX-VB (\citealt{jaakkola2007parameter}) to hierarchical models. The procedure is re-formulated below in the notation of this paper:

\begin{lemma}[Parameter Expanded Variational Bayes - \citealt{jaakkola2007parameter}]
	\label{lemma:pxvb}
	Given some factorization assumption $\mathcal{X}$, the following procedure converges no slower than the associated CAVI algorithm and maintains a monotonic improvement of the $\mathrm{ELBO}$.
	
	\begin{enumerate}
		\item Perform one step of CAVI (e.g. Algorithm~\ref{alg:CAVI}, Steps 1-4) giving $q(\bm{\theta})$ and $\mathrm{ELBO}_{q(\bm{\theta})}$.
		\item Noting $q(\bm{\theta}) \sim^d q(\bm{\theta}^X)$ when $\bm{\xi} = \bm{\xi}_{\mathrm{Null}}$ and thus $\mathrm{ELBO}^{X-\bm{\xi}_{\mathrm{Null}}}_{q(\bm{\theta})} = \mathrm{ELBO}_{q(\bm{\theta})}$, maximize the $\mathrm{ELBO}^{X-\bm{\xi}}_{q(\bm{\theta})}$ over $\bm{\xi}$.
		
		$$\hat{\bm{\xi}} = \argmax_{\bm{\xi}} \mathrm{ELBO}^{X-\bm{\xi}}_{q(\bm{\theta})} = \argmax_{\bm{\xi}} E_{q(\bm{\theta})}[\ln p^X(\bm{y}, \bm{\theta} | \bm{\xi})] - E_{q(\bm{\theta})}[\ln q(\bm{\theta})] $$
		
		Note that $\mathrm{ELBO}^{X-\hat{\bm{\xi}}}_{q(\bm{\theta})} \geq \mathrm{ELBO}_{q(\bm{\theta})}$.
		\item Apply the reduction function to recover a distribution on the original, non-expanded space. Equivalently, transform $q(\bm{\theta})$ by applying a change-of-variables using $t_{\hat{\bm{\xi}}}(\bm{\theta})$.
		
		$$q'(\bm{\theta}) = \int t_{\hat{\bm{\xi}}}(\bm{\theta}) q(\bm{\theta}) d\bm{\theta}$$
		
		Note that $\mathrm{ELBO}_{q'(\bm{\theta})} = \mathrm{ELBO}^{X-\hat{\bm{\xi}}}_{q(\bm{\theta})}$ and $\mathrm{ELBO}_{q'(\bm{\theta})} \geq \mathrm{ELBO}_{q(\bm{\theta})}$.
	\end{enumerate}
\end{lemma}

The intuition is very similar to why parameter-expanded EM (PX-EM; \citealt{liu1998pxem}) is guaranteed to produce weakly faster convergence. After performing one step of CAVI (i.e. one set of updates to all approximating distributions in the original model, e.g. Algorithm~\ref{alg:CAVI}), one notes that the model has been estimated \emph{assuming} the expansion parameter is at its null value ($\bm{\xi} = \bm{\xi}_{\mathrm{Null}}$), e.g. zero in the case of a location transformation. Thus, by optimizing over $\bm{\xi}$, one must weakly improve the objective and by applying the ``reduction'' function, one returns to a $q'(\bm{\theta})$ in the original variational family with an expansion parameter that is (implicitly) $\bm{\xi}_{\mathrm{Null}}$. Being able to ``move'' in the unidentified space allows for faster convergence. As long as the optimal $\hat{\bm{\xi}} \neq \bm{\xi}_{\mathrm{Null}}$ (e.g. not equal $\bm{0}$ for the location transformation), then applying PX-VB will increase the objective function and thus ensure faster convergence by decreasing the number of iterations required.

The specific algorithm for Scheme I is shown in Algorithm~\ref{alg:PXVB}. Note that the only terms that involve $\bm{\mu}_j$ is the prior on $\bm{\alpha}^X_{j,g}$ as shown by the expanded model (Definition~\ref{def:expanded_hier}). Pleasingly, the closed form solution is quite simple: Center the random effects $\{\bm{\alpha}_{j,g}\}_{g=1}^{g_j}$ to be mean zero and adjust $\bm{\beta}$ correspondingly.

\begin{algorithm}[!ht]
	\caption{Accelerating CAVI for Scheme I (Mean-Expansion)}
	\label{alg:PXVB}
	\begin{algorithmic}
		\State{\textbf{Initialize as per Algorithm 1}}
		\For{$t$ in $1, \cdots, T$}
		\State{1. Perform one step of CAVI (i.e. Updates 1-4 from Algorithm 1) to get $q(\bm{\theta})$}
		\State{2. Maximize $\mathrm{ELBO}^{X-\bm{\xi}}_{q(\bm{\theta})}$ over the expansion parameter $\{\bm{\mu}_j\}_{j=1}^J$.}
		
		$$\{\hat{\bm{\mu}}_j\}_{j=1}^J = \argmax_{\{\bm{\mu}_j\}_{j=1}^J} E_{q(\bm{\alpha}),q(\bm{\Sigma}_j)}\left[\sum_{j=1}^J\left[\sum_{g=1}^{g_j} -\frac{1}{2} \left(\bm{\alpha}_{j,g} - \bm{\mu}_j\right)^T \bm{\Sigma}_j^{-1} \left(\bm{\alpha}_{j,g} - \bm{\mu}_j\right) \right]\right]$$
		$$\hat{\bm{\mu}}_j = \frac{1}{g_j} \sum_{g=1}^{g_j} E_{q(\bm{\alpha}_{j,g})}\left[\bm{\alpha}_{j,g}\right]$$
		\State{3. ``Reduce'' the variational parameters; note that this is guaranteed to result in a weakly higher ELBO than $q(\bm{\theta})$ from Step 1:}
		
		$$\left[\tilde{\bm{\mu}}_{\alpha,j}\right]_g \longleftarrow \left[\tilde{\bm{\mu}}_{\alpha,j}\right]_g - \hat{\bm{\mu}}_j, \quad \tilde{\bm{\mu}}_\beta \longleftarrow \tilde{\bm{\mu}}_\beta + \sum_{j=1}^J \bm{M}_j \hat{\bm{\mu}}_j $$
		
		\State{4. Check for convergence.}
		\EndFor
	\end{algorithmic}
\end{algorithm}

A second way to accelerate the algorithm is to update certain parameter blocks jointly. Note that Algorithm~\ref{alg:CAVI} implies a cyclical rotation through each random effect $j$ to update $\tilde{\bm{\mu}}_{\alpha,j}$ and $\tilde{\bm{\Lambda}}_{\alpha,j}$. Initial experiments showed that dramatic gains could be achieved by updating \emph{all} $\tilde{\bm{\mu}}_{\alpha,j}$ and $\tilde{\bm{\mu}}_\beta$ jointly. This can be done very quickly using a sparse Cholesky decomposition. Given those mean parameters, the $\tilde{\bm{\Lambda}}_{\alpha,j}$ can be updated cyclically. For models considered in this paper, this added little computational cost but dramatically improved convergence.

\subsection{Impact of Acceleration}

\begin{figure}[!ht]
	\caption{ELBO After Acceleration}
	\label{fig:accelerate_CAVI}
	\includegraphics[width=\textwidth]{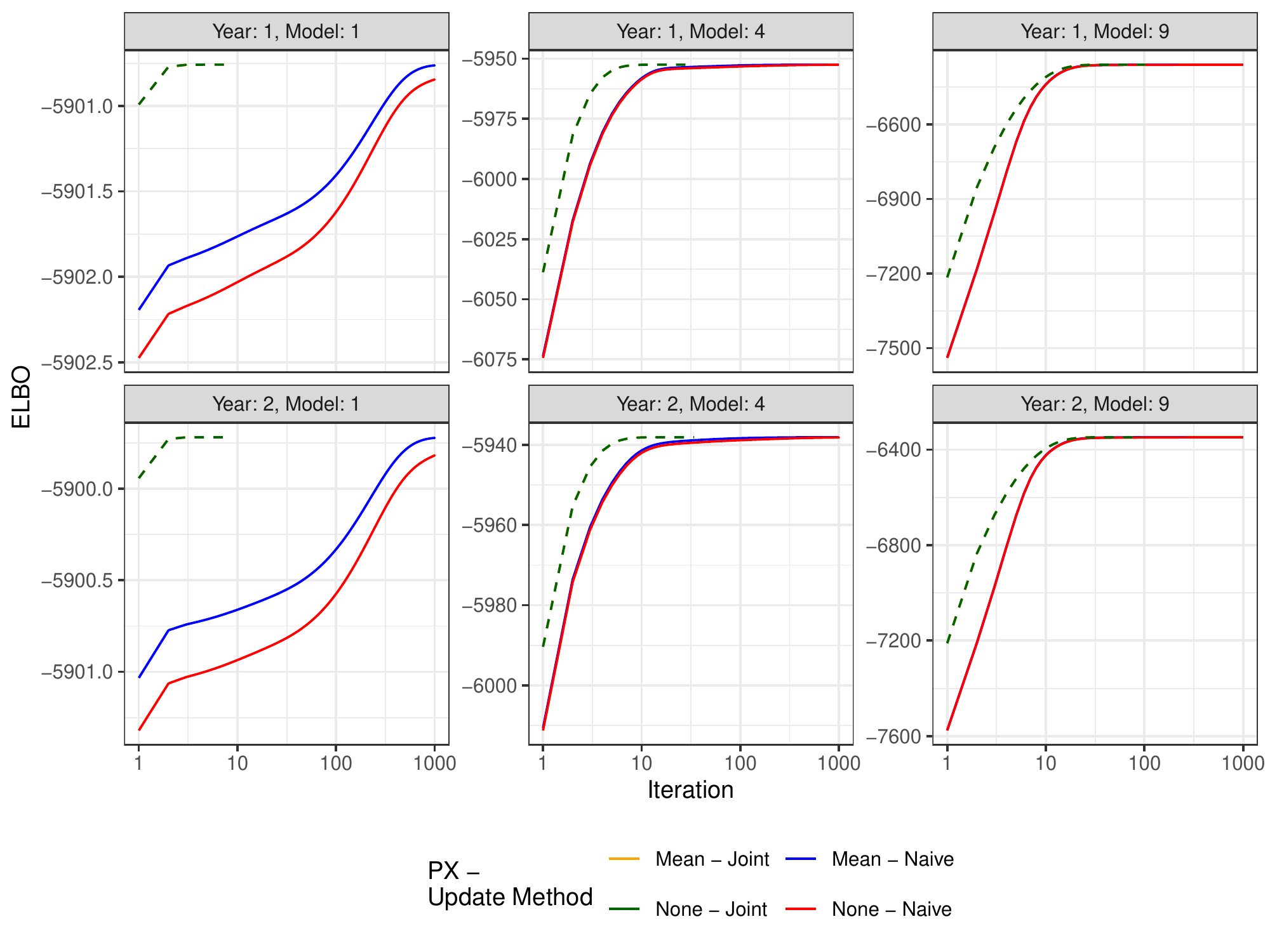}
	\caption*{\footnotesize \emph{Note}: The horizontal axis (number of iterations) is shown on a logged scale to illustrate the differences more clearly. Dashed lines refer to an algorithm that converged within 1,000 iterations.}
\end{figure}

Figure~\ref{fig:accelerate_CAVI} shows the major limitation of the naive CAVI in Algorithm~\ref{alg:CAVI}. I focus on Models 1, 4, and 9 for clarity. Even given 1,000 iterations, it still has not obtained convergence using the threshold of the ELBO changing by less than $10^{-8}$ or all parameters changing by less than $10^{-5}$. Note that only the lines that are \emph{dashed} reached convergence. It compares the two strategies for accelerating discussed above: (i) PX-VB (``Mean'', vs ``None'') and (ii) jointly updating the mean parameters of $q(\bm{\beta}, \bm{\alpha})$ (``Joint'', vs ``Naive'').

The PX-VB algorithm noted above does help performance by a faster increasing ELBO in most models. Jointly updating the mean parameters, however, has a much more considerable impact in all models leading to convergence in sometimes orders of magnitude fewer iterations.

Table~\ref{tab:pxvb} shows the change in the ELBO at the final iteration, i.e. the 1000th iteration or convergence. We see that using PX-VB as defined above after jointly updating the parameters does little, while there is a clear improvement from using PX-VB when applying the naive cyclical updates in Algorithm~\ref{alg:CAVI}. Even in the cases where PX-VB is insufficient to allow convergence, the final change in the ELBO is around x10 smaller (e.g. $10^{-5}$ vs $10^{-4}$) after 1000 iterations.

\begin{table}
	\caption{Final Change in ELBO by Estimation Method}	
	\label{tab:pxvb}	
	\begin{tabular}{lllll}
		\hline\hline
		Estimation Method & Year & Model 1 & Model 4 & Model 9\\	
		\hline	
		None - Joint & 2004 & -7.65 & -8.15 & -8.07 \\
None - Joint & 2008 & -7.61 & -8.05 & -8.09 \\
Mean - Joint & 2004 & -7.65 & -8.15 & -8.07 \\
Mean - Joint & 2008 & -7.61 & -8.05 & -8.09 \\
None - Naive & 2004 & -4.08 & -4.19 & -3.74 \\
None - Naive & 2008 & -3.99 & -4.02 & -3.72 \\
Mean - Naive & 2004 & -4.61 & -5.32 & -5.35 \\
Mean - Naive & 2008 & -4.67 & -5.34 & -5.25 \\

		\hline\hline
		\multicolumn{5}{l}{
			\begin{minipage}{0.7\textwidth}%
				\footnotesize \emph{Note}: The final change in the ELBO (i.e. at convergence or the 1000th iteration) is shown. All numbers can be interpreted as the final change is $10^{x}$, i.e. $\log_{10}$ of the change in the ELBO. Estimation Method is described in the main text and is the type of parameter expansion - the update method for the mean parameters of $q(\bm{\beta}, \bm{\alpha})$.
			\end{minipage}
		}
	\end{tabular}
\end{table}

\begin{figure}[!ht]
	\caption{Drift in $\bar{\bm{\alpha}}_j$}
	\label{fig:drift_CAVI}
	\includegraphics[width=\textwidth]{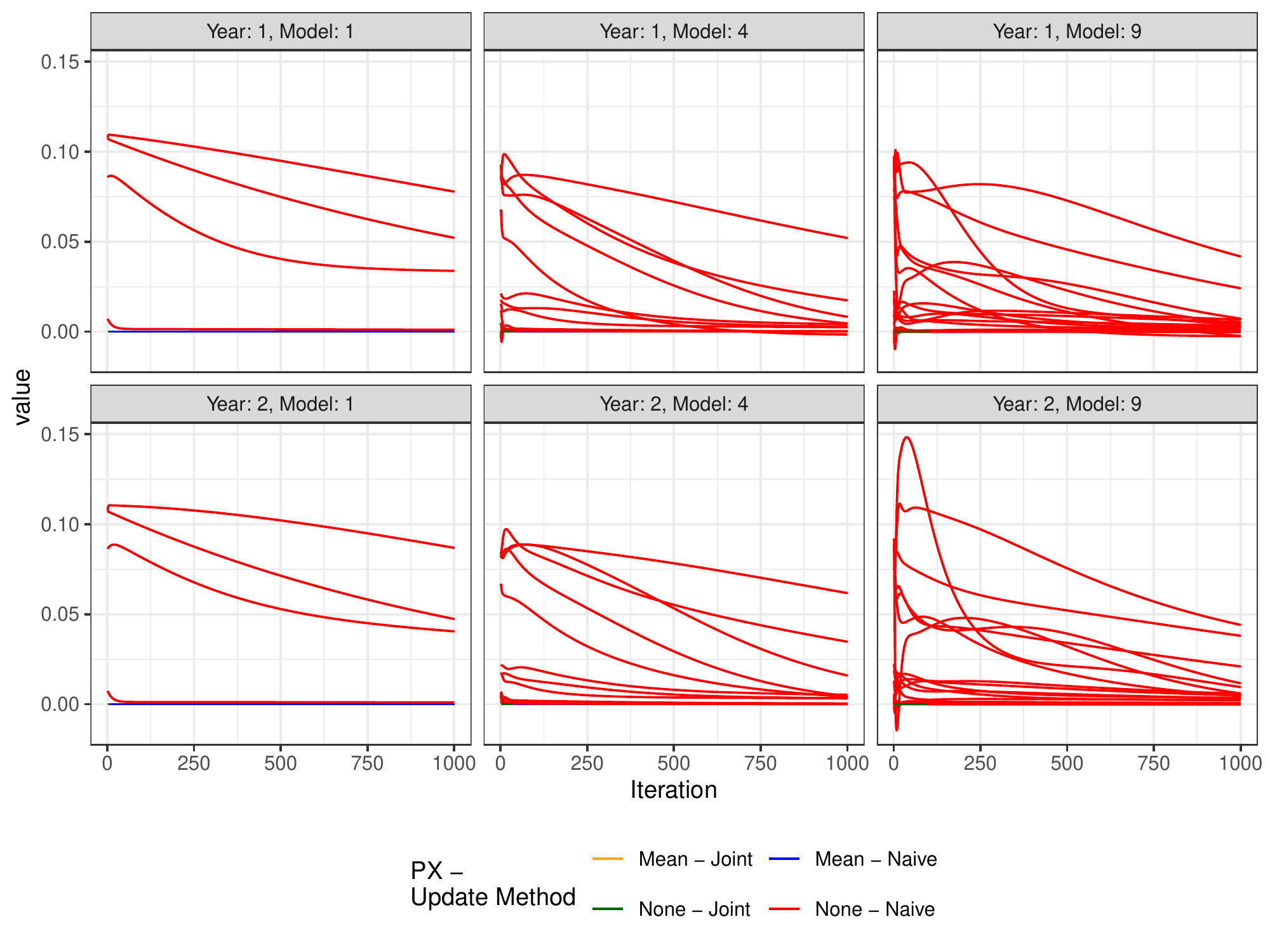}
\end{figure}

Figure~\ref{fig:drift_CAVI} explains why this is the case; it plots the trajectory of the mean of the random effects at each iteration, i.e. $1/g_j \sum_{g=1}^{g_j} [\tilde{\mu}_{\alpha,j}]_g$. It shows starkly that there is a very slow decline towards zero for the naive cyclical algorithm without parameter expansion. By contrast, using PX-VB (Algorithm~\ref{alg:PXVB}) or the joint updates resolves this problem as they are centered at every iteration.

\section{Additional Simulations}
\label{section_app:simulations_extra}

\subsection{Varying Magnitude of Parameters}
I show the model in the main text where the standard deviation of the fixed effects (i.e. $\bm{\beta} \sim N(\bm{0}, \sigma^2_\beta \bm{I})$) and the random effects (i.e. $\alpha_{1,g[i]} \sim N(0, \sigma^2_\alpha)$) varies. In the main text, I only report results where $\sigma_\beta = 0.2$ and $\sigma_\alpha = 1$. The accuracy and coverage are shown below as they are more discriminating across methods.

\begin{figure}[!ht]
	\caption{Varying Size of Parameters}
	\begin{subfigure}[b]{\textwidth}
		\caption{Accuracy}
		\includegraphics[width=\textwidth]{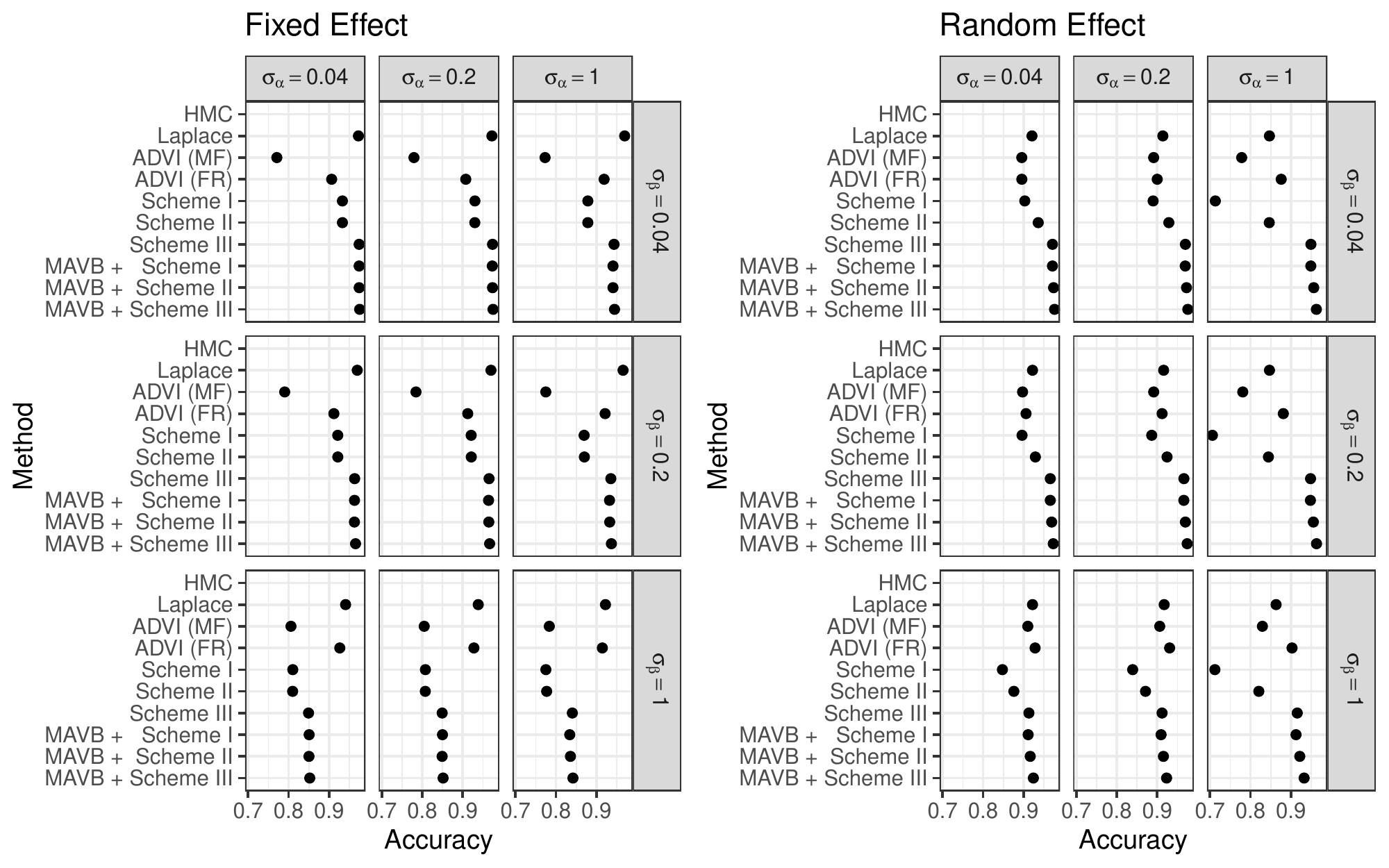}
	\end{subfigure}
	
	\begin{subfigure}[b]{\textwidth}
		\caption{Coverage}
		\includegraphics[width=\textwidth]{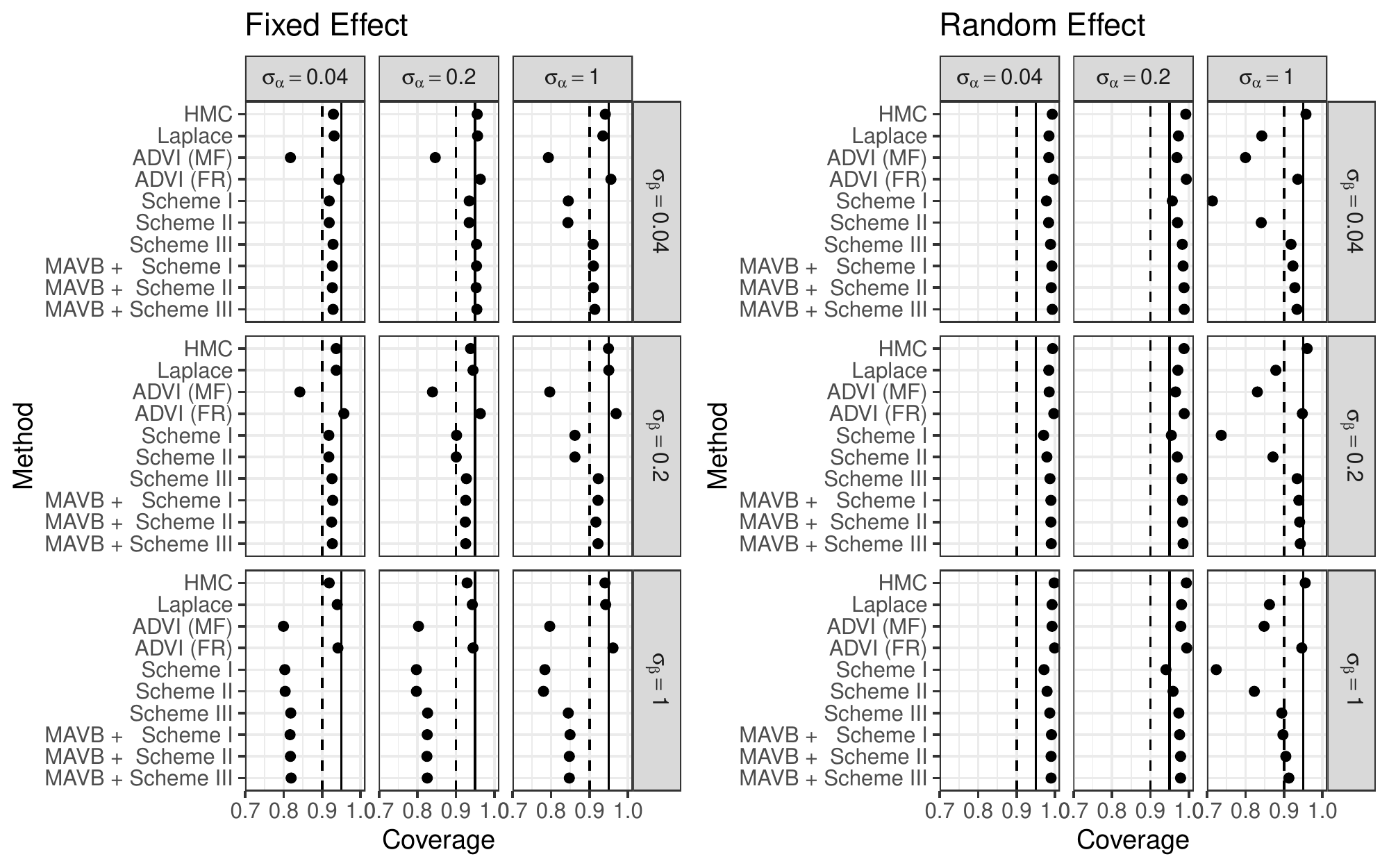}
		\caption*{\footnotesize \emph{Note}: The solid line indicates 0.95 and the dashed line indicates 0.90.}
	\end{subfigure}
\end{figure}

The story is broadly similar to the main text; in cases where $\sigma_\alpha$ is large, Schemes I and II undercover. Applying MAVB improves this and puts coverage to near nominal levels. It similarly has large benefits to accuracy. The one exception is when the distribution of the true fixed effects is wide (i.e. $\sigma_\beta = 1$). In this case, even after applying MAVB, the coverage of the fixed effects is somewhat poor (0.90) although above ADVI. Accuracy also remains noticeably below the Laplace approximation. In those cases (i.e. $\sigma_\beta = 1$), it is worth noting that the linear predictor has a very wide distribution ($\psi_i$ has a 5-95\% interval of around -5.9 to 5.5) and the probability distribution of $p_i$ is highly bimodal around 0 and 1. When $\sigma_\beta = 0.2$, as in the main text, the scale is more plausible for a logistic model with the 5-95\% interval of the linear predictor being roughly between -2.7 and 2.3---with a much more uniform distribution of probabilities $p_i$. This is roughly the scale of the linear predictors observed in bootstrapped data.

\subsection{Bootstrapping Ghitza and Gelman}

To address this and see how the variational schemes perform on real data, I conducted a simulation based on \citet{ghitza2013mrp}. I fit Model 1---$J = 4$ with random effects for age, income, ethnicity and state---and take the parameter estimates from the Laplace approximation as the ``ground truth''. Given those estimates, I generate 100 datasets with the observed covariates from 2004 and simulate a binary outcome for each of the 4,080 observations. Table~\ref{tab:app_gg_sims} reports three measures of the performance; and reports the same figures as in Table~\ref{tab:sims}.

\begin{table}[!ht]
	\caption{Results from Bootstrapped Simulations}
	\label{tab:app_gg_sims}
	\begin{center}
		\begin{tabular}{rlrr|rr|rr|rr}
			\hline\hline
			& & \multicolumn{2}{c}{Bias} & \multicolumn{2}{c}{RMSE} & \multicolumn{2}{c}{Accuracy} & \multicolumn{2}{c}{Coverage} \\
			& & FE & RE & FE & RE & FE & RE \\
			\hline
			 & Laplace & 0.004 & -0.003 & 0.017 & 0.015 & 0.929 & 0.916 & 0.977 & 0.992 \\
 & HMC &  &  &  &  &  &  & 0.977 & 0.994 \\
 & ADVI (MF) & -0.005 & 0.002 & 0.083 & 0.048 & 0.631 & 0.829 & 0.802 & 0.974 \\
 & ADVI (FR) & -0.074 & 0.027 & 0.295 & 0.159 & 0.814 & 0.861 & 0.940 & 0.975 \\
 & Scheme I & 0.003 & -0.003 & 0.016 & 0.014 & 0.659 & 0.836 & 0.913 & 0.985 \\
 & Scheme II & 0.003 & -0.003 & 0.016 & 0.013 & 0.659 & 0.918 & 0.912 & 0.991 \\
 & Scheme III & 0.003 & -0.003 & 0.016 & 0.011 & 0.943 & 0.959 & 0.973 & 0.992 \\
\hline
\multirow{3}{*}{MAVB +} & Scheme I & 0.003 & -0.003 & 0.016 & 0.015 & 0.791 & 0.936 & 0.932 & 0.991 \\
 & Scheme II & 0.002 & -0.003 & 0.016 & 0.013 & 0.793 & 0.943 & 0.930 & 0.991 \\
 & Scheme III & 0.003 & -0.003 & 0.016 & 0.011 & 0.951 & 0.965 & 0.970 & 0.992 \\

			\hline\hline
			\multicolumn{10}{l}{
				\begin{minipage}{\textwidth}
					\footnotesize \emph{Note}: This reports the bias (Bias), root mean squared error (RMSE) of the estimated posterior means against those estimated from HMC. The distance between the distributions (Accuracy) and frequentist coverage (Coverage) are reported; see the main text for an explanation of these measures. The statistics are disaggregated by fixed (FE) and random effects (RE). All results are created using all relevant parameters in each simulation and then averaged across one hundred simulations. ADVI (MF) uses the mean-field approximation; ADVI (FR) uses the full rank approximation in \citet{kuckelbir2017advi}.
				\end{minipage}
			}
		\end{tabular}
	\end{center}
\end{table}

The results are similar to those in the main text. Scheme I without MAVB out-performs ADVI (Mean Field) in terms of Bias, RMSE, and coverage while having approximately the same accuracy. The comparison against ADVI (Full Rank) is more complex; it has a noticeably higher RMSE and bias, although its accuracy is noticeably better than Scheme I. Applying MAVB to Scheme I results in a large improvement in accuracy (around 10\%) and makes it comparable to ADVI (Full Rank). Scheme III continues to perform well having a high accuracy (94-95\%) and out-performs all other approximate methods including both versions of ADVI. Its performance is slightly improved (1-2\%) by MAVB.

\subsection{Checking for Non-Convexities}

Models estimated by variational inference have the possibility of getting stuck in local optima. The default settings in the variational algorithms use a deterministic initialization based on an EM algorithm where the random effects are replaced by a fixed ridge prior. I examine a (reasonable) random initialization where (i) the variance parameters are set to zero (i.e. $\tilde{\bm{\Lambda}}_{\alpha-\beta} = \bm{0}$) and (ii) the mean parameters are randomly sampled from standard independent Gaussians and $\tilde{\bm{\Phi}}_j$ is drawn from a standard Inverse Wishart.

\begin{figure}[!ht]
	\caption{Examining Different Initializations}
	\label{fig:nonconvex}
	\includegraphics[width=\textwidth]{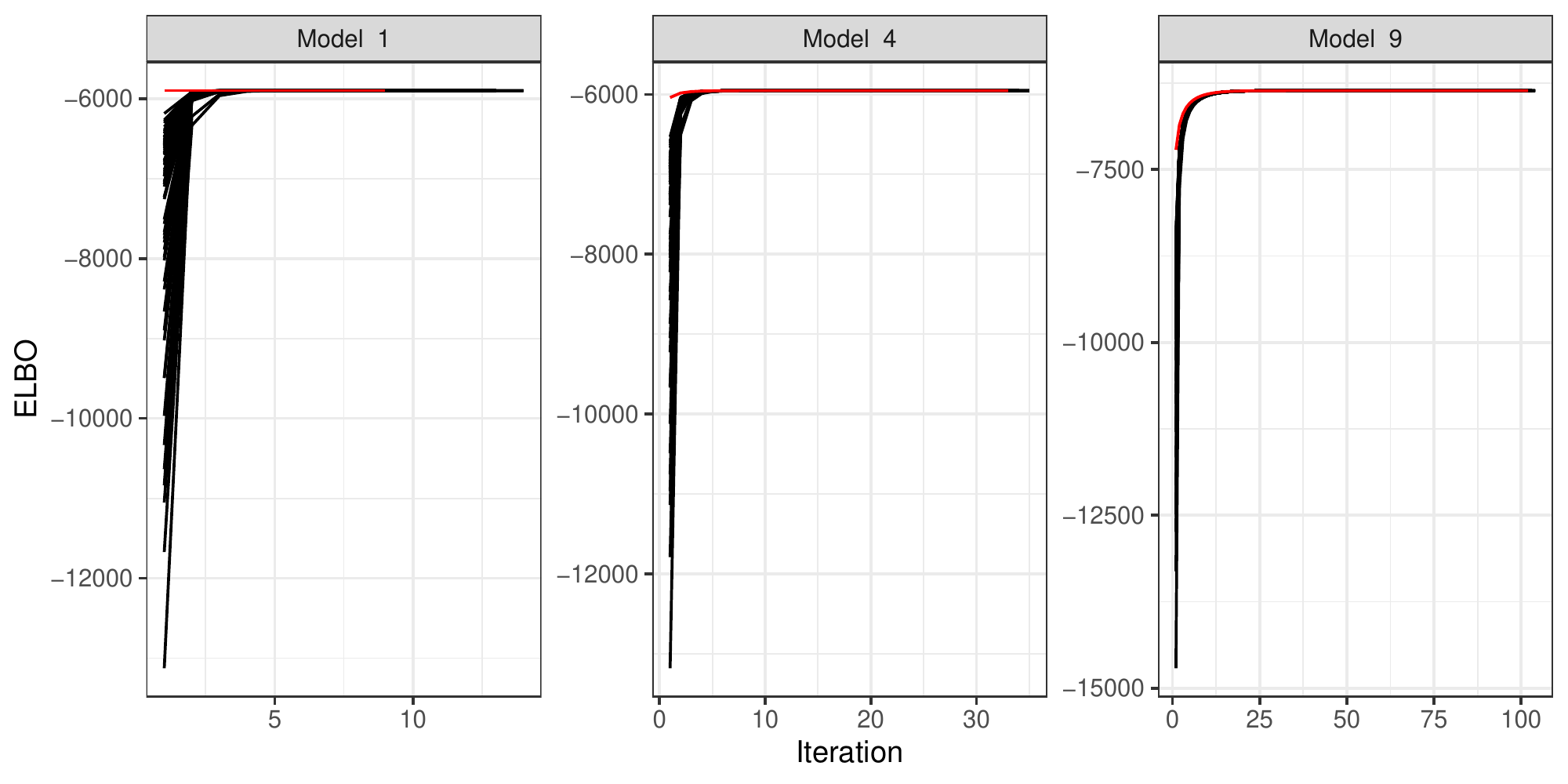}
	\caption*{\footnotesize \emph{Note}: The red line indicates a model initialized via EM; the black lines indicate random initializations.}
\end{figure}

Figure~\ref{fig:nonconvex} shows the results of the EM starting values (in red) versus 100 random initializations using Scheme I on Models 1, 4, and 9. It is clear that all converge to the same value of the ELBO. The maximum difference between any of the random initializations and the EM initialized model is less than $10^{-7}$.

In my other experiments, I did not observe issues with non-convexity although it might occur in other settings or with poor starting values. Thus, it is reasonable for the researcher to check for this in their domain-specific setting. 

\section{Additional Analyses for Ghitza \& Gelman}
\label{section_app:gg_extra}

\begin{figure}[!ht]
	\caption{Disaggregating Run Time between CAVI and MAVB}
	\label{fig:gg_time}
	\includegraphics[width=\textwidth]{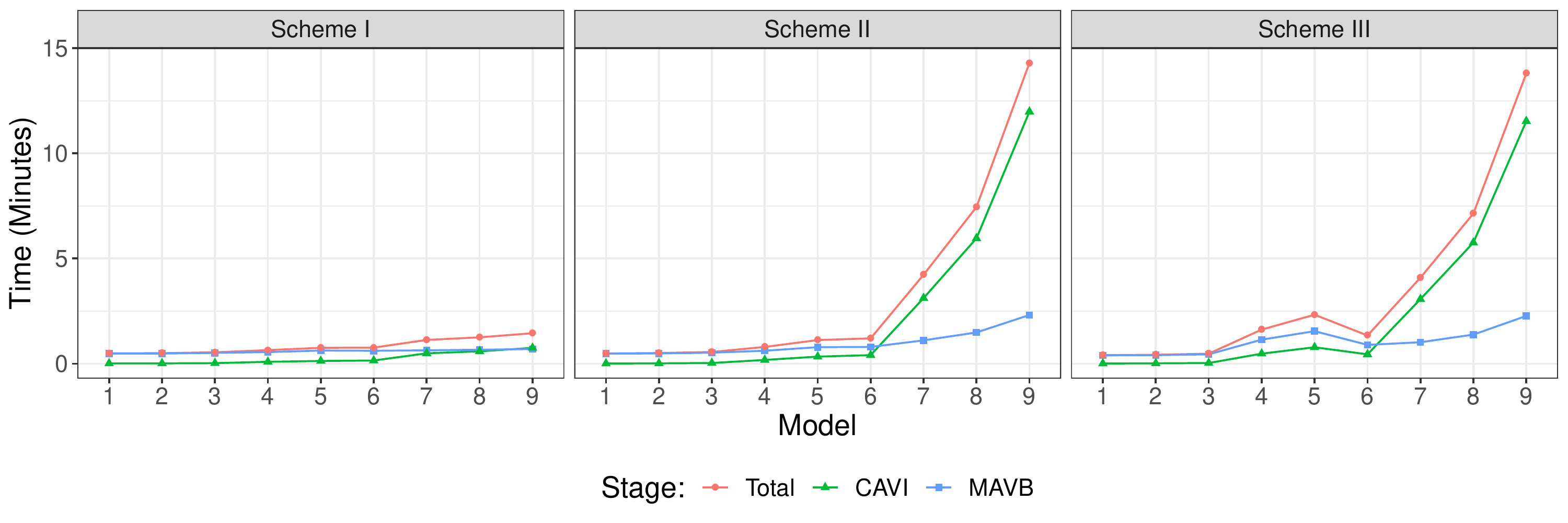}
	\caption*{\footnotesize \emph{Note}: Each figure plots the run-time of Schemes I-III. The reported times are averaged across the 2004 and 2008 elections. The time is broken up into Total, CAVI (parameter estimation), and MAVB (4,000 samples plus transformation). Model 1-9 are described in Table~\ref{tab:gg_models}. All models are fit on a computer with 16 GB of RAM.}
\end{figure}

Figure~\ref{fig:gg_time} shows disaggregated results on the run time for the models shown in Figure~\ref{fig:speed} by breaking the time into that used for estimating the parameters via CAVI and drawing samples and transforming them using MAVB. As noted in the main text, the time for MAVB is the time to draw 4,000 samples to make it comparable to the output of HMC. It shows that MAVB never consumes more than a minute or two of time---although its cost does grow in Schemes II and III when the number of parameters is huge.

\begin{figure}[!ht]
	\caption{Time of CAVI by Scheme and Stage}
	\label{fig:gg_time_stage}
	\includegraphics[width=\textwidth]{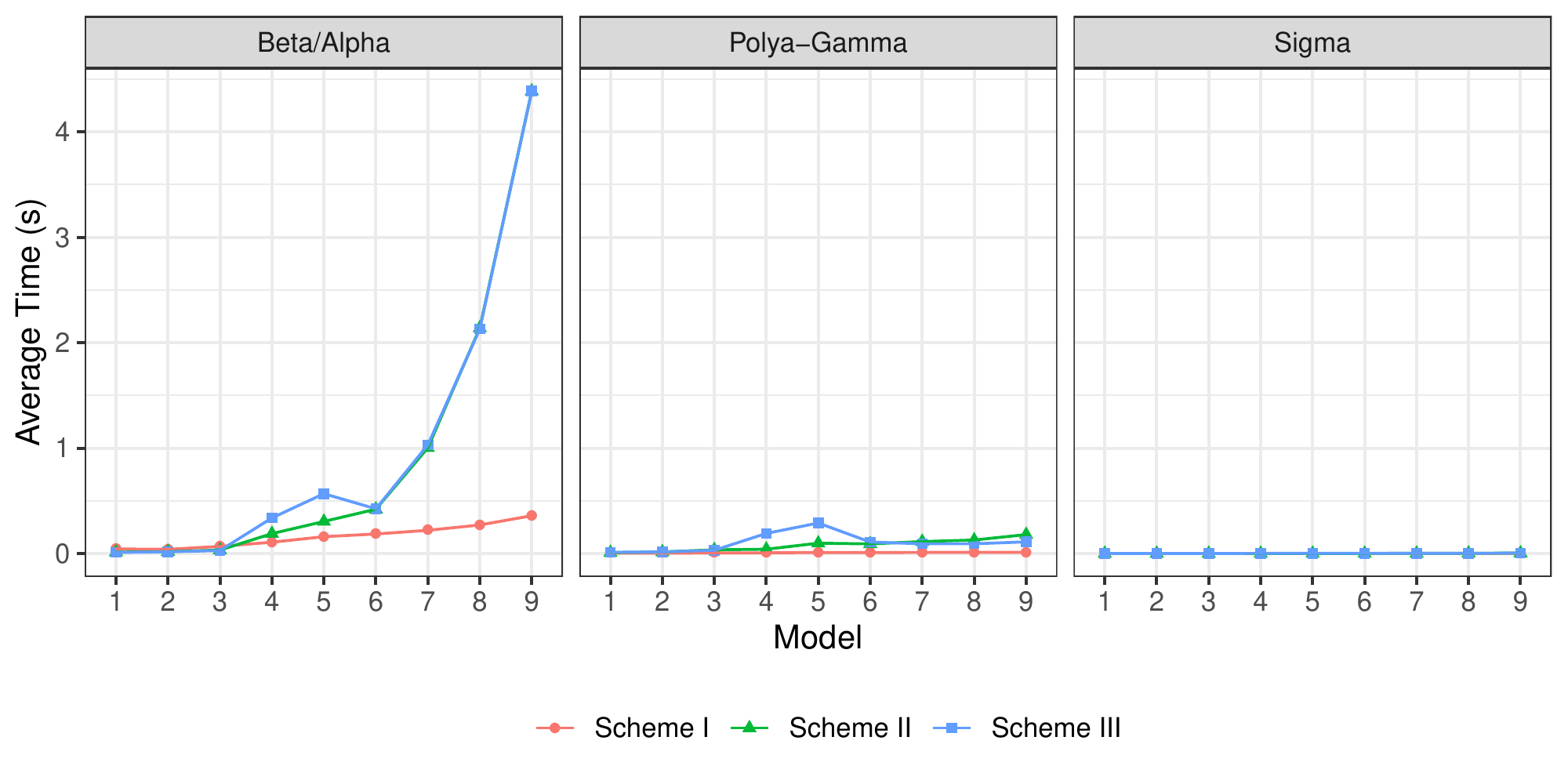}
	\caption*{\footnotesize \emph{Note}: The average time in seconds per each stage of CAVI is noted. ``Beta/Alpha'' refers to updating $q(\bm{\beta},\bm{\alpha})$, ``Polya-Gamma'' refers to $q\left(\{\omega_i\}_{i=1}^N\right)$, and ``Sigma'' refers to $q\left(\{\bm{\Sigma}_j\}_{j=1}^J\right)$.}
\end{figure}

Figure~\ref{fig:gg_time_stage} shows the mean time of each stage, averaged across years, for the nine models in the main paper. It shows that the main increase in run-time for Schemes I, II, and III comes from updating $q(\bm{\beta}, \bm{\alpha})$---although there is a slight increase in the Polya-Gamma update time. Further, it shows that the increase in the size of $\bm{\alpha}$ imposes noticeably more significant costs on Schemes II and III.

\begin{figure}[!ht]
	\caption{Improvements from MAVB on Mean}
	\label{fig:gg_mavb_mean}
	\includegraphics[width=\textwidth]{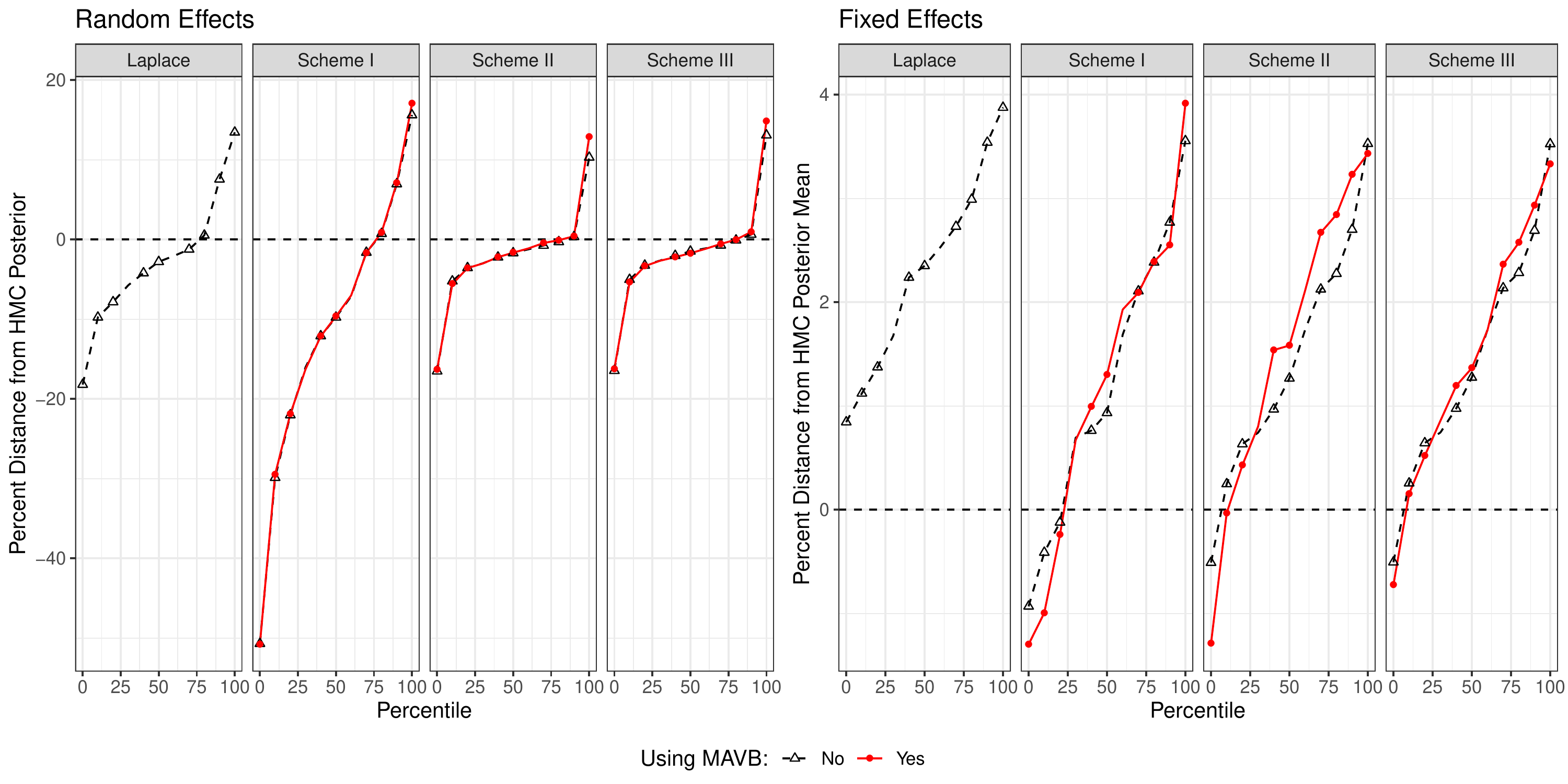}
	\caption*{\footnotesize \emph{Note}: This figure plots the percentile of the percentage gap between the absolute value of the means estimated via Hamiltonian Monte Carlo [HMC] and the approximate methods. A negative value on the vertical axis indicates that the corresponding percentile has a smaller variance than HMC. A vertical shift upward of the line indicates the variance of the parameters has increased. The solid markers indicate the deciles and extremes of the distribution. The dashed line with hollow triangles represents the estimates without using MAVB. The red line with solid circles represents the results after using MAVB.}
\end{figure}

Figure~\ref{fig:gg_mavb_mean} shows the results of applying MAVB to the percentage gap on the absolute value of the means between the approximate methods and Hamiltonian Monte Carlo. As before, the distribution of values is over the statistic aggregated across groups. Note that the scale is much smaller than the corresponding figure on standard deviations in the main text (Figure~\ref{fig:gg_post_var}) especially for the fixed effects.

\begin{figure}[!ht]
	\caption{Disaggregated Improvements from MAVB}
	\label{fig:gg_mavb_disagg}
	\includegraphics[width=\textwidth]{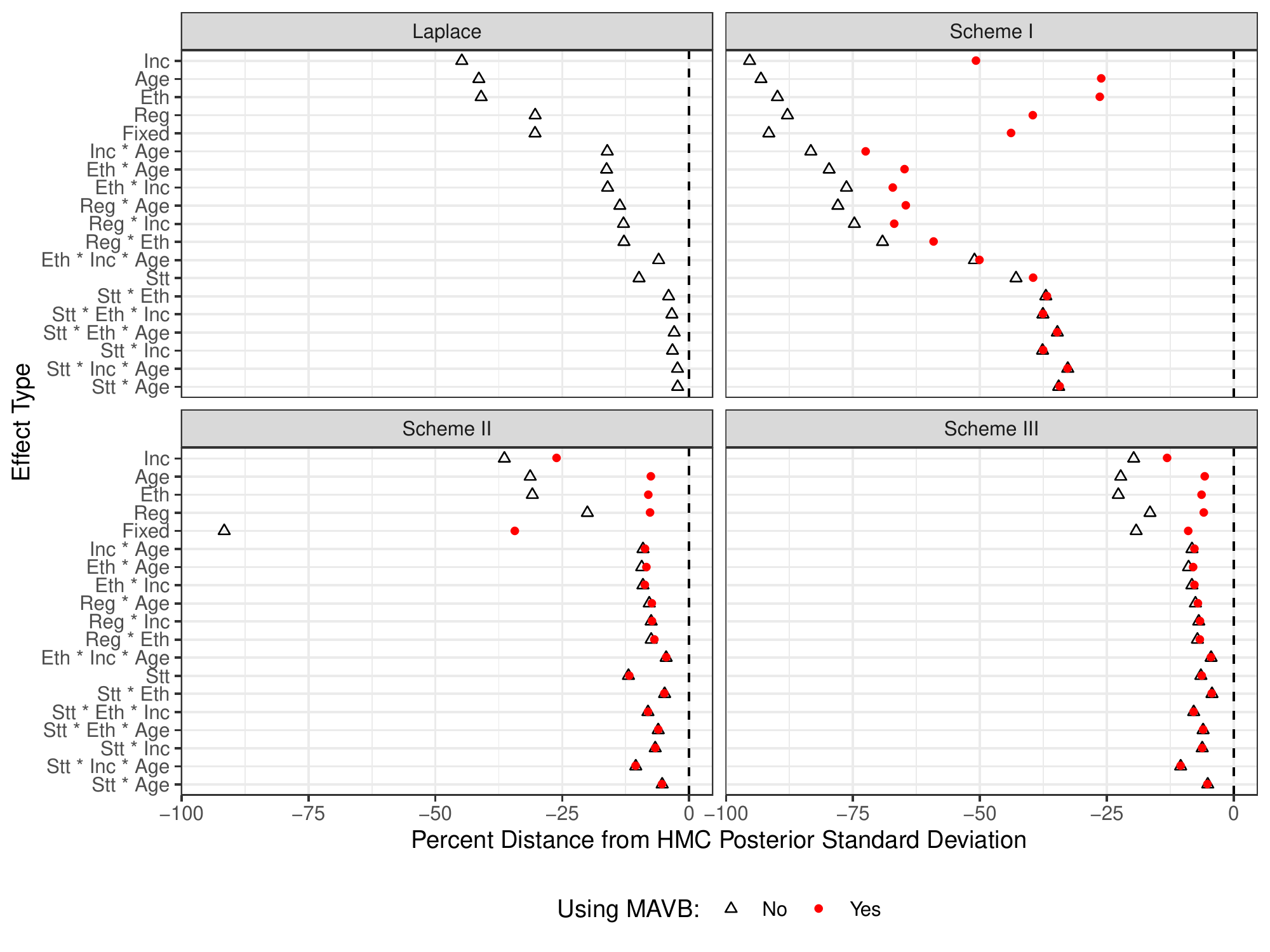}
	\caption*{\footnotesize \emph{Note}: This figure plots the average discrepancy for the standard deviation of each random effect estimated by an approximation method from that coming from Hamiltonian Monte Carlo [HMC]. Negative values indicate that HMC has a larger standard deviation. Hollow triangles represent the estimates without MAVB. Filled circles represents the estimates using MAVB.}
\end{figure}

Figure~\ref{fig:gg_mavb_disagg} shows the average (percentage) discrepancy for each type of random effect. As noted in the main text, it shows that there is severe underestimation of variability for random effects with small numbers of groups---especially age, income and ethnicity. Fortunately, applying MAVB to Scheme I corrects some of this discrepancy giving it variability that is closer to the baseline of Hamiltonian Monte Carlo and beating the Laplace approximation on certain blocks. Scheme III remains systematically close to HMC and clearly out-performs the Laplace approximation on this measure.

\begin{figure}[!ht]
	\caption{Analysis of Linear Predictor}  
	\label{fig:vis_lp}
	\includegraphics[width=\textwidth]{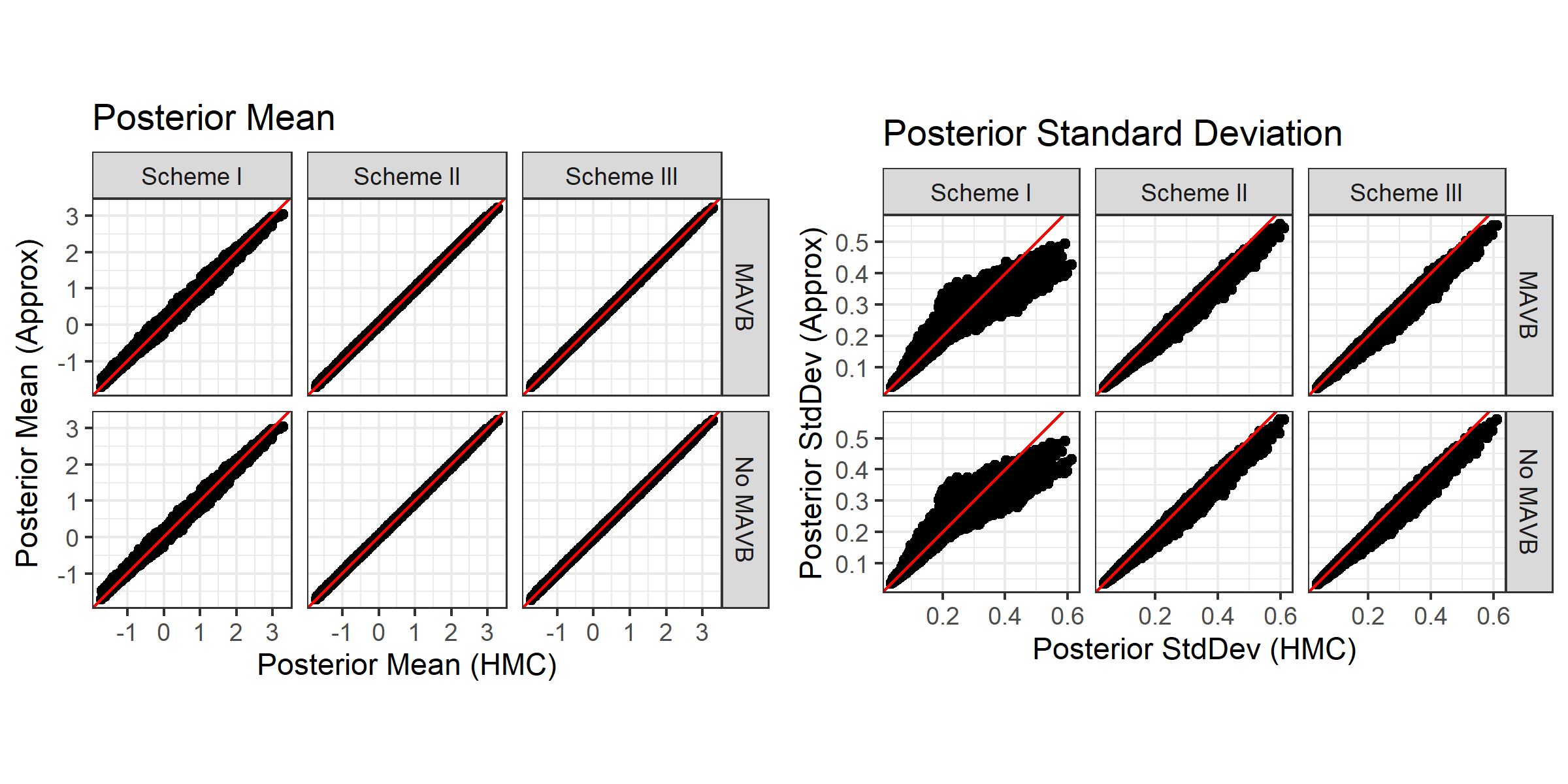}
\end{figure}

Figure~\ref{fig:vis_lp} shows the correlation for the linear predictor (i.e. $\bm{x}_i^T\bm{\beta} + \bm{z}_i^T\bm{\alpha}$) against HMC---showing both mean and standard deviation. I focus only on the observations where $n_i > 0$ as \texttt{brms} does not use observations where $n_i = 0$ in estimating the model. For the mean, there is very close correspondence as should be expected from the tight correspondence of the means shown elsewhere. For the standard deviation, there is more variable performance. Scheme I has a small bias -0.013 (or 2\%), a high correlation (0.96) but clearly quite wide variability. Schemes II and III are much closer (0.99 correlation; visually quite tight) if slightly systematically under-estimating variability. MAVB adds little benefit here.

\begin{figure}[!ht]
	\caption{``Gold Standard'' CV}
	\label{fig:gg_gold_CV}
	\includegraphics[width=\textwidth]{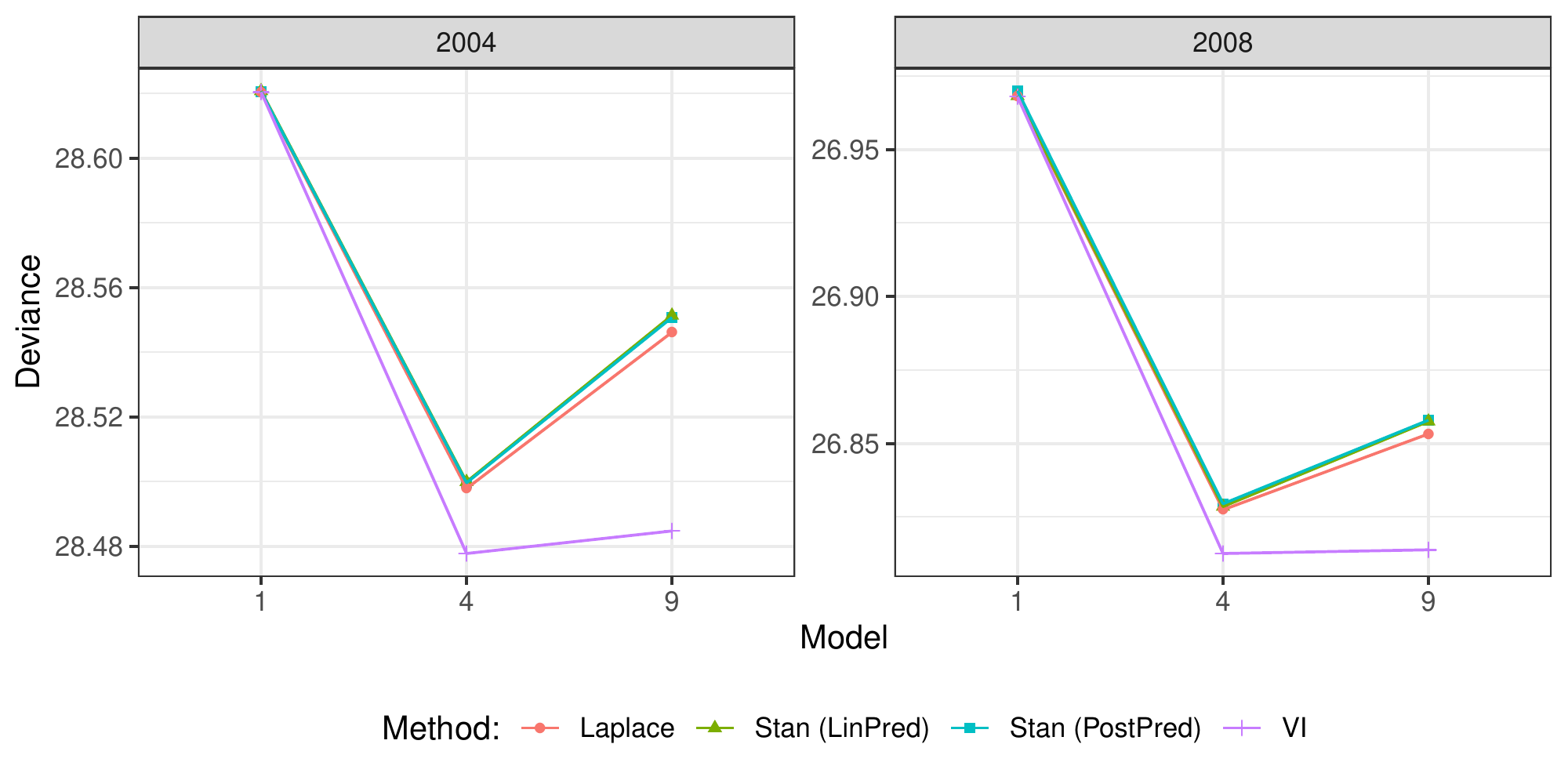}
	\caption*{\footnotesize \emph{Note}: This plots the mean out-of-sample deviance.}
\end{figure}

Figure~\ref{fig:gg_gold_CV} shows the result of a second cross-validation where all models (HMC, Laplace, VI [Scheme I]) are fit on ten folds and out-of-sample predictions are generated. It reports the deviance, as defined in the main text. All methods select Model 4 over Model 9 although less clearly in the case of Scheme I in 2008. Note that the folds here are separate from those in the main text; thus, it is reassuring that Model 4 is again selected by cross-validation. ``Stan (LinPred)'' refers draws of the expected outcome (i.e. the linear predictor draws pushed through the logistic link and divided by $n_i$ to get a probability; \texttt{posterior\_linpred(..., transform = TRUE)}). ``Stan(PostPred)'' refers to averaged draws of the posterior predictive distribution (\texttt{posterior\_predict(...)}). The two measures are nearly identical with a mean absolute error of 0.003 on the probability scale.

\begin{figure}[!ht]
	\caption{Correlation of Out-of-Sample Predictions}
	\label{fig:gg_gold_cor}
	\includegraphics[width=\textwidth]{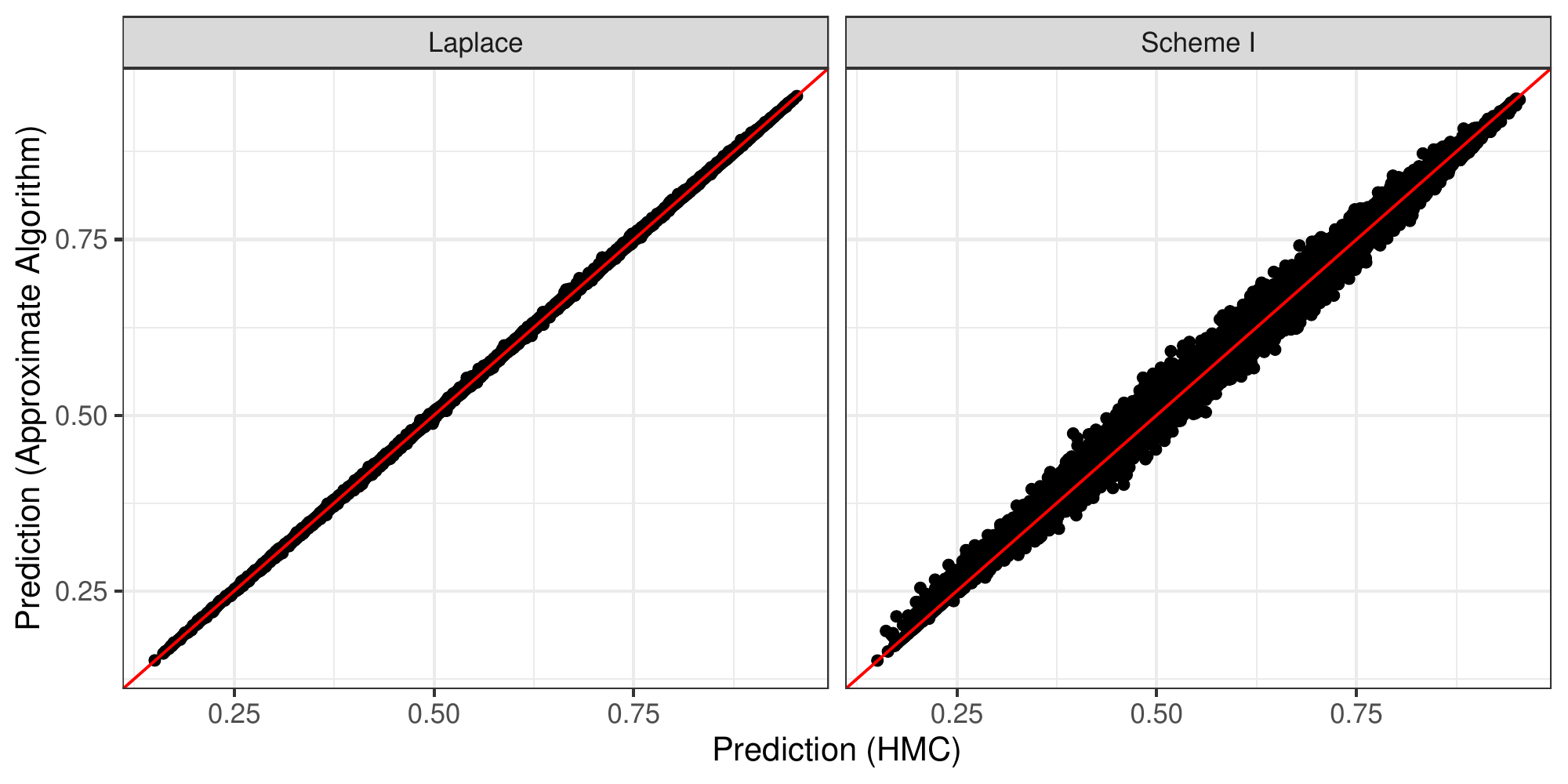}
	\caption*{\footnotesize \emph{Note}: The out-of-sample predicted probabilities $p_i$ are shown. A red line demarks the 45-degree line.}
\end{figure}

Finally, Figure~\ref{fig:gg_gold_cor} shows the correlation between the out-of-sample predictions between HMC and the Laplace and Scheme I methods. As expected from the main results, the Laplace predictions are nearly identical to the HMC predictions while the Scheme I are highly correlated ($\rho = 0.998$) but clearly somewhat more noisy.

\end{document}